\documentclass[fleqn,usenatbib]{mnras}
\usepackage{amsmath}
\usepackage{newtxmath}
\usepackage[T1]{fontenc}
\usepackage[latin9]{inputenc}
\usepackage{graphicx}
\usepackage[authoryear]{natbib}

\makeatletter

\providecommand{\tabularnewline}{\\}
\newcommand{\lyxdot}{.}

\providecommand{\tabularnewline}{\\}

\usepackage{newtxtext}

\def\kms{km~s$^{-1}$}

\def\etal{{et al. }}

\DeclareRobustCommand{\VAN}[3]{#2}
\let\VANthebibliography\thebibliography
\def\thebibliography{\DeclareRobustCommand{\VAN}[3]{##3}\VANthebibliography}

\title[Short title, max. 45 characters]{MNRAS \LaTeXe\ template -- title goes here}

\title{The galaxy cluster AC114. II. Stellar populations and the mass-metallicity relation}

\author[I. Saviane et al.]{
Ivo Saviane,$^{1}$\thanks{E-mail: ivo.saviane@eso.org}
Irina Yegorova,$^{2}$
Dominique Proust$^{3}$
\\
$^{1}$ European Southern Observatory, Alonso de Cordova 3107, Vitacura, Casilla 19001, Santiago de Chile 19, Chile. \\
$^{2}$ Departamento de Ciencias Fisicas, Universidad Andres Bello, Fernandez Concha 700, Las Condes, Santiago, Chile\\
$^{3}$ GEPI, Observatoire de Paris, F92195 Meudon Principal CEDEX, France
}

\date{Accepted XXX. Received YYY; in original form ZZZ}

\pubyear{2022}

\makeatother

\begin{document}
\label{firstpage} \pagerange{\pageref{firstpage}--\pageref{lastpage}}
\maketitle

\begin{abstract}
We investigate the mass-metallicity relation for galaxies in the Abell
cluster AC114 from 7~hours of VIMOS/MR data collected at the ESO-VLT
telescope in 2009. The dynamical analysis completed in our previous
paper 
allowed us to select cluster members, whose spectra are here analyzed
with stellar population synthesis models. 
Active and passive galaxies are identified based on the presence/absence
of the {[}\ion{O}{II}{]}$\lambda3727$, {[}\ion{O}{III}{]}$\lambda\lambda4959,5007$
and/or H$\beta$ emission lines, depending on the galaxy redshift.
We find that active galaxies have lower average masses than passive
ones, and have lower average metallicities. The mass-metallicity relation
(MZR) of the cluster is found to be steeper than that for galaxies
in the local universe. 
\end{abstract}



\section{Introduction}

The mass-metallicity relation (MZR) is one of the fundamental constraints
to galaxy evolution, and for this reason it is the subject of a substantial
body of literature \citep[see ][for a review]{Maiolino2019DeReMetallicatheCosmicChemicalEvolutionofGalaxies},
with a section of that literature looking at its evolution with redshift:
indeed, studies of the MZR of galaxies are now available up to $z\sim8$
\citep{Jones2020TheMassMetallicityRelationatZ?8DirectMethodMetallicityConstraintsandnearFutureProspects}.

However, metallicities at different redshifts are obtained with different
methods, owing to abundance-sensitive spectral features moving across
the optical-infrared range. This makes a comparison of both absolute
and relative trends across cosmic time rather difficult. Furthermore,
spectra of high-redshift objects cannot be obtained at high resolution,
so metallicity indices calibrated with nearby galaxies must be used.
For example in the case of star forming galaxies, abundance indicators
are calibrated based on local \ion{H}{II} regions, but physical
conditions of the interstellar medium (ISM) at high-redshift might
be different than those at present day, thus adding further uncertainty
to the process.

Interestingly, up to $z\sim0.5$ emission lines that are used to compute
abundances and physical conditions of the ISM, still fall in the optical
range, and sufficiently good spectra can be obtained with 10m-class
telescopes, with reasonable exposure times. Thus a few years ago we
identified the galaxy cluster AC114 as an interesting target to obtain
emission-line spectra of galaxies spanning a range of masses: the
cluster sits at $z\sim0.32$, and as explained in \citet{Saviane2014Themass-metallicityrelationofgalaxiesuptoredshift0.35},
we expect that star-forming galaxies at that epoch are a factor $\sim~1.4$
more metal-poor than those in the local universe. Therefore, obtaining
gas-phase metallicities of AC114 cluster members gives us the possibility
to check the evolution of the MZR in the last $\sim~4$~Gyr, with
abundances obtained with the exact same method used for local star-forming
regions.

The substantial body of work on the cluster has been reviewed in our
previous paper \citep{Proust2015}, nevertheless we summarize here
the key features that make it relevant for the current study. The
cluster is classified as Bautz-Morgan type II-III \citep{Abell1989}
with a galaxy distribution that tends to be diffuse. Indeed, X-ray
emission has an irregular morphology, with two components dominating
the soft part of the spectrum (below 0.5 keV): a tail extends about
400 kpc from the central emission to the southeast \citep[see][]{DeFilippis2004}.
Its compact core is dominated by a cD galaxy and has strong lensing
power with several bright arcs and multiple image sources \citep{Smail1995d,Natarajan1998h,Campusano2001}.
Note however that it is also classified as a ``non-cool core'' cluster
by \citet{Zhang2016}, meaning that a central supermassive black hole
shall not be present. AC114 has a higher fraction of blue, late-type
galaxies compared with lower redshift clusters (up to 60\% outside
the core region; \citealt{Couch1998,Sereno2010c}, making it a prototypical
Butcher-Oemler cluster. Having many active members is a crucial property
when studying the MZR based on \ion{H}{II} regions.

In the course of data reduction, we realised that abundance-sensitive
absorption features can also be reliably measured in our VIMOS spectra,
therefore a study of the stellar component of AC114 galaxies is presented
in this work, while ISM abundances will be the subject of a forthcoming
paper. It shall be noted that a recent study of stellar populations
can also be found in \citet{RodriguezDelPino2014}, who targeted thirteen
disk galaxies with integral field spectroscopy. While most galaxies
in their sample to not display emission lines, they still host a young
stellar population; this is not centrally concentrated therefore it
is suggested that star formation was recently truncated by gradual
processes such as ram-pressure stripping or weak galaxy-galaxy interactions.

The present study builds on a dynamical analysis of the velocities
of galaxies in our sample, which was performed in \citet{Proust2015}
hereafter referred to as Paper~I. As explained in that paper, thanks
to a sample that reaches 1~magnitude fainter ($R\simeq21$), and
which is about 30\% larger than those in previous studies, we could
revisit the structure and dynamics of the cluster, which helped us
select cluster members.

\subsection{Main results from Paper~I}

\label{subsec:Main-results-from}

The mean redshift obtained in Paper~I is $z=0.31665\pm0.0008$ and
the velocity dispersion is $\sigma=1893_{-82}^{+73}$ \kms ~based
on a catalogue of 524~velocities. The distribution in redshift of
all galaxies is shown in Fig.\ref{fig:Distribution-in-redshift} with
bins $z=0.01$ widths (magenta histogram). The cluster has a very
elongated main radial filament spanning 12000~\kms~in redshift
space. In addition, a radial foreground structure was detected within
the central 0.5/h~Mpc radius, recognizable as a redshift group at
the same central redshift value. A background structure could also
be identified (see also Fig. \ref{fig:Three-dimensional-representation}).
Two hundred and sixty five galaxy members were identified, which yield
a dynamical mass $M_{200}=(4.3\pm0.7)\times10^{15}$M$_{\odot}$/h
for AC114 and $M_{v}=(5.4\pm0.7\pm0.6)\times10^{15}~$M$_{\odot}$/h
from the intrinsic velocity dispersion out to a radius of 3.98/h~Mpc.

\subsection{Analysis of the stellar populations based on Paper~I and new data}

\label{new-analysis}

Once the redshift is known, the spectrum can be corrected to the rest-frame:
this exercise led to the identification of a number of galaxies whose
redshift value needed revision, mainly because of their low S/N values,
in particular when the R~value of the correlation peak is lower or
close to 3, except for emission-line objects (see Sect.~2). A few
redshifts could be determined from the position of the most prominent
emission and absorption lines: \ion{O}{II} and CaI \ion{H}
and \ion{K}.

\begin{figure}
\includegraphics[width=1\columnwidth]{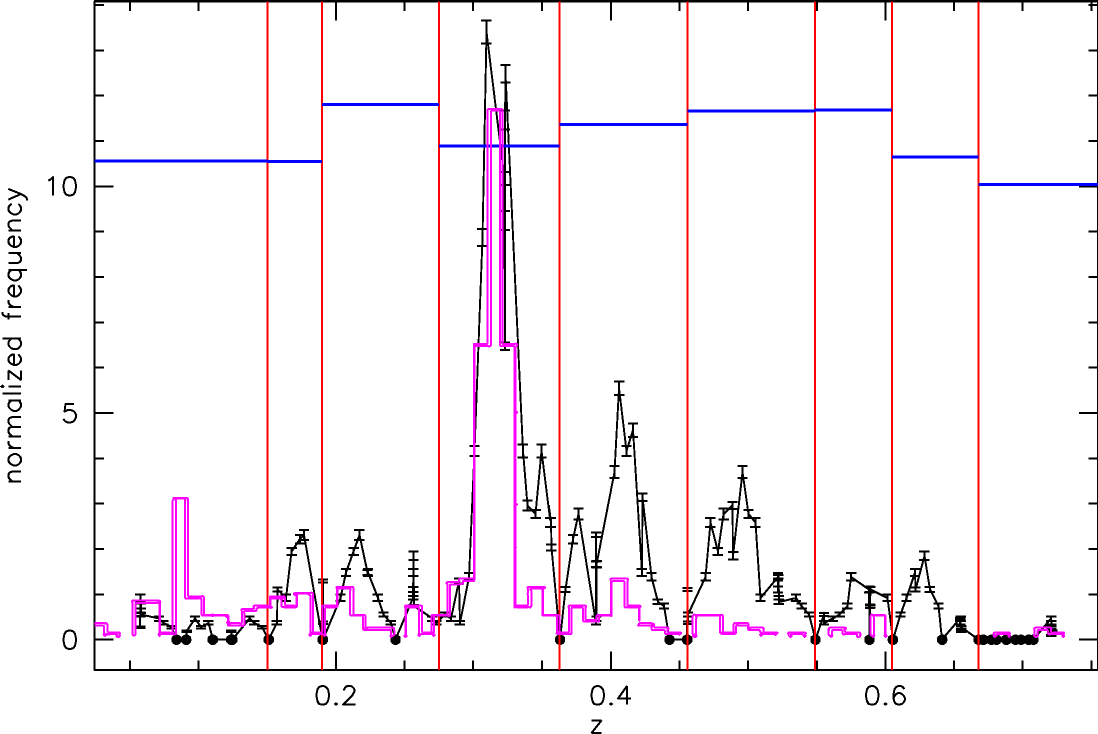}

\caption{Distribution in redshift of all galaxies. The vertical and horizontal
segments mark the limits of redshift ranges that are used later in
the analysis. These limits are consistent with redshift clustering
presented in Paper I, which are based on a larger sample including
literature data (magenta histogram). \label{fig:Distribution-in-redshift}}
\end{figure}

\begin{figure*}
\begin{centering}
\includegraphics[width=0.9\textwidth]{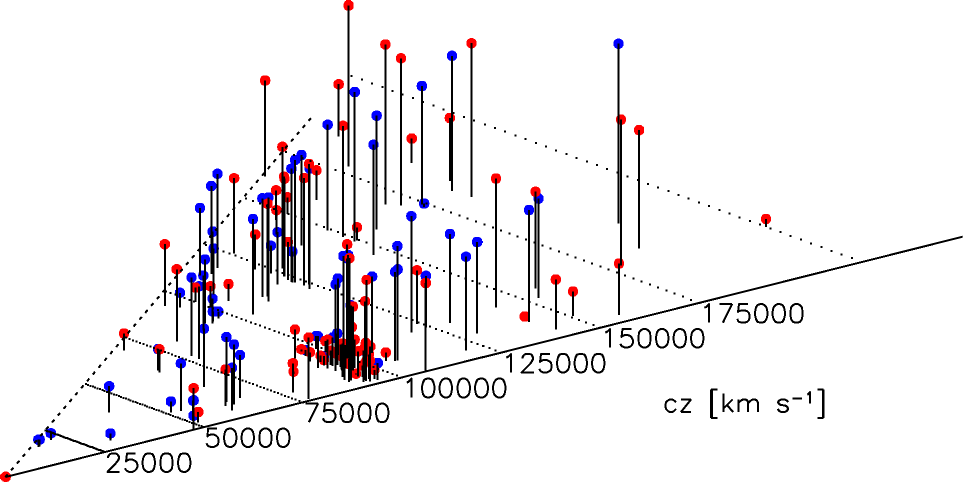} 
\par\end{centering}
\caption{Three-dimensional representation of objects in the sky area covered
by our VIMOS observations, with blue dots identifying active galaxies,
and red dots identifying passive ones. AC 114 is the clump of objects
near $cz=100,000\,{\rm km\,s^{-1}}$. The transversal scale has been
greatly amplified to better show structures surrounding the cluster:
at the redshift of the cluster, the 16 arcmin of RA span of VIMOs,
convert into 4.8 Mpc linear distance, while the linear distance spanned
by the redshift limits of the cluster is $\sim300$ Mpc \citep[see][for the calculations]{Wright2006ACosmologyCalculatorfortheWorldWideWeb}.
\label{fig:Three-dimensional-representation}}
\end{figure*}

The new distribution in redshift of all galaxies in shown in Fig.\ref{fig:Distribution-in-redshift}
(black lines) where objects are clearly clustered in a few redshift
intervals whose limits are identified by the vertical lines and horizontal
segments. The limits are listed in Table \ref{tab:Redshift-ranges-used}.
Galaxies belonging to the cluster are defined as those within the
lower and upper limits of rows 3 and 4 of Table \ref{tab:Redshift-ranges-used}:
they span a $\sim$5$\sigma$ range of velocities.

%
In this paper, we reanalyse the data obtained in Paper~I. Section~2
presents the observations and reduction of the data and section~3
analyses the stellar populations properties; these properties are
used in section~4 to construct and interpret the mass-metallicity
relation, and finally section~5 presents the conclusions based on
the global results obtained for AC114.

\begin{table}
\caption{Redshift limits used to group galaxies. Galaxies belonging to AC114
are defined as those within the lower and upper limits of rows 3 and
4. \label{tab:Redshift-ranges-used}}

\centering{}%
\begin{tabular}{lr@{\extracolsep{0pt}.}lrc}
 & \multicolumn{2}{c}{} &  & \tabularnewline
\hline 
\hline 
\#  & \multicolumn{2}{c}{$z$} & \multicolumn{1}{c}{$c\,z$} & Notes\tabularnewline
 & \multicolumn{2}{c}{} & \kms  & \tabularnewline
\hline 
1  & 0&1503  & 45,059  & \tabularnewline
2  & 0&1900  & 56,961  & \tabularnewline
3  & 0&2840  & 85,141  & lower limit\tabularnewline
4  & 0&3310  & 99,231  & upper limit\tabularnewline
5  & 0&4558  & 136,645  & \tabularnewline
6  & 0&5487  & 164,496  & \tabularnewline
7  & 0&6048  & 181,314  & \tabularnewline
8  & 0&6678  & 200,201  & \tabularnewline
\hline 
\end{tabular}
\end{table}

\begin{figure}
\begin{centering}
\includegraphics[width=0.9\columnwidth]{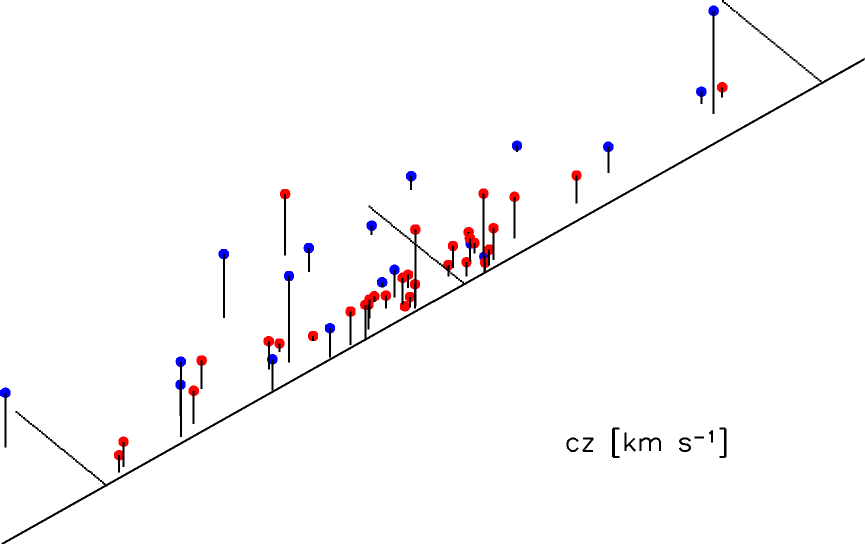} 
\par\end{centering}
\caption{A zoom into the three-dimensional representation of AC114. The three
segments show velocity values from 90,000 \kms ~to 100,000 \kms
~in steps of 5,000 \kms. The transversal scale has been amplified
by a factor $\sim10$ compared to the radial scale, but it is still
clear that the cluster is a very elongated structure. Note that gaps
in the object density are artefacts of the partial sky coverage of
the VIMOS footprint. \label{fig:A-zoom-into}}
\end{figure}

\section{Observations and reductions}

\begin{table}
\caption{Observations log (OB = Observing Block). \label{tab:Observations-log-(OB}}
{\small{}{}{}{}}%
\begin{tabular}{ccccl}
\hline 
{\small{}{}{}{}Date}  & {\small{}{}{}{}OB start$^{{\rm a}}$}  & {\small{}{}{}{}Exp. time$^{{\rm b}}$}  & {\small{}{}{}{}Filter}  & {\small{}{}{}{}Grism}\tabularnewline
\hline 
{\small{}{}{}{}2009-09-17}  & {\small{}{}{}{}02:37}  & {\small{}{}{}{}2250}  & {\small{}{}{}{}GG475}  & {\small{}{}{}{}HR red}\tabularnewline
{\small{}{}{}{}2009-09-17}  & {\small{}{}{}{}03:32}  & {\small{}{}{}{}2250}  & {\small{}{}{}{}GG475}  & {\small{}{}{}{}HR red}\tabularnewline
{\small{}{}{}{}2009-09-17}  & {\small{}{}{}{}04:11}  & {\small{}{}{}{}2250}  & {\small{}{}{}{}GG475}  & {\small{}{}{}{}HR red}\tabularnewline
{\small{}{}{}{}2009-09-17}  & {\small{}{}{}{}04:50}  & {\small{}{}{}{}2250}  & {\small{}{}{}{}GG475}  & {\small{}{}{}{}HR red}\tabularnewline
{\small{}{}{}{}2009-09-17}  & {\small{}{}{}{}05:28}  & {\small{}{}{}{}2250}  & {\small{}{}{}{}GG475}  & {\small{}{}{}{}HR red}\tabularnewline
{\small{}{}{}{}2009-09-17}  & {\small{}{}{}{}06:07}  & {\small{}{}{}{}2250}  & {\small{}{}{}{}GG475}  & {\small{}{}{}{}HR red}\tabularnewline
{\small{}{}{}{}2009-09-21}  & {\small{}{}{}{}03:54}  & {\small{}{}{}{}2250}  & {\small{}{}{}{}GG475}  & {\small{}{}{}{}HR red}\tabularnewline
{\small{}{}{}{}2009-08-16}  & {\small{}{}{}{}05:25}  & {\small{}{}{}{}2250}  & {\small{}{}{}{}GG475}  & {\small{}{}{}{}MR}\tabularnewline
{\small{}{}{}{}2009-08-16}  & {\small{}{}{}{}06:15}  & {\small{}{}{}{}2250}  & {\small{}{}{}{}GG475}  & {\small{}{}{}{}MR}\tabularnewline
{\small{}{}{}{}2009-08-16}  & {\small{}{}{}{}06:55}  & {\small{}{}{}{}2250}  & {\small{}{}{}{}GG475}  & {\small{}{}{}{}MR}\tabularnewline
{\small{}{}{}{}2009-08-21}  & {\small{}{}{}{}07:24}  & {\small{}{}{}{}2250}  & {\small{}{}{}{}GG475}  & {\small{}{}{}{}MR}\tabularnewline
{\small{}{}{}{}2009-08-21}  & {\small{}{}{}{}06:11}  & {\small{}{}{}{}2250}  & {\small{}{}{}{}GG475}  & {\small{}{}{}{}MR}\tabularnewline
{\small{}{}{}{}2009-09-25}  & {\small{}{}{}{}04:02}  & {\small{}{}{}{}2250}  & {\small{}{}{}{}GG475}  & {\small{}{}{}{}MR}\tabularnewline
{\small{}{}{}{}2009-09-25}  & {\small{}{}{}{}04:53}  & {\small{}{}{}{}2250}  & {\small{}{}{}{}GG475}  & {\small{}{}{}{}MR}\tabularnewline
\hline 
\end{tabular}

{\small{}{}{}{}$^{{\rm a}}$ }{\scriptsize{}{}{}{}beginning
of the observation block, UT time}{\scriptsize\par}

{\small{}{}{}{}$^{{\rm b}}$ }{\scriptsize{}{}{}{}exposure time
of the science part, in seconds}{\scriptsize\par}

\end{table}

A complete account of the observations and data reduction is given
in paper~I, so just a summary is given here. VIMOS \citep{LeFevre2003CommissioningandPerformancesoftheVLTVIMOSInstrument}
was used to carry out observations in service mode at the Very Large
Telescope, under program 083.A-0566, between August 16 and September
25, 2009 (see Table~\ref{tab:Observations-log-(OB}). Seven exposures
for each grism HR-red and MR were obtained, for a total shutter time
of $\sim4.4$h for each setup. The MR grism has a spectral resolution
of 580 for a $1^{\prime\prime}$ slit, over the 500 - 1000~nm spectral
range, while the HR-red grism has a spectral resolution of 2500 for
a $1^{\prime\prime}$ slit, over the 630 - 870~nm spectral range. 

The galaxy selection was made from the pre-imaging frames of the cluster.
To construct the masks, initially, the SIMBAD catalogue was used without
imposing any restriction criteria. As a first step we identified already
known galaxies of the cluster. Then we selected the non-stellar objects
in this region by eye in order to punch a maximum number of slits
in each of the four quadrants. Such a visual inspection allows to
discriminate between extended objects and stars.

\subsection{Data reduction}

The VIMOS pipeline was used to reduce the data. We reduced separately
the data taken with the grism HR red and MR. Each scientific frame
was bias subtracted and flat field corrected, the cosmic rays were
removed, and sky emission lines subtracted (the same procedure was
done for the standard stars). The spectra were wavelength calibrated
and seven frames for each VIMOS quadrant were combined. This gave
us S/N = 20 per pixel on the average at the continuum level on the
final spectra. Since the continuum of most of the galaxies is too
weak we did not use the spectra extracted by the pipeline. Instead,
we extracted one-dimensional spectra with MIDAS. Flux calibration
was obtained with spectrophotometric standards from \citep{Hamuy1992SouthernSpectrophotometricStandards.I.,Hamuy1994SouthernSpectrophotometricStandards.II}.

In this work, only MR data are used, while HR spectroscopy is used
in the forthcoming paper on the nebular abundances, to resolve \ion{N}{II}
from H$\alpha$ emission lines. 

Apparent magnitudes in the R-band were obtained by computing the integrated
flux of our spectra in the range 4897.5~Å\ to 9785.5~Å and calibrating
it using objects in common with the superCOSMOS database \citep{Maddox1990TheAPMGalaxySurveyI.APMMeasurementsandStarGalaxySeparation.,Maddox1990TheAPMGalaxySurvey.IIPhotometricCorrections.a}.
Photometry could also have been obtained from pre-imaging frames,
but these are observed under variable sky conditions, so magnitudes
would have to be calibrated on the superCOSMOS system anyway. The
precision of our calibration was estimated in paper~I as $\simeq0.5$
magnitudes, which is comparable with that of superCOSMOS. Furthermore,
synthetic photometry can be obtained with vastly more efficient means
compared to reducing pre-imaging frames from scratch, and we show
in the sections below that luminosities and masses obtained this way
yield well defined mass-metallicity relations.  As shown in Paper
I, our galaxy sample reaches luminosities as faint as $R\sim21$,
but it starts to suffer from incompleteness at $R\sim20$.

We obtained a total of 163~redshifts, combining and checking those
already obtained in Paper~I. The complete redshift values with their
individual error measurements are published in Table~5 and correspond
to the highest R-value obtained from the cross-correlations \citep{Tonry1979}.
UK-J $B_{j}$ and UK-J $R_{j}$ magnitudes are from superCOSMOS. 

The content of this table is as follows:
\begin{itemize}
\item number of the object in each quadrant from slit position;\smallskip{}
\item right ascension (J2000);\smallskip{}
\item declination (J2000);\smallskip{}
\item UK-J $B_{j}$ magnitude from superCOSMOS;\smallskip{}
\item UK-J $R_{f}$ magnitude from superCOSMOS;\smallskip{}
\item $R_{comp}$ computed magnitude from spectra;\smallskip{}
\item redshift;\smallskip{}
\item redshift error;\smallskip{}
\item R value from the cross-correlation (Tonry \& Davis 1979);\smallskip{}
\item notes 
\end{itemize}
In the course of data analyses, we identified two objects that were
removed from the table, as their spectra tricked the cross-correlation
algorithm, which returned wrong redshift values; 
indeed objects Q4/39 and Q4/59 were identified as M-dwarf stars, after
finding that their $D_{n}(4000)$ values were too high for their supposed
mass (see below).

Using data from Table~\ref{tab:Positions,-photometric-data}, Fig.~\ref{fig:Three-dimensional-representation}
shows the spatial distribution of all galaxies in our VIMOS sample
with a measured spectrum, with blue and red dots marking active and
passive galaxies respectively. The cluster is clearly visible as the
dense agglomeration of objects near $cz=100,000$ \kms, and it appears
that most of its galaxies are not actively forming stars, although
in the cluster periphery there are some that do. \citet{Couch2001}
note that in AC114 the star formation rate from the {H$\alpha$}
emission is not exceeding 4~M$_{\odot}$ yr$^{-1}$ and that the
{H$\alpha$} luminosity function is an order of magnitude below
that observed for field galaxies at the same redshift. As anticipated
in Sec. \ref{subsec:Main-results-from},  figure~2 also suggests
the existence of two other structures, which appear as filaments extending
for tens thousands \kms, one of them seemingly connected to the cluster
core. Figure~3 shows a zoom into the three-dimensional representation
of the cluster. It is clear that the cluster has a very elongated
structure spanning $\simeq10,000$ \kms.

\subsection{Classification of galaxies \label{subsec:Classification-of-galaxies}}

\begin{figure}
\includegraphics[width=1\columnwidth]{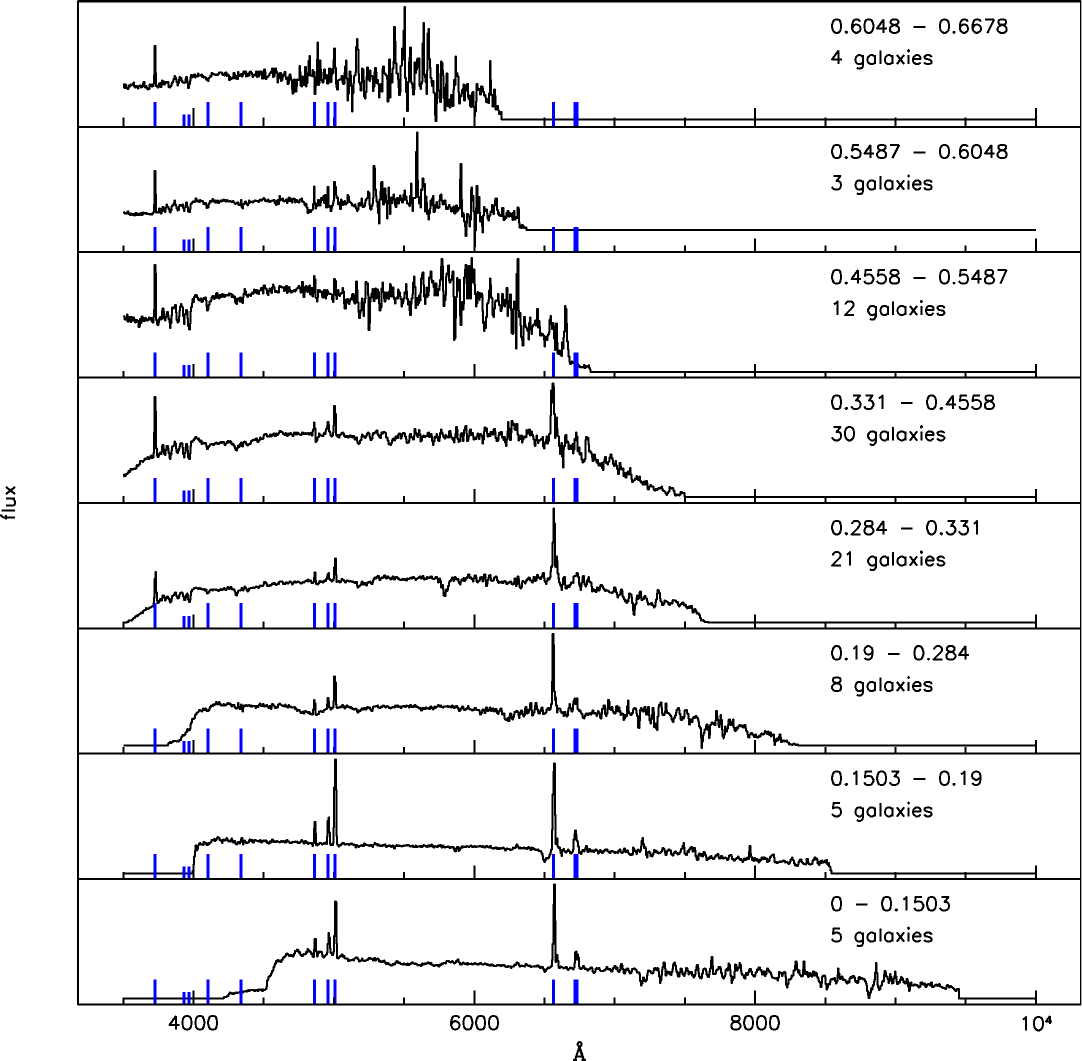}

\caption{For each redshift range from table \ref{tab:Redshift-ranges-used},
this figure shows the co-added spectra of active galaxies. The vertical
blue bars mark the position of the following emission lines: {[}\ion{S}{ii}{]}~$\lambda\lambda$6716,6731,
H$\alpha$, {[}\ion{O}{iii}{]}~$\lambda\lambda$4959,5007, H$\beta$,
H$\gamma$, H$\delta$, {[}\ion{O}{ii}{]}~$\lambda$; and Ca
H and K absorption lines (3968Å, 3934Å).\label{fig:For-all-redshift-active}}
\end{figure}

\begin{figure}
\includegraphics[width=1\columnwidth]{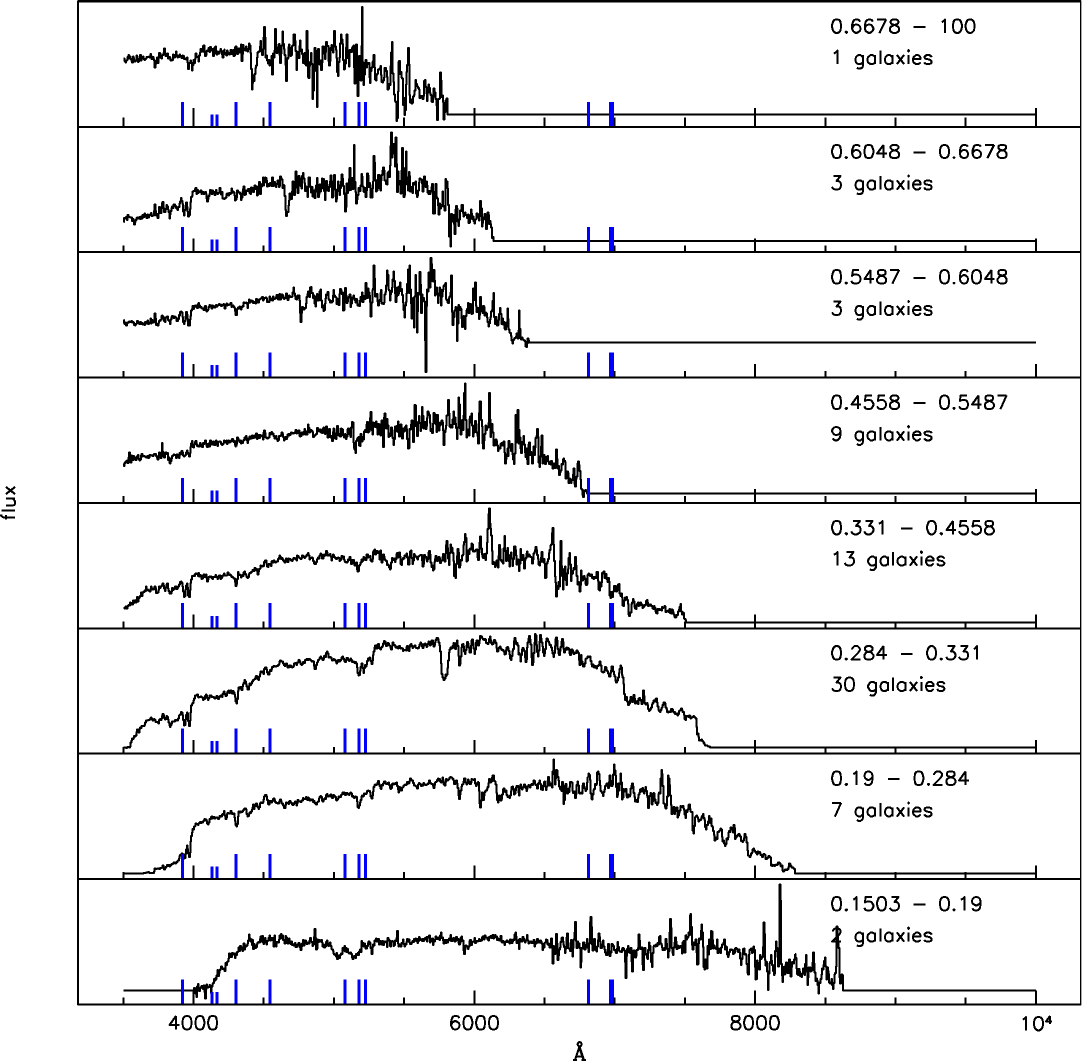}

\caption{Same as Fig. \ref{fig:For-all-redshift-active} for co-added spectra
of passive galaxies. \label{fig:For-all-redshift-passive}}
\end{figure}

After correcting spectra to rest-frame the 163~galaxies were classified
as `active' or `passive' based on the presence/absence of {[}\ion{O}{II}{]}$\lambda3727$,
{[}\ion{O}{III}{]}$\lambda\lambda4959,5007$ and/or H$\beta$
emission lines, depending on the galaxy redshift, with the following
procedure. 

The first step consisted in blindly scanning all spectra for the presence
of {[}\ion{O}{III}{]}$\lambda5007$ and/or H$\beta$: a linear
fit to the continuum was subtracted, and a maximum in flux was searched
in the window $4800\leq\lambda\leq4841$ for H$\beta$ and in the
window $4980\leq\lambda\leq5090$ for {[}\ion{O}{III}{]}$\lambda5007$.
A Gaussian fit was then attempted, provided that the peak in flux
is three times larger than the continuum noise (assumed to be the
semi-quartile of the flux variation in the spectral window). To estimate
errors, the fit was repeated for the nominal continuum level, and
by assuming a higher or lower continuum by the amount of noise. The
result of this procedure was that, for each spectrum where a fit is
possible, and for each of the two lines, the central wavelength, its
FWHM and error, and the area of the Gaussian and its error are obtained.
After this first screening, a quality parameter $q$ is assigned to
each spectrum, defined as the sum for the two lines, of their Gaussian
area divided by its error. Spectra are then sorted by fit quality,
and plotted one by one to visually check the quality of the fit. We
find that spectra with $q\geq4$ show unambiguous presence of emission
lines, so galaxies that satisfy this criterion were classified as
`active', while the rest were classified as `passive'. While performing
the visual checks, a few galaxies preliminarily classified as `passive'
were noticed having a clear {[}\ion{O}{II}{]}$\lambda3727$ emission
line, and were added to the list of `active' objects: their spectra
are shown in Fig.~\ref{fig:Spectra-of-galaxies-3727} of Appendix~A.
While some passive galaxies might have very weak emission lines that
are not detected by our methods, the sections below show that the
two galaxy classes have very different properties, thus our classification
must be broadly correct. 

Table \ref{tab:Classification-of-galaxies} summarises the results,
and adds further remarks on some galaxies, such as those actually
identified as local M-dwarf stars, and those having low-quality spectra
that are not further considered in this paper. 

In order to show the redshift evolution of their spectral energy distribution
(SED), Figures \ref{fig:For-all-redshift-active} and \ref{fig:For-all-redshift-passive}
show the co-added spectra of the two galaxy classes, in the redshift
ranges defined in Table \ref{tab:Redshift-ranges-used}. 

It can be seen that, in the redshift range of AC114, the majority
of galaxies are passive: there are 30 galaxies classified as such
(59\%), versus 21 classified as active (41\%). This confirms the visual
impression given by Fig.~\ref{fig:Three-dimensional-representation}.

\section{Stellar populations}

\subsection{Representative age and metallicity of active and passive galaxies
in AC114\label{subsec:Representative-age}}

\begin{figure}
\includegraphics[width=1\columnwidth]{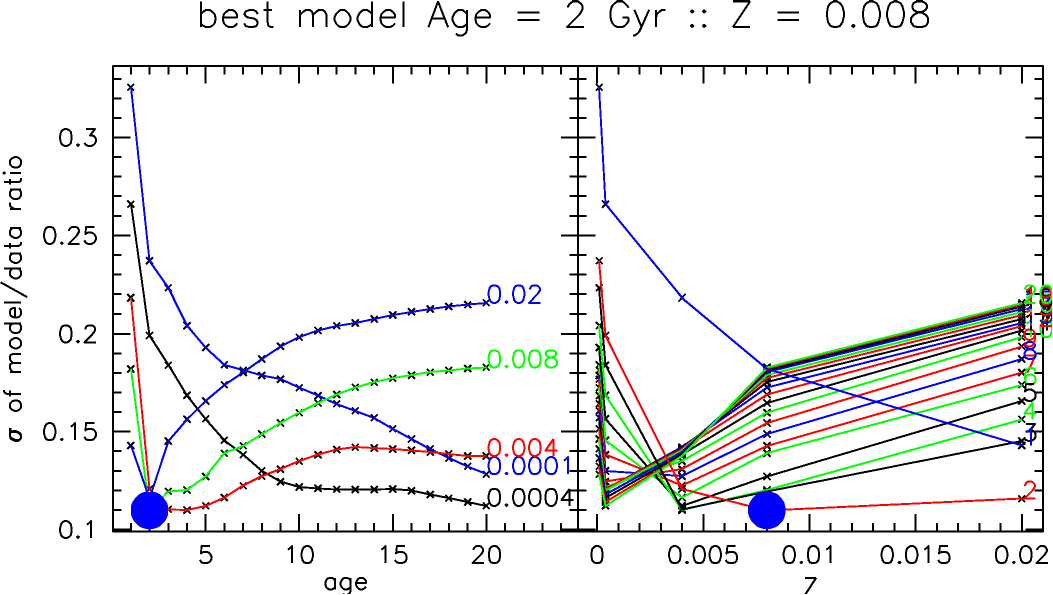}

\caption{To select the best simple stellar population (SSP) model that fits
the co-added spectrum of active galaxies in AC114, the dispersion
of the model-to-data ratio is plotted as a function of age and metallicity:
this figure shows that the minimum value is reached for a model age
of 2~Gyr, and a metallicity Z=~0.008. \label{fig:To-select-the-active-best-model}}
\end{figure}

\begin{figure}
\includegraphics[width=1\columnwidth]{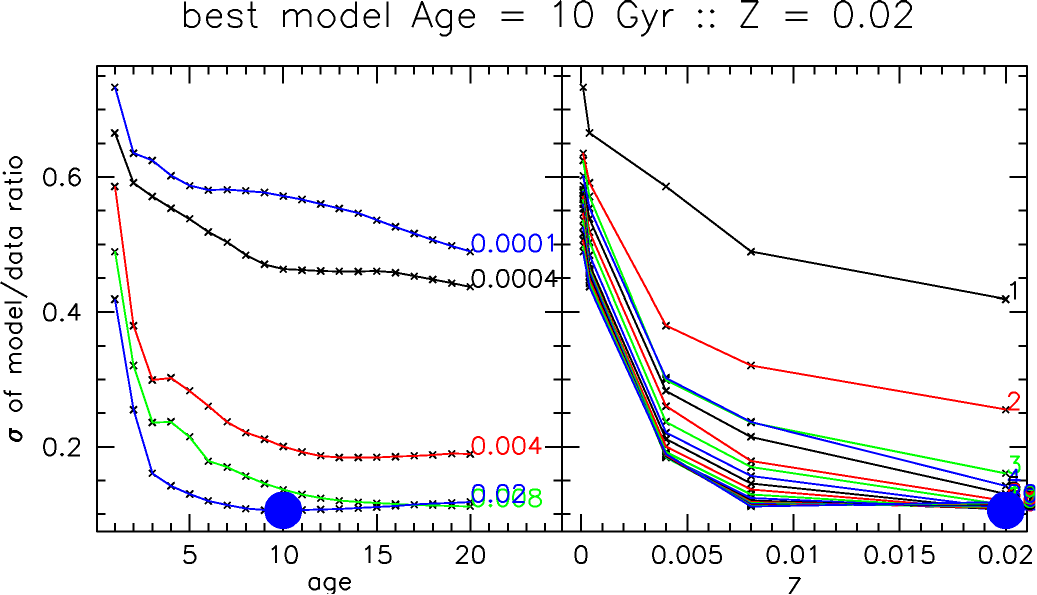}

\caption{Same as Fig. \ref{fig:To-select-the-active-best-model}, for the co-added
spectrum of passive galaxies in AC114. In this case, the best model
fit is reached for an age of 10~Gyr, and a metallicity Z=~0.02.
\label{fig:To-select-the-passive-best-model}}
\end{figure}

\begin{figure}
\includegraphics[width=1\columnwidth]{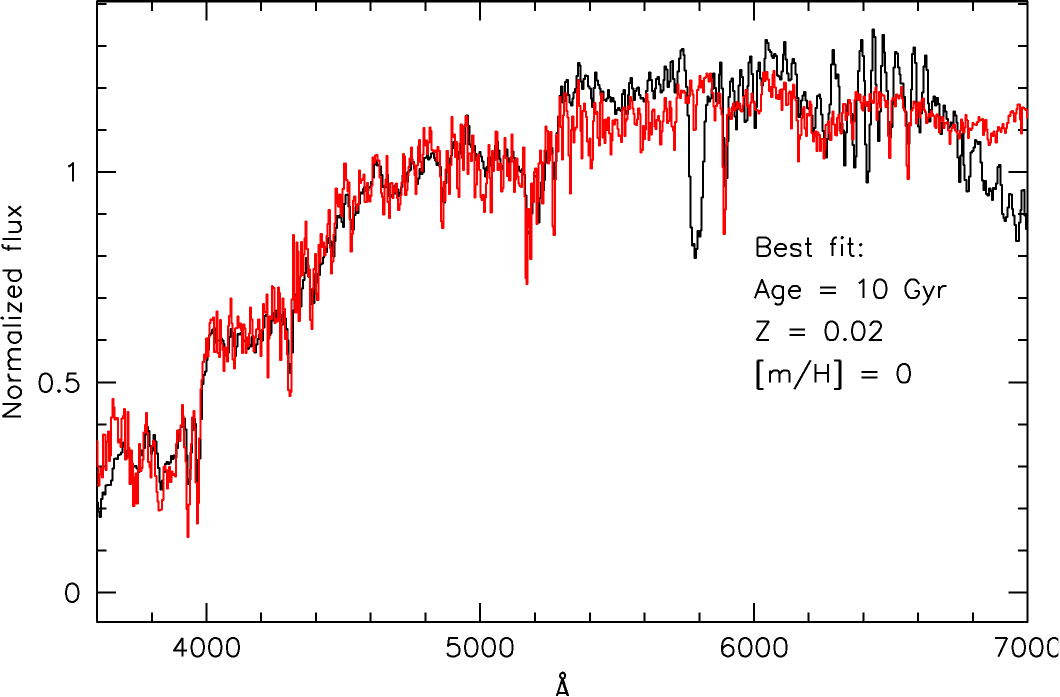}

\caption{The co-added spectrum of all passive galaxies in AC114 redshift range
(black) fitted with the best model of \citet{Bruzual2003} in red.
This gives the typical age and metallicity of the underlying stellar
population. \label{fig:The-summed-spectrum-passive-bc03}}
\end{figure}

\begin{figure}
\includegraphics[width=1\columnwidth]{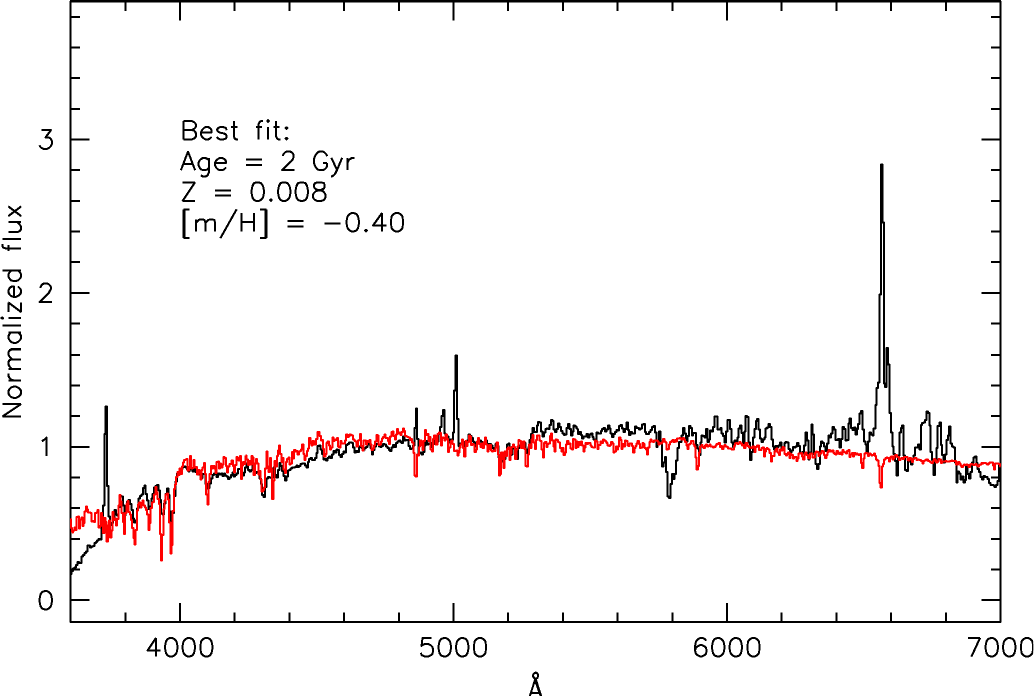}

\caption{Same as figure \ref{fig:The-summed-spectrum-passive-bc03} for active
galaxies in the cluster. \label{fig:The-summed-spectrum-active-bc03}}
\end{figure}

As a first step towards the characterisation of galaxies in AC114,
we fitted the co-added spectra obtained above in the redshift range
of the cluster, with stellar population synthesis models from \citet{Bruzual2003},
following the prescriptions of \citet{Saviane2007_NGC5011C:AnOverlookedDwarfGalaxyintheCentaurusAGroup}.
Briefly, model spectra were degraded to the resolution of our VIMOS
data, and then for each model spectrum, the steps were the following: 
\begin{itemize}
\item select the spectral range from 3800 to 6000Å; 
\item remove emission lines from co-added spectra, before fitting: i.e.,
remove regions near H$\beta$ and {[}\ion{O}{III}{]} ; 
\item normalise both model and co-added spectra to maximum flux within the
fit range; 
\item compute ratio of model to observed spectrum; 
\item compute average and dispersion of spectral ratio; 
\item select the best fit model as the one that gives the lowest dispersion
of the spectral ratio. 
\end{itemize}
The results of this procedure are shown in Figures \ref{fig:To-select-the-active-best-model}
and \ref{fig:To-select-the-passive-best-model}, for active and passive
galaxies, respectively: the typical age of active galaxies in the
cluster is 2~Gyr, and their metallicity is Z~=~0.008; while in
the case of passive galaxies, the best fit is reached for an age of
10~Gyr, and a metallicity Z=~0.02. Galaxies cannot be older than
the universe, so it is reassuring to see that the age of the universe
at the redshift of AC114 is indeed no less than 10 Gyr (see Figures
\ref{fig:Mass-of-passive-vs-z} and \ref{fig:Mass-of-active-vs-z}).
The best-fit models are shown overlaid to the co-added spectra in
figures \ref{fig:The-summed-spectrum-passive-bc03} and \ref{fig:The-summed-spectrum-active-bc03}.
Figures \ref{fig:To-select-the-active-best-model} and \ref{fig:To-select-the-passive-best-model}
show that adopting model ages and metallicities closest to the best
fits, would also give good results: this uncertainty is included in
the calculation of M/L values in Sect.~\ref{subsec:Mass-to-Light-ratios}. 

This exercise reveals that passive galaxies must have formed the bulk
of their population early on, in a strong burst that quickly enriched
their ISM to solar metallicities: indeed, the remarkably good fit
of a simple stellar population model (SSP) indicates that those stars
were formed in a relatively short time span. On the contrary, the
low stellar metallicity of active galaxies is an indication that star
formation (SF) proceeded at a low pace in these systems, which are
indeed still forming stars. Still, the good fit of a SSP to the observed
spectrum is telling us that most stars in active galaxies were also
formed rapidly, although later than those in passive galaxies.

Passive galaxies are in general more massive than active ones (see
below), but physical sizes do no vary too much between the two classes;
therefore one expects a higher gas density in passive galaxies, which
would explain their higher SFR, according to the Schmidt-Kennicutt
law \citep{Kennicutt1998TheGlobalSchmidtLawinStarFormingGalaxies}.
Figures \ref{fig:Three-dimensional-representation} and \ref{fig:A-zoom-into}
show that passive galaxies are concentrated near the cluster centre,
where mass must have been accumulating in the early formation epochs,
thus easing the formation of the largest objects (see also Fig. 6
in \citealt{Proust2015}). In addition, SF in the densest regions
of galaxy clusters is quenched more effectively than in their outskirts
\citep[see, e.g., ][]{Deshev2017GalaxyEvolutioninMergingClustersthePassiveCoreoftheTrainWreckClusterofGalaxies<ASTROBJ>A520</ASTROBJ>}:
therefore the active/passive dichotomy can also be explained, at least
partially, by an effect of the environment.

\subsection{Comparison with \citet{Couch1987a}}

\begin{figure}
\includegraphics[width=1\columnwidth]{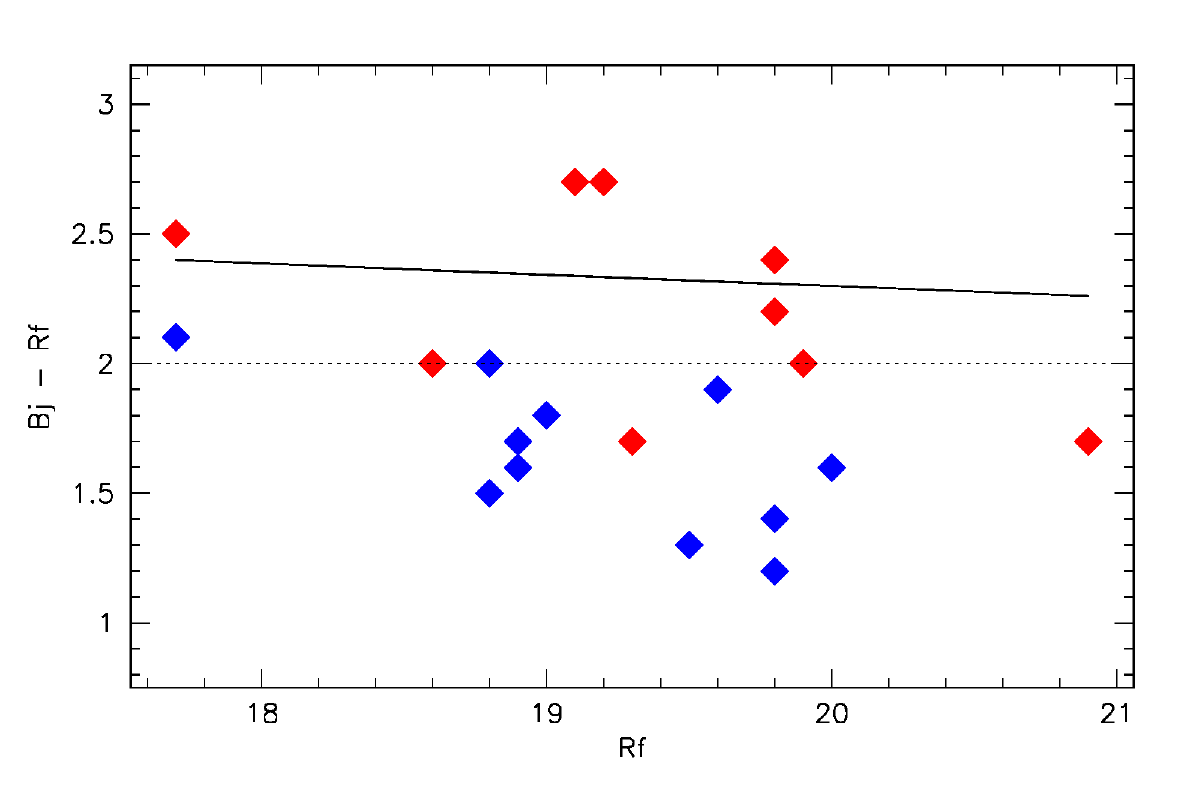}

\caption{Colour-magnitude diagram for AC114 galaxies with photometry in superCOSMOS
database \citep{Maddox1990TheAPMGalaxySurvey.IIPhotometricCorrections.a}.
The solid line is the red sequence as defined in Paper~I, and the
horizontal dotted line represents the separation between red and blue
galaxies as in CS87. \label{fig:Colour-magnitude-diagram-for}}
\end{figure}

To confirm the results above, it is instructive to compare them to
\citet[ hereafter CS87]{Couch1987a}. They classify their galaxy sample
into `red' and `blue' objects, based on their position in the
colour-magnitude plot, which is shown in Fig.~\ref{fig:Colour-magnitude-diagram-for}
for the subset of galaxies having published colours in superCOSMOS.
In CS87, blue galaxies are defined has having colours smaller than
$B_{{\rm J}}-R_{{\rm F}}=2$, which is represented by the dotted line
in the figure; the solid line represents the `red sequence' as defined
in paper~I, which is similar to the one found by CS87. Passive and
active galaxies are plotted with red or blue symbols, so the figure
demonstrates that most active galaxies would be classified as `blue'
by CS87, while passive galaxies gather around the red sequence. 

Most red galaxies in CS87 have spectra comparable to nearby E/S0 galaxies,
and indeed their comparison to models of population synthesis yields
a typical age of 10~Gyr. They also find 15\% of red galaxies having
strong H$\delta$ absorption, which are interpreted as hosting a short
0.5 Gyr burst with an age of 1.5 Gyr. These results are consistent
with our findings above, where the typical age of passive galaxies
was indeed found to be 10~Gyr, with a few exceptions identified in
Sec.~\ref{subsec:Special-d4000}. This age is used in Sect.~\ref{subsec:Mass-to-Light-ratios}
to compute M/L ratios, therefore we can expect that masses of passive
galaxies are on the average correct: indeed the tight sequence found
in Sect.~\ref{subsec:d4000--vs.-mass} and Fig.~\ref{fig:d4000-galaxies}
validate this conclusion. 

As expected from their spread in colour, blue galaxies in CS87 have
a more diverse SF history and interestingly, almost half of them are
pure absorption-line objects. By comparison to local galaxies and
models, their emission line objects are split between having spectra
comparable to nearby spiral galaxies (their type 2) and hosting a
current SF burst (their type 1). Finally, blue galaxies with no emission
lines are interpreted as having experienced a SF burst of varying
strength, up to 1.5 Gyr prior to the epoch of observation. The stacked
spectrum of our active galaxies, which by definition are emission-line
objects, must be a mix of type 1 and type 2 of CS87, therefore the
age of 2~Gyr found above is a reasonable guess. Once more, this is
reinforced by Fig.~\ref{fig:d4000-galaxies}, which shows a well-defined
trend of metallicity index versus masses. Note also that some objects
do require a small correction to their ages, as discussed in Sect.~\ref{subsec:Special-d4000}. 

Only four galaxies are in common between this work and CS87, and they
all belong to the red sequence, with normal H$\delta$ absorption.
A larger overlap in the two sets would be needed for a thorough comparison,
nevertheless in Appendix~C we show that our spectra match the classification
of CS87. 

In the context of this discussion, it is also interesting to examine
`postage stamps' of our targets as shown in figures~\ref{fig:Images-of-galaxies-passive}
and \ref{fig:Images-of-galaxies-active}, for passive and active galaxies,
respectively. As expected for an E/SO type, most passive galaxies
have regular shapes, but there are some galaxies that appear to be
interacting (at least in projection) with nearby smaller objects (see
image caption). It is interesting to note that two of these potentially
interacting galaxies (44 and 49) present an anomalous $D_{n}(4000)$
index, as discussed in Section~\ref{subsec:Special-d4000}, which
means that their ages deviate from the typical one of their class.
Conversely, galaxies 64 and 21 also have a deviant $D_{n}(4000)$
index, but they appear to be isolated in Fig.~\ref{fig:Images-of-galaxies-passive}.
Active galaxies appear more diffuse than passive ones, and contrary
to passive ones, objects with anomalous $D_{n}(4000)$ index do not
have close companions but rather appear irregular in shape (numbers
36, 56, 48, 40, and 26). The implications of galaxy morphology will
be examined in more detail in a forthcoming paper where nebular abundances
will be obtained (Andrade et al., in preparation), and contrasted
with the stellar abundances obtained in this work.

\subsection{Mass-to-Light ratios \label{subsec:Mass-to-Light-ratios}}

\begin{figure}
\includegraphics[width=1\columnwidth]{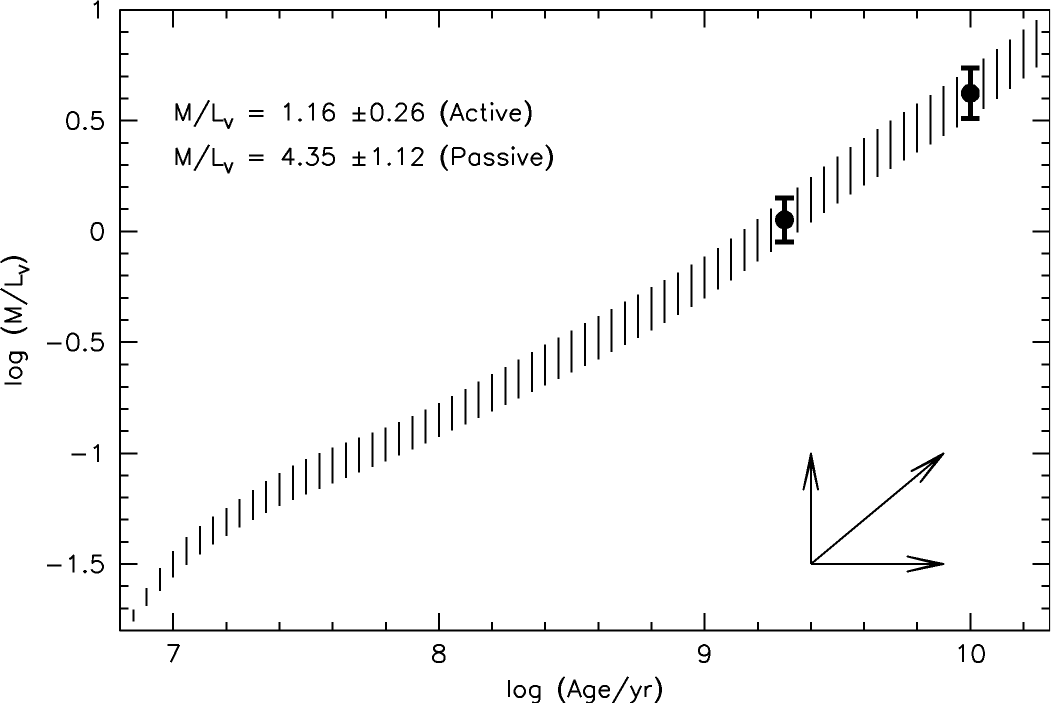}

\caption{V-band mass-to-light ratio as a function of the age of the population,
from BC2003: the band represents uncertainties introduced by the choice
of different initial IMF, spectral library, theoretical evolutionary
tracks, and metallicities. Based on the age returned by the best-fit
models, representative M/L ratios can be estimated for passive and
active galaxies, which are shown by the dots and error bars. Arrows
in the lower right corner show that, for ages greater than $\sim0.1$
Gyr, a change in age of 0.5~dex translates into a similar 0.5~dex
change in M/L ratio. \label{fig:V-band-mass-to-light-ratio}}
\end{figure}

Having in hand the age of the stellar populations, the next step is
the computation of their mass-to-light ratios. From the plots in BC03
we obtained Fig.~\ref{fig:V-band-mass-to-light-ratio}, which shows
M/L$_{V}$ vs. age: as discussed in BC03, the mass-to-light ratio
is affected by the choice of the initial mass function (IMF), spectral
library, theoretical evolutionary tracks, and metallicity, so the
shaded band in the figure represents those uncertainties, estimated
as $1/4$ of the full variation interval.

As shown in Table \ref{tab:Observations-log-(OB}, the most complete
photometric catalogue for our galaxies is in the R-band, therefore
mass-to-light ratios were converted from V- to R-band using the expression:
$M/L_{R}=M/L_{V}\times10^{-0.4\,(V-R)}$, where the $V-R$ colour
of an old population was taken as the average of globular clusters
in the \citet{Harris1996ACatalogofParametersforGlobularClustersintheMilkyWay}
catalogue: $V-R=0.64\pm0.17$. The resulting ratios are then $M/L_{R}=0.64\pm0.26$
and $M/L_{R}=2.41\pm1.12$ for active and passive galaxies, respectively.

\subsection{Galaxy distances, luminosities, and masses}

\begin{figure}
\includegraphics[width=1\columnwidth]{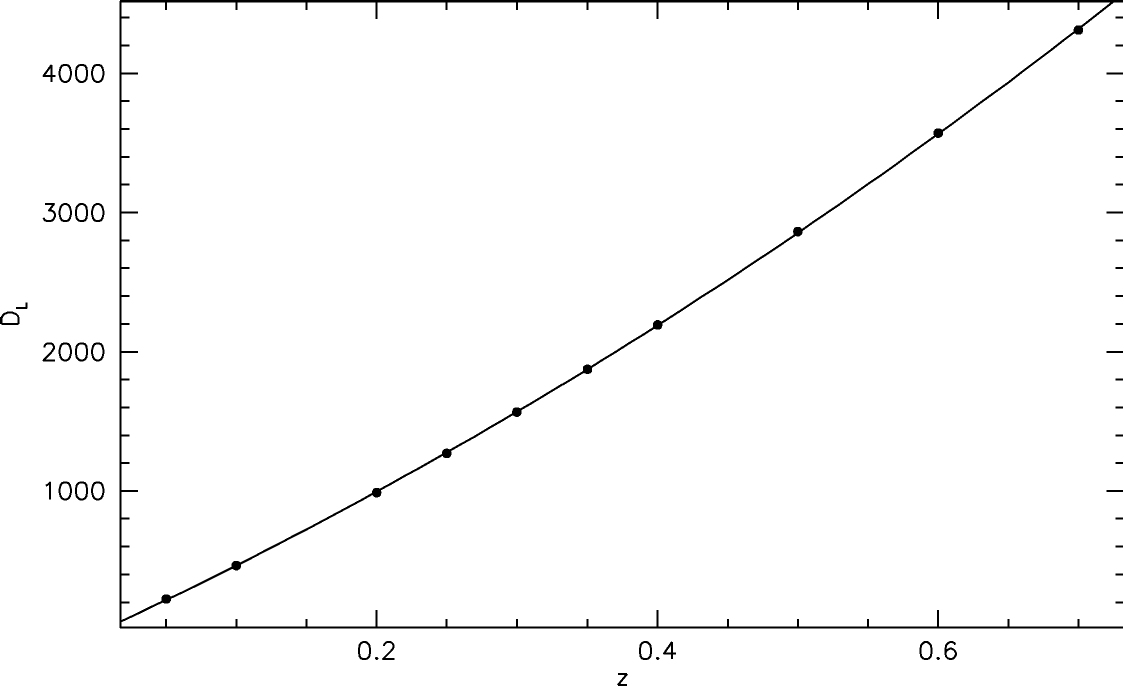}

\caption{Luminosity distance, expressed in Mpc, vs. redshift: the curve is
a quadratic interpolation of selected values calculated using the
cosmology calculator of \citet{Wright2006ACosmologyCalculatorfortheWorldWideWeb}.
\label{fig:Luminosity-distance-vs.}}
\end{figure}

\begin{figure}
\includegraphics[width=1\columnwidth]{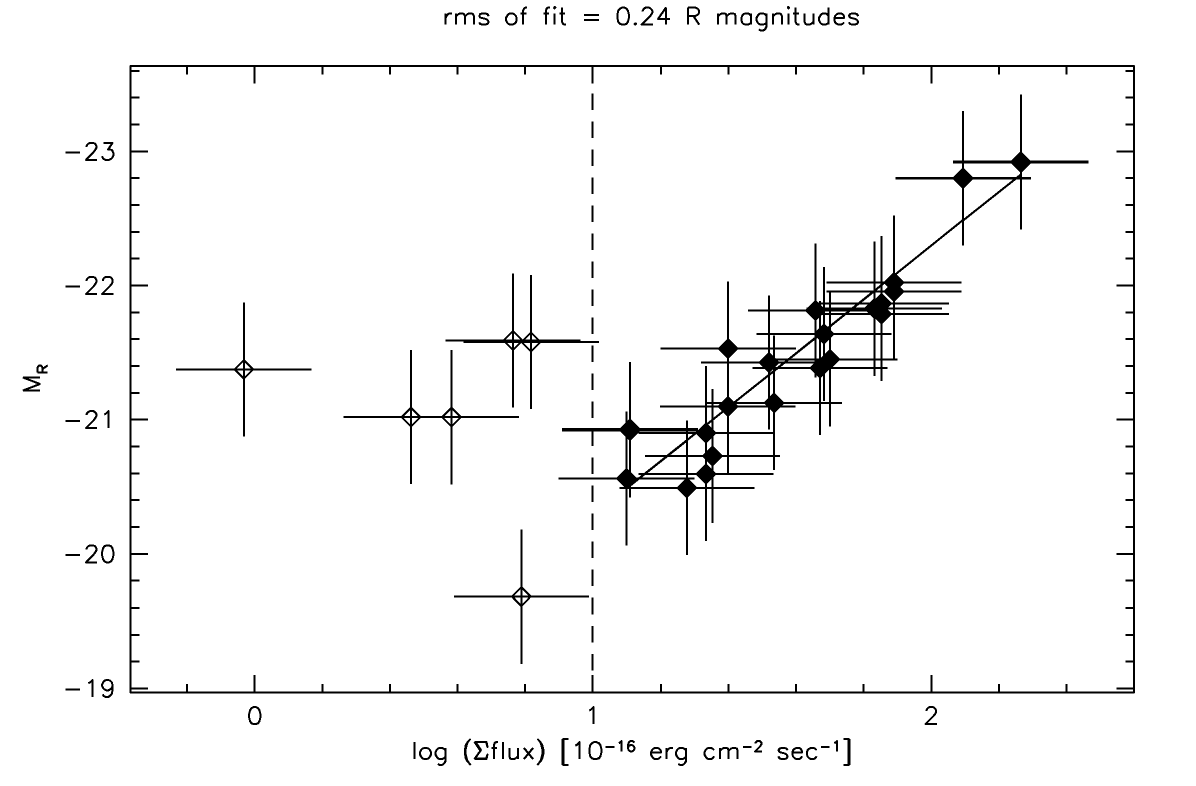}

\caption{Comparing absolute $R$ magnitudes vs. integrated flux, for passive
galaxies in AC114 redshift range. As expected, the two quantities
are proportional to each other, but the correlation breaks down at
the lower end of total fluxes. An uncertainty of 0.5 magnitudes has
been assumed both for the total flux and the R-band photometry. \label{fig:Comparing-absolute-mag-fluxes-passive}}
\end{figure}

\begin{figure}
\includegraphics[width=1\columnwidth]{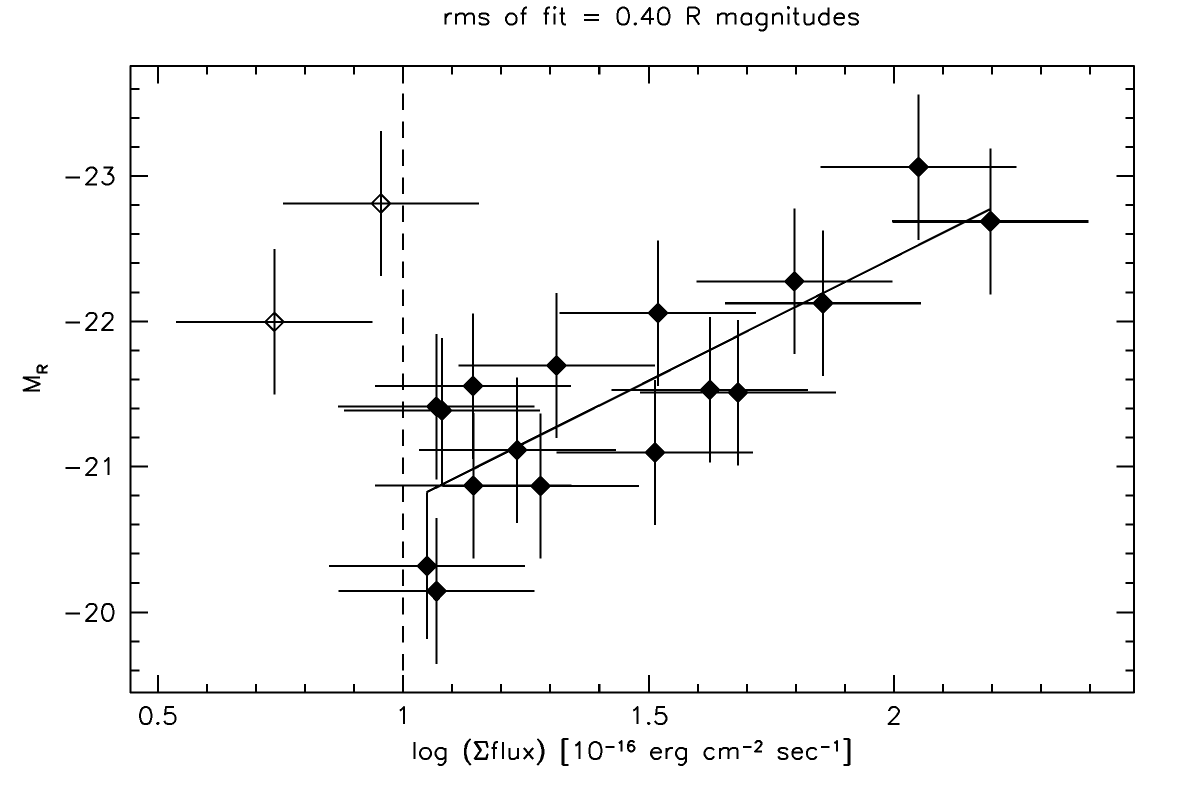}

\caption{Same as Fig. \ref{fig:Comparing-absolute-mag-fluxes-passive}, for
active galaxies in AC114 redshift range. \label{fig:Comparing-absolute-mag-fluxes-active}}
\end{figure}

\begin{figure}
\includegraphics[width=1\columnwidth]{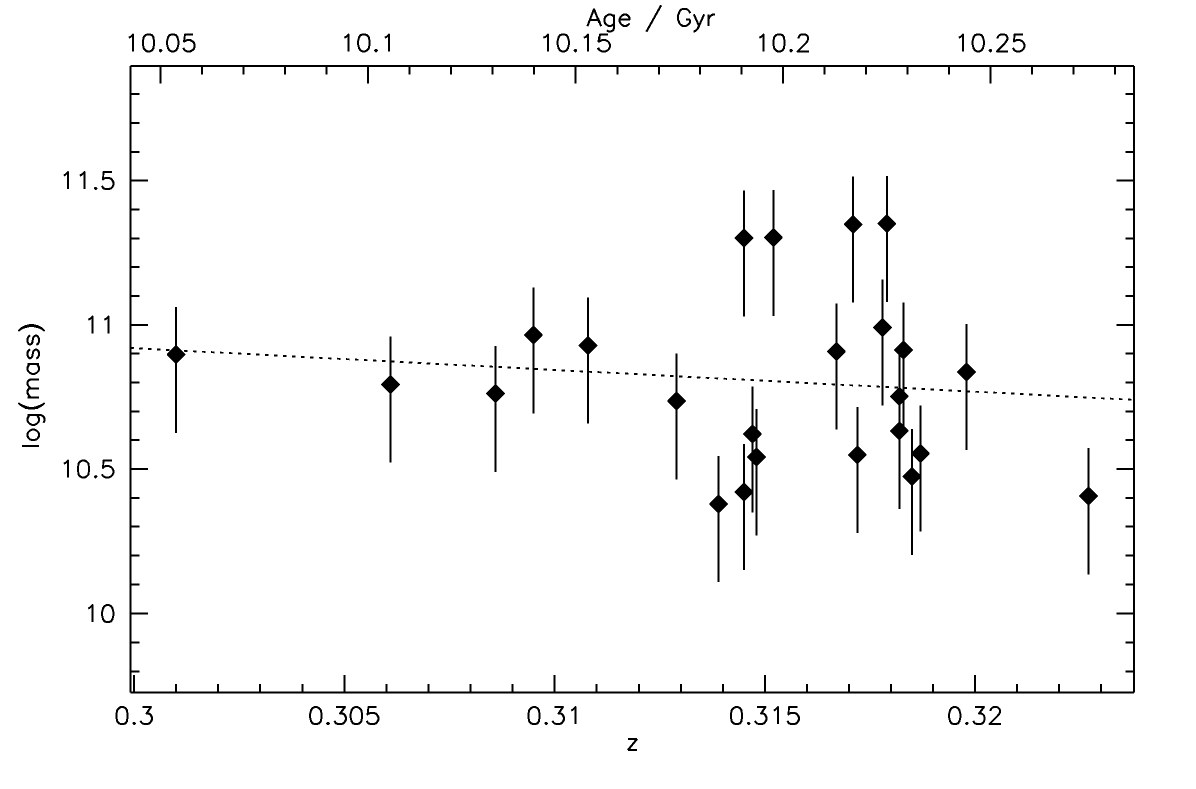}

\caption{Mass of passive galaxies as a function of redshift and age of the
universe. The dotted line represents a linear fit through the data.
Errors in mass are taken from Table~\ref{tab:Main-characteristics-of-passive},
while errors in redshift are smaller than the symbols. \label{fig:Mass-of-passive-vs-z}}
\end{figure}

\begin{figure}
\includegraphics[width=1\columnwidth]{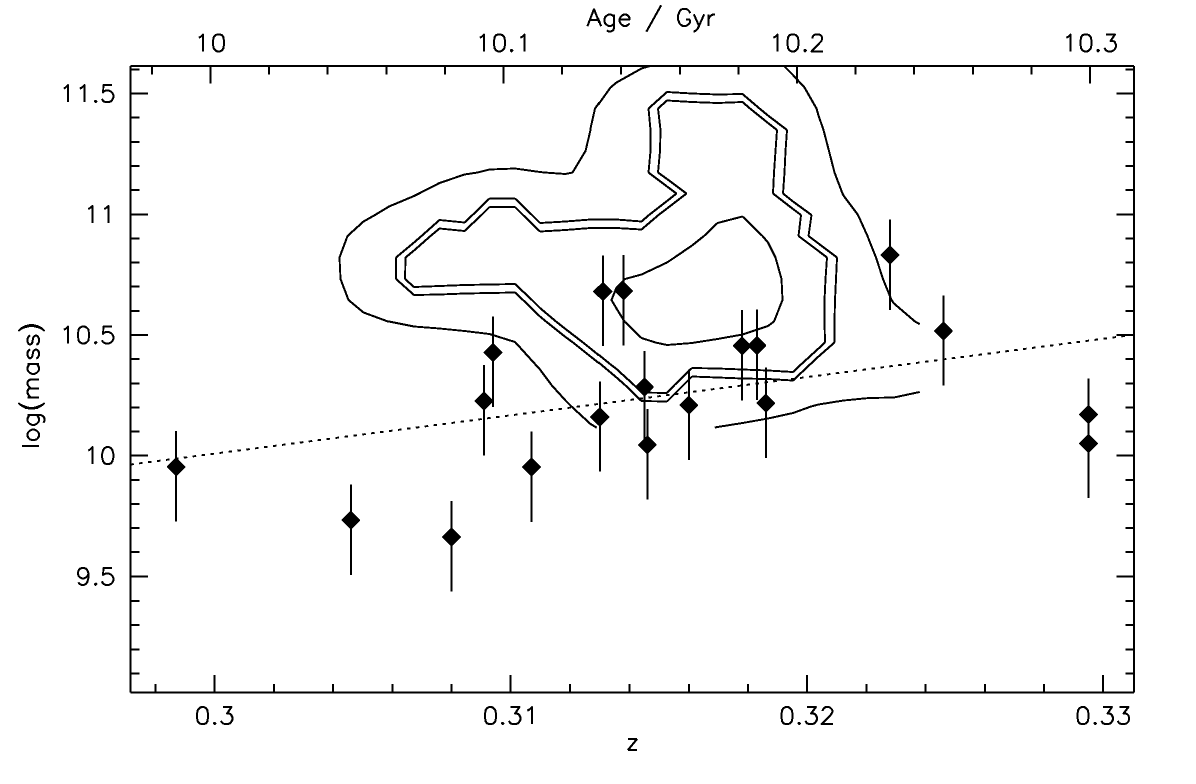}

\caption{Mass of active galaxies as a function of redshift. The dotted line
represents a linear fit through the data, while the contours represent
the general distribution of passive galaxies in the same plot of Fig.
\ref{fig:Mass-of-passive-vs-z}. Errors in mass are taken from Table~\ref{tab:Main-characteristics-of-passive},
while errors in redshift are smaller than the symbols.\label{fig:Mass-of-active-vs-z}}
\end{figure}

To compute galaxy luminosities, apparent magnitudes were converted
to absolute ones with the usual $M_{R}=R-\mu$, where the distance
modulus is $\mu=5\,\log D_{L}-5$. The luminosity distance $D_{L}$
is a function of redshift, and it was evaluated following \citet{Wright2006ACosmologyCalculatorfortheWorldWideWeb},
with input parameters $H_{0}=69.6$, $\Omega_{{\rm matter}}=0.286$,
and $\Omega_{{\rm vac}}=0.714$ (flat universe). After computing the
relation for a few selected values, it was interpolated by a quadratic
function, as shown in Fig.~\ref{fig:Luminosity-distance-vs.}. We
find:

\noindent $D_{L}=2245.7\,z^{2}+4628.4\,z-20.838$

\noindent which is valid for $0.05\leq z\leq0.7$. At the mean redshift
quoted in section 1.1, the distance is $D_{L}=~1670\pm17$~Mpc 

Absolute luminosities were then computed as $L_{R}=10^{-0.4\,(M_{R}-M_{R_{\odot}})}$,
with $M_{R_{\odot}}=4.5$. Finally, luminosities were converted into
masses with the mass-to-light ratios computed in Sect.~\ref{subsec:Mass-to-Light-ratios}.
Uncertainties on mass values were computed by adding in quadrature
uncertainties from photometry and from mass-to-light ratios. 

As a consistency check between photometric and spectral properties,
the integrated spectral flux is compared to absolute $R$ magnitudes
in Figures \ref{fig:Comparing-absolute-mag-fluxes-passive} and \ref{fig:Comparing-absolute-mag-fluxes-active}
for the two classes of galaxies. It is reassuring to see that there
is a good correlation between the two quantities: the correlation
coefficient is $-0.94$, with a dispersion of $\sim0.24$ magnitudes
for passive galaxies, and the same quantities are $-0.85$ and $\sim0.4$
magnitudes for active ones. However the correlation breaks down for
galaxies at the faint end of the luminosity range: therefore for the
next analysis we retained only objects with $\Sigma({\rm flux)>10^{-15}\,{\rm erg\,cm^{-2}sec^{-1}}}$.

Figure \ref{fig:Mass-of-passive-vs-z} shows that passive galaxies
in AC114 span an order of magnitude in mass; interestingly, looser
cluster members at higher and lower redshift tend to define a mild
tendency of having masses increasing while moving to more recent times:
from 9.8 to 10.3 Gyr ago, their mass increases from $0.4\times10^{11}\,M_{\odot}$
to $10^{11}M_{\odot}$. It could be inferred that some galaxies continued
forming stars both before and after a cluster-wide main SF episode.

On the contrary, Fig. \ref{fig:Mass-of-active-vs-z} shows that active
galaxies closer to us have lower masses than objects located in the
main body of the cluster, which also span a smaller mass range of
$\sim0.5$~dex, and are generally less massive than passive galaxies
in the same region. This confirms the idea that active galaxies formed
later than passive ones, and did not participate to the initial strong
galaxy formation epoch.

\subsection{Metallicities \label{subsec:Metallicities}}

\begin{table*}
\caption{Main characteristics of passive galaxies entering the mass-metallicity
relation. Luminosities and masses are given in $10^{10}$ solar units.
Galaxies are sorted by luminosity, from fainter to brighter. Typical
error on $R$ magnitudes is 0.5, which yields $\sim50\%$ errors on
luminosities and masses, after accounting for the uncertainty in mass-to-light
ratios. \label{tab:Main-characteristics-of-passive}}

\begin{tabular}{lllllllllllll}
 &  &  &  &  &  &  &  &  &  &  &  & \tabularnewline
\hline 
\hline 
Q\#-sl\#  & RA  & DEC  & $z$  & $R$  & $M_{R}$  & $L_{R}$  & \multicolumn{2}{c}{$\mathcal{M}/\mathcal{M_{\odot}}$ $\pm$} & \multicolumn{2}{c}{$D_{n}(4000)$ $\pm$} & \multicolumn{2}{c}{{[}m/H{]} $\pm$}\tabularnewline
\hline 
 &  &  &  &  &  &  &  &  &  &  &  & \tabularnewline
Q4-45  & 344.652  & -34.781  & 0.31390  & 20.600  & -20.490  & 0.990  & 2.390  & 1.110  & 1.589  & 0.099  & -0.668  & 0.308\tabularnewline
Q4-23  & 344.646  & -34.743  & 0.32270  & 20.600  & -20.560  & 1.060  & 2.550  & 1.180  & 1.387  & 0.095  & -1.355  & 0.402\tabularnewline
Q4-24  & 344.638  & -34.744  & 0.31450  & 20.500  & -20.600  & 1.090  & 2.630  & 1.220  & 1.438  & 0.076  & -1.160  & 0.296\tabularnewline
Q4-16  & 344.623  & -34.732  & 0.31850  & 20.400  & -20.730  & 1.230  & 2.970  & 1.380  & 1.660  & 0.121  & -0.481  & 0.336\tabularnewline
Q4-21  & 344.711  & -34.738  & 0.30610  & 19.500  & -21.530  & 2.580  & 3.117  & 1.448  & 1.402  & 0.086  & -1.022  & 0.445\tabularnewline
Q3-26  & 344.618  & -34.614  & 0.31480  & 20.200  & -20.900  & 1.440  & 3.480  & 1.620  & 1.494  & 0.096  & -0.962  & 0.349\tabularnewline
Q4-43  & 344.620  & -34.778  & 0.31720  & 20.200  & -20.920  & 1.470  & 3.540  & 1.650  & 1.394  & 0.084  & -1.327  & 0.350\tabularnewline
Q4-63  & 344.719  & -34.811  & 0.31870  & 20.200  & -20.930  & 1.490  & 3.580  & 1.660  & 1.386  & 0.082  & -1.359  & 0.345\tabularnewline
Q4-28  & 344.612  & -34.751  & 0.31470  & 20.000  & -21.100  & 1.730  & 4.180  & 1.940  & 1.592  & 0.090  & -0.659  & 0.276\tabularnewline
Q2-35  & 344.919  & -34.651  & 0.31290  & 19.700  & -21.380  & 2.260  & 5.440  & 2.530  & 1.642  & 0.092  & -0.525  & 0.257\tabularnewline
Q4-41  & 344.616  & -34.774  & 0.31820  & 19.700  & -21.430  & 2.350  & 5.650  & 2.630  & 1.626  & 0.081  & -0.567  & 0.231\tabularnewline
Q4-20  & 344.664  & -34.739  & 0.30860  & 19.600  & -21.450  & 2.400  & 5.780  & 2.690  & 1.704  & 0.099  & -0.379  & 0.244\tabularnewline
Q4-2  & 344.646  & -34.707  & 0.31980  & 19.500  & -21.640  & 2.850  & 6.880  & 3.200  & 1.581  & 0.074  & -0.690  & 0.228\tabularnewline
Q4-38  & 344.618  & -34.767  & 0.30100  & 19.200  & -21.790  & 3.270  & 7.890  & 3.670  & 1.576  & 0.136  & -0.705  & 0.447\tabularnewline
Q4-47  & 344.645  & -34.785  & 0.31670  & 19.300  & -21.810  & 3.350  & 8.080  & 3.760  & 1.624  & 0.092  & -0.572  & 0.266\tabularnewline
Q3-40  & 344.642  & -34.648  & 0.31830  & 19.300  & -21.830  & 3.390  & 8.180  & 3.800  & 1.753  & 0.090  & -0.278  & 0.195\tabularnewline
Q4-58  & 344.674  & -34.802  & 0.31080  & 19.200  & -21.870  & 3.520  & 8.490  & 3.940  & 1.694  & 0.096  & -0.401  & 0.241\tabularnewline
Q4-49  & 344.665  & -34.788  & 0.31820  & 20.000  & -21.130  & 1.780  & 8.560  & 3.971  & 1.895  & 0.164  & -0.185  & 0.255\tabularnewline
Q4-51  & 344.693  & -34.792  & 0.30950  & 19.100  & -21.960  & 3.830  & 9.220  & 4.280  & 1.748  & 0.151  & -0.288  & 0.357\tabularnewline
Q4-31  & 344.652  & -34.755  & 0.31780  & 19.100  & -22.020  & 4.060  & 9.800  & 4.550  & 1.825  & 0.094  & -0.155  & 0.166\tabularnewline
Q4-32  & 344.658  & -34.757  & 0.31710  & 18.200  & -22.920  & 9.260  & 22.330  & 10.380  & 1.947  & 0.103  & -0.014  & 0.114\tabularnewline
Q4-52  & 344.609  & -34.794  & 0.31790  & 18.200  & -22.920  & 9.320  & 22.460  & 10.440  & 1.925  & 0.104  & -0.033  & 0.129\tabularnewline
Q4-64  & 344.631  & -34.815  & 0.31450  & 18.300  & -22.800  & 8.290  & 39.865  & 18.536  & 2.077  & 0.136  & -0.072  & 0.067\tabularnewline
Q4-44  & 344.671  & -34.779  & 0.31520  & 18.300  & -22.800  & 8.330  & 40.065  & 18.616  & 2.063  & 0.133  & -0.075  & 0.075\tabularnewline
 &  &  &  &  &  &  &  &  &  &  &  & \tabularnewline
\hline 
\end{tabular}
\end{table*}

\begin{table*}
\caption{Main characteristics of active galaxies entering the mass-metallicity
relation. Luminosities and masses are given in $10^{10}$ solar units.
Galaxie are sorted by luminosity, from fainter to brighter. Typical
error on $R$ magnitudes is 0.5, which yields $\sim50\%$ errors on
luminosities and masses, after accounting for the uncertainty in mass-to-light
ratios.\label{tab:Main-characteristics-of-active}}

\begin{tabular}{lllllllllllll}
 &  &  &  &  &  &  &  &  &  &  &  & \tabularnewline
\hline 
\hline 
Q\#-sl\#  & RA  & DEC  & $z$  & $R$  & $M_{R}$  & $L_{R}$  & \multicolumn{2}{c}{$\mathcal{M}/\mathcal{M_{\odot}}$ $\pm$} & \multicolumn{2}{c}{$D_{n}(4000)$ $\pm$} & \multicolumn{2}{c}{{[}m/H{]} $\pm$}\tabularnewline
\hline 
 &  &  &  &  &  &  &  &  &  &  &  & \tabularnewline
Q4-15  & 344.608  & -34.731  & 0.30800  & 20.900  & -20.140  & 0.720  & 0.460  & 0.190  & 1.117  & 0.063  & -2.497  & 0.399\tabularnewline
Q4-14  & 344.673  & -34.730  & 0.30460  & 20.700  & -20.320  & 0.840  & 0.540  & 0.220  & 0.931  & 0.071  & -3.674  & 0.449\tabularnewline
Q2-38  & 344.846  & -34.663  & 0.29870  & 20.100  & -20.870  & 1.400  & 0.900  & 0.370  & 1.013  & 0.053  & -3.155  & 0.335\tabularnewline
Q4-19  & 344.609  & -34.737  & 0.31070  & 20.200  & -20.870  & 1.400  & 0.900  & 0.360  & 0.996  & 0.048  & -3.262  & 0.304\tabularnewline
Q4-46  & 344.712  & -34.784  & 0.32950  & 20.100  & -21.110  & 1.760  & 1.130  & 0.460  & 1.155  & 0.056  & -2.256  & 0.354\tabularnewline
Q1-18  & 344.846  & -34.751  & 0.31300  & 19.700  & -21.380  & 2.260  & 1.450  & 0.590  & 1.137  & 0.073  & -2.370  & 0.462\tabularnewline
Q3-1  & 344.673  & -34.555  & 0.32950  & 19.800  & -21.410  & 2.320  & 1.480  & 0.600  & 1.175  & 0.063  & -2.129  & 0.399\tabularnewline
Q1-37  & 344.849  & -34.791  & 0.31600  & 19.600  & -21.510  & 2.530  & 1.620  & 0.660  & 1.290  & 0.056  & -1.402  & 0.354\tabularnewline
Q4-26  & 344.675  & -34.748  & 0.32460  & 18.900  & -22.280  & 5.130  & 1.644  & 1.330  & 1.071  & 0.031  & -2.410  & 0.230\tabularnewline
Q1-30  & 344.901  & -34.779  & 0.31860  & 19.600  & -21.530  & 2.580  & 1.650  & 0.670  & 1.251  & 0.066  & -1.648  & 0.418\tabularnewline
Q2-32  & 344.854  & -34.643  & 0.30910  & 19.500  & -21.550  & 2.640  & 1.690  & 0.690  & 1.174  & 0.055  & -2.136  & 0.348\tabularnewline
Q4-22  & 344.666  & -34.741  & 0.31450  & 19.400  & -21.700  & 3.010  & 1.930  & 0.780  & 1.137  & 0.059  & -2.370  & 0.373\tabularnewline
Q4-48  & 344.621  & -34.787  & 0.31460  & 20.000  & -21.100  & 1.730  & 2.215  & 0.450  & 1.470  & 0.088  & -0.726  & 0.392\tabularnewline
Q3-12  & 344.654  & -34.583  & 0.30940  & 19.000  & -22.060  & 4.190  & 2.680  & 1.090  & 1.227  & 0.051  & -1.800  & 0.323\tabularnewline
Q4-40  & 344.604  & -34.772  & 0.31780  & 19.000  & -22.120  & 4.460  & 2.850  & 1.160  & 1.225  & 0.037  & -1.813  & 0.234\tabularnewline
Q4-60  & 344.685  & -34.806  & 0.31830  & 19.000  & -22.130  & 4.470  & 2.860  & 1.160  & 1.224  & 0.039  & -1.819  & 0.247\tabularnewline
Q1-40  & 344.848  & -34.799  & 0.32280  & 18.100  & -23.060  & 10.580  & 3.393  & 2.750  & 1.081  & 0.047  & -2.335  & 0.348\tabularnewline
Q4-36  & 344.650  & -34.764  & 0.31310  & 18.400  & -22.690  & 7.480  & 9.557  & 1.950  & 1.757  & 0.094  & -0.025  & 0.132\tabularnewline
Q4-56  & 344.684  & -34.800  & 0.31380  & 18.400  & -22.690  & 7.520  & 9.597  & 1.960  & 1.743  & 0.095  & -0.039  & 0.149\tabularnewline
 &  &  &  &  &  &  &  &  &  &  &  & \tabularnewline
\hline 
\end{tabular}
\end{table*}

\subsubsection{Introduction of the D4000 index \label{subsec:Introduction-of-the-d4000}}

Having measured the masses of AC114 galaxies in the previous sections,
metallicities were then estimated using the $D_{n}(4000)$ index,
which measures the ratio of spectral continuum at the red and blue
side of the 4000Å~discontinuity. As in \citet{Balogh1999DifferentialGalaxyEvolutioninClusterandFieldGalaxiesatZ0.5ex0.3},
the average of the continuum and its dispersion were calculated within
the bands 3850--3950Å~and 4000--4100Å, both for the galaxy sample
and the models. If we call $\overline{F_{{\rm red}}}$ and $\overline{F_{{\rm blue}}}$
the average continuum level in the two bands computed over $n_{{\rm pix}}$
spectral elements, and $\sigma_{{\rm red}}$ and $\sigma_{{\rm blue}}$
their dispersion, then $D_{n}(4000)$ and its uncertainty are computed
as ${\rm D_{n}(4000)}=\overline{{\rm F_{red}}}/\overline{{\rm F_{blue}}}$
and $\sigma_{{\rm D_{n}(4000)}}=D_{n}(4000)\times(\sigma_{{\rm F,blue}}/\overline{{\rm F_{blue}}}+\sigma_{{\rm F,red}}/\overline{{\rm F_{red}}})$,
where $\sigma_{{\rm {\rm F}}}=\sigma_{{\rm px}}/\sqrt{n_{{\rm px}}}$.

\subsubsection{$D_{n}(4000)$ vs. mass for galaxy spectra \label{subsec:d4000--vs.-mass}}

\begin{figure}
\includegraphics[width=1\columnwidth]{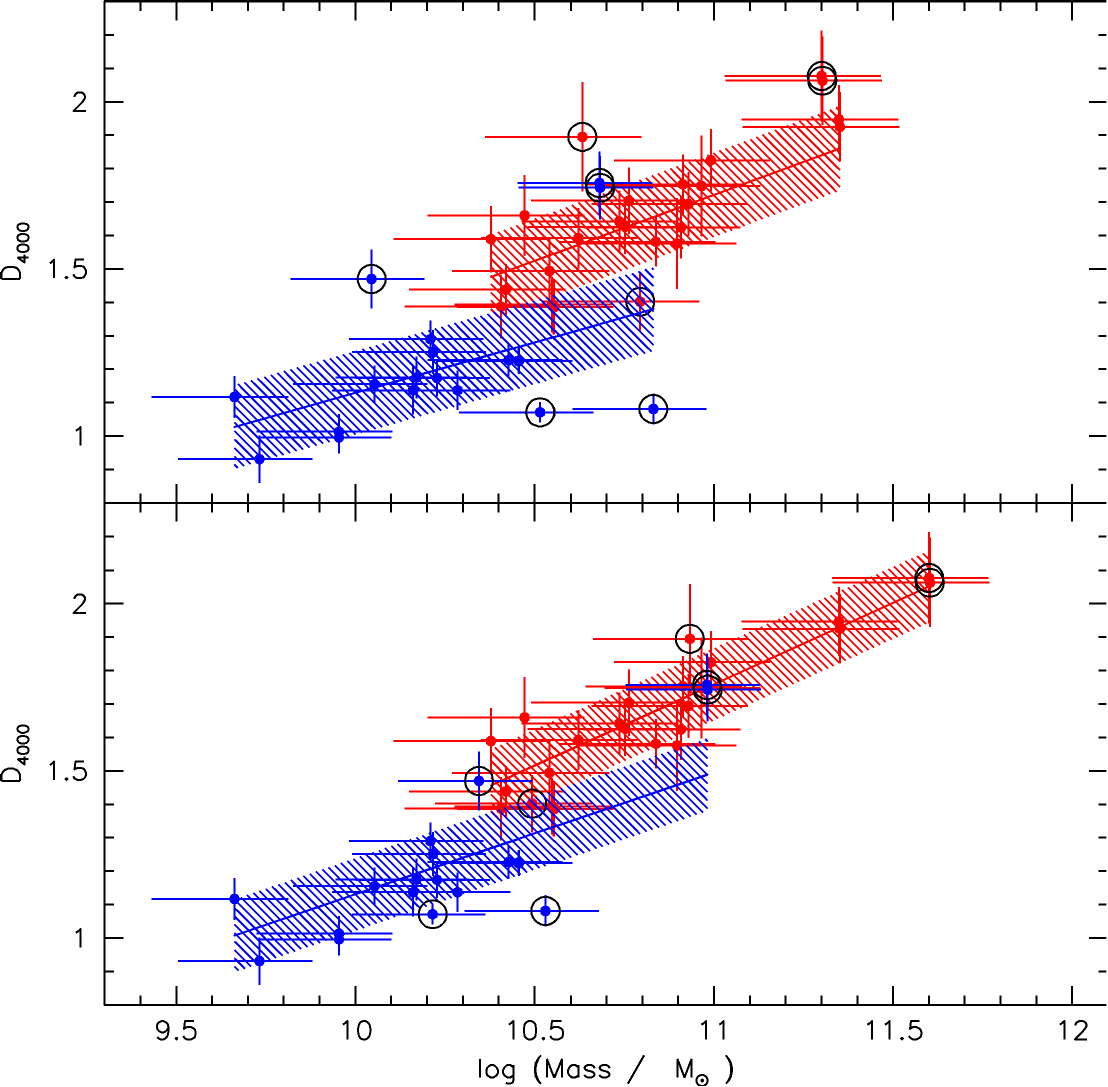}

\caption{Relationship between $D_{n}(4000)$ and galaxy mass, for passive (red
dots) and active (blue dots) galaxies. The straight lines and shaded
areas represent weighted linear fits to the data, and the dispersion
around the fit. The top panel shows the original measurements, with
the most evident outliers marked with large circles. To these objects,
a correction in mass has been applied, as explained in the text, and
their new position can be seen in the bottom panel. Their spectra
are plotted in Fig. \ref{fig:Spectra-of-passive-active-d4000-out}.
\label{fig:d4000-galaxies}}
\end{figure}

Figure \ref{fig:d4000-galaxies} shows the results of our measurements,
with $D_{n}(4000)$ plotted as a function of galaxy mass. It appears
that passive and active galaxies occupy distinct places in such diagram,
with active galaxies generally with lower masses and smaller $D_{n}(4000)$
values, compared with passive galaxies. In the hypotheses that galaxies
within a class have a common age (Sec.~\ref{subsec:Representative-age}),
the variation of the index must be mainly due to metallicity differences,
thus the graph can be interpreted as passive galaxies having higher
metallicities than active ones.

Figure \ref{fig:d4000-galaxies} also shows that some objects lay
outside the general trend defined by galaxies of comparable mass,
which can be interpreted by using Fig.~\ref{fig:Relationship-between-d4000-mh-theo}:
it is likely that galaxies with larger/smaller values of $D_{n}(4000)$
have typical ages that are larger/smaller than the typical age inferred
in Sec.~\ref{subsec:Representative-age}.

\subsubsection{Special cases of of $D_{n}(4000)$\label{subsec:Special-d4000}}

\begin{figure*}
\noindent \begin{centering}
\includegraphics[height=0.3\textheight]{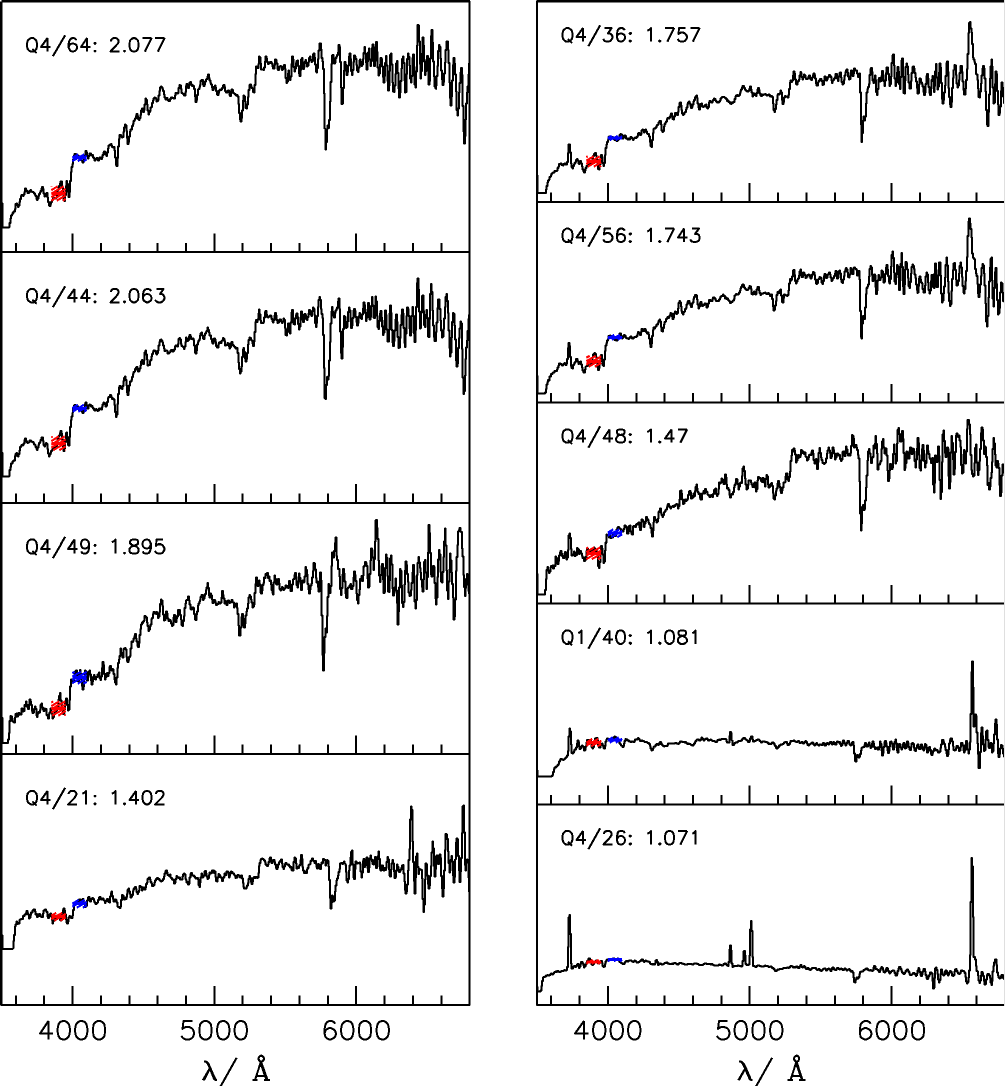}\includegraphics[height=0.3\textheight]{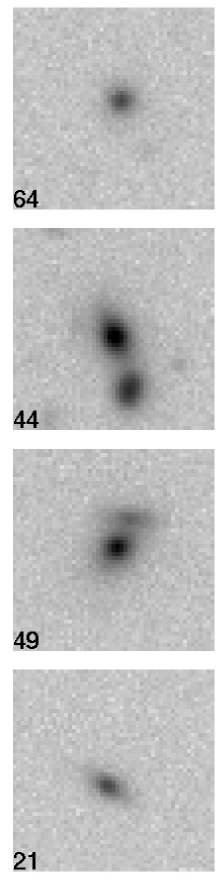}\includegraphics[height=0.3\textheight]{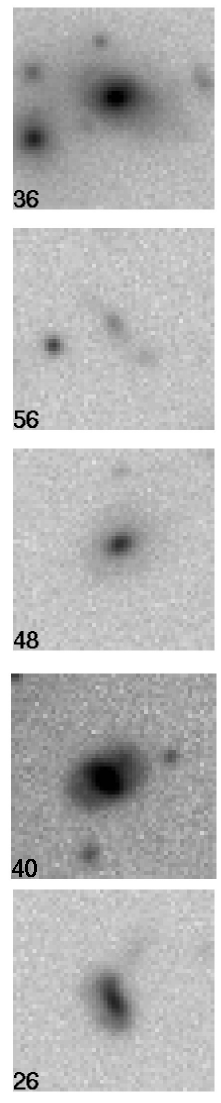} 
\par\end{centering}
\caption{In the left two panels, spectra of passive galaxies having $D_{n}(4000)$
values that are larger or smaller than those of galaxies of comparable
mass, are plotted in the left-hand panels, with index values increasing
from bottom to top. The right-hand panels plot spectra of `deviant'
active galaxies. Within each panel, the $D_{n}(4000)$ value is printed
in the top left corner, following the galaxy identification. In each
spectrum, the blue and red bands highlight the wavelength range that
is used to compute the ratio of continuum near the 4000~Å~break.
In the right two panels, images of the same galaxies, extracted from
the VIMOS preimaging are displayed. \label{fig:Spectra-of-passive-active-d4000-out}}
\end{figure*}

An inspection of Fig. \ref{fig:Spectra-of-passive-active-d4000-out}
shows that such interpretation must be correct: active galaxies with
large $D_{n}(4000)$ values have red SEDs which resemble those of
passive galaxies of similar index values, thus having ages greater
than the typical 2~Gyr. Furthermore, their spectra have very weak
{[}\ion{O}{II}{]}$\lambda3727$ and H$\alpha$ emission lines,
which indicates a low SFR. At the opposite side, active galaxies with
relatively low index values, have blue SEDs and very prominent emission
lines, indicative of an age younger than 2~Gyr. For these galaxies,
a second possibility is a metallicity which is intrinsically lower
than galaxies of comparable mass, for example due to accretion of
fresh gas: indeed the same Fig.~\ref{fig:Spectra-of-passive-active-d4000-out}
shows a low-luminosity tail emerging from object Q4/26, so this option
cannot be excluded.

As Fig.~\ref{fig:V-band-mass-to-light-ratio} demonstrates, an age
increase/decrease translates into a larger/smaller M/L ratio, with
a change of 0.1~dex in M/L for every change of 0.1~dex in age. Thus,
if galaxies with large $D_{n}(4000)$ are older than the others, their
luminosities must be converted into larger masses: from Fig.~\ref{fig:d4000-galaxies},
it can be evinced that corrections of $\sim\pm0.3$ dex in mass, i.e.
in M/L, would bring deviant galaxies back into the general trend.
Such corrections are within the uncertainties of the measurements.

Such a change would bring the age of active galaxies with large $D_{n}(4000)$
close to that of passive galaxies, in agreement with the conclusion
inferred from their spectra. At the other side, active galaxies with
a low discontinuity index would re-enter the general trend, if their
ages were $\sim1$~Gyr.

Looking at passive galaxies, object Q4/49 has a $D_{n}(4000)$ value
and SED close to that of objects Q4/44 and Q4/64: again, a correction
of $\sim+0.3$~dex in M/L, corresponding to a correction of $\sim+0.3$~dex
in age, would bring its mass in agreement with the general trend.
Conversely, galaxies with relatively low index values have bluer SEDs
than galaxies with masses initially estimated to be the same: a reduction
in age of a few Gyr would thus move them to the left in the figure,
and reconcile $D_{n}(4000)$ measurements with estimated masses.

The lower panel of Fig. \ref{fig:d4000-galaxies} shows how the mass
and hence age corrections proposed above improve the definition of
the loci occupied by the two classes of galaxies.

\subsubsection{$D_{n}(4000)$ for model spectra}

\begin{figure}
\includegraphics[width=1\columnwidth]{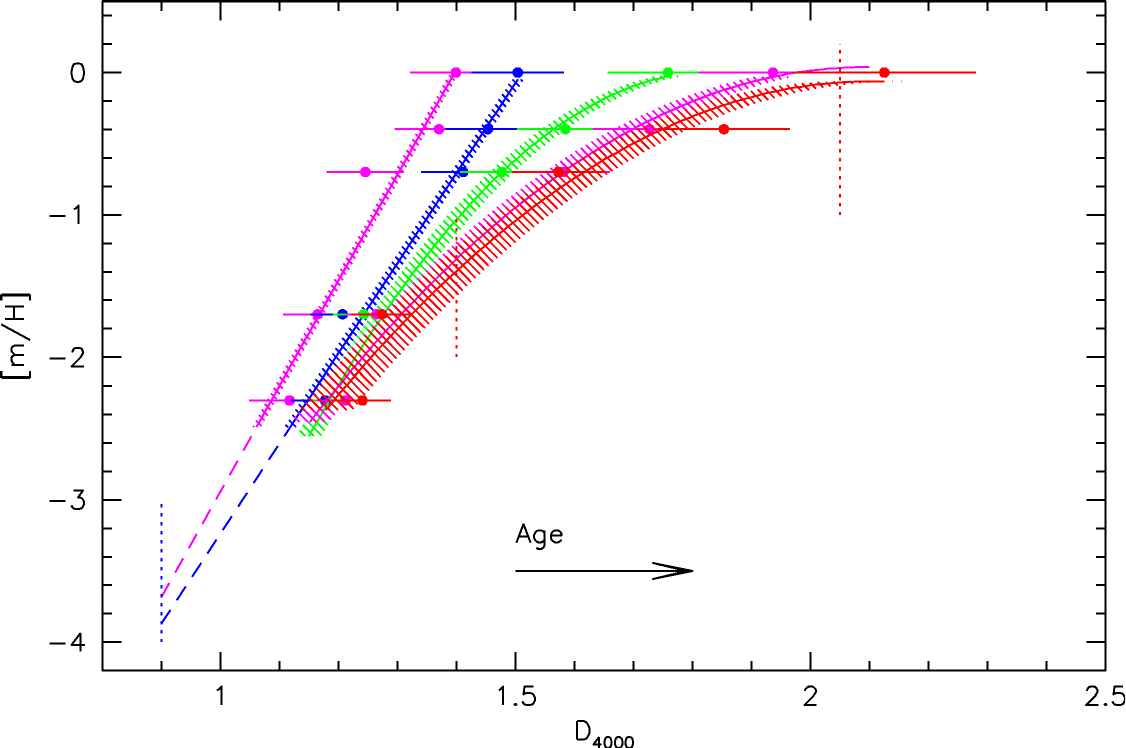}

\caption{Relationship between $D_{n}(4000)$ and {[}m/H{]} from the BC03 model
spectra, for ages 1, 2, 5, 10, and 20 Gyr, going from left to right.
The shaded areas represent linear (for ages $\protect\leq2$ Gyr)
and quadratic fits to the points, and the dispersion around the fits.
The vertical blue dotted segment marks the smaller value of $D_{n}(4000)$
for active galaxies, while the vertical red dotted segments mark the
range of $D_{n}(4000)$ for passive galaxies. \label{fig:Relationship-between-d4000-mh-theo}}
\end{figure}

To convert $D_{n}(4000)$ values to metallicities, a calibration of
the index was obtained from BC03 models, as shown in Fig.~\ref{fig:Relationship-between-d4000-mh-theo}.
It can be seen that a given value of $D_{n}(4000)$ can correspond
to a range of metallicities, depending on the age of the population:
for most active and passive galaxies, the calibration for ages 2 Gyr
and 10 Gyr were used. Calibrations for 5 Gyr and 20 Gyr (in a relative
extrapolation) were used for passive galaxies needing correction to
their masses as explained in Sect. \ref{subsec:Special-d4000}, while
for active galaxies calibrations for 1 Gyr and 5 Gyr were used.

The figure also shows that $D_{n}(4000)$ measured for active galaxies
reaches values lower than those spanned by theoretical models. Although
the linear extrapolation looks relatively safe, metallicities lower
than {[}m/H{]}$\sim-2.5$ must be regarded as uncertain.

Metallicities can now be added to the main parameters of passive and
active galaxies in AC114, which are summarised in Tables~\ref{tab:Main-characteristics-of-passive}
and \ref{tab:Main-characteristics-of-active}. It is also useful to
be able to inspect the morphology of each object, therefore Figs.~\ref{fig:Images-of-galaxies-passive}
and~\ref{fig:Images-of-galaxies-active} in Appendix~B show images
of the same galaxies, extracted from the VIMOS pre-imaging.

\section{Discussion}

Taking the key quantities from the aforementioned tables, we can now
use them to define the evolutionary status of galaxies in AC~114.
In particular, we can construct the mass-metallicity relation, and
interpret it by way of a simple model of chemical evolution: it will
tell us that both the mass of a galaxy, and its location within the
cluster, are key drivers of evolution.

\subsection{Mass-metallicity relation }

\begin{figure}
\includegraphics[width=1\columnwidth]{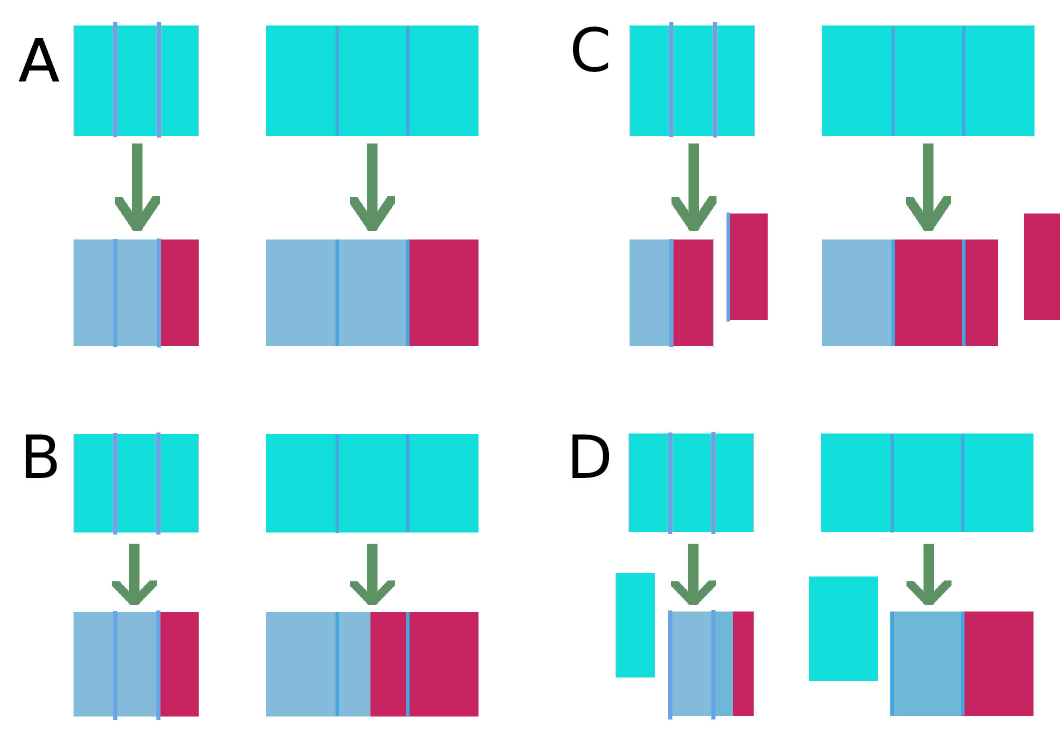}

\caption{Illustration of different ways to obtain an MZR, with boxes representing
two galaxies of different mass, and arrows connecting the start and
end points of their evolution, assumed to happen in the same time
interval. Light blue is pristine gas, darker blue is enriched gas,
and red represents metals. In case (A), two galaxies of different
mass produce the same fraction of metals over pristine gas, so no
MZR is generated. In case (B), the more massive galaxy generates a
higher fraction of metals. In case (C), both galaxies produce the
same fraction of metals, but the low-mass objects ejects more of them
into the IGM. Finally in case (D), gas is lost before chemical evolution
is complete, in such a way that the less massive galaxy will have
generated a smaller fraction of metals, compared with the more massive
one.\label{fig:Illustration-of-different-mzr}}
\end{figure}

As subsequent stellar generations are created within a galaxy, they
enrich the ISM and build the stellar mass at the expense of gas mass.
It is therefore expected that mass and metallicity grow in time, and
if galaxies were able to perform the gas conversion at the same rate,
isolated galaxies of the same age would have converted the same fraction
of gas into stars, and thus would have the same metallicity, irrespective
of mass. But galaxies show an ubiquitous mass-metallicity relation,
as testified by a vast literature on the subject \citep[see, e.g.][ for a review]{Maiolino2019DeReMetallicatheCosmicChemicalEvolutionofGalaxies}.
This fact can have several explanations, which are illustrated in
Fig.~\ref{fig:Illustration-of-different-mzr}: galaxies of larger
mass might be able to convert larger fractions of gas into stars,
in the same amount of time (higher SF efficiency); their stellar generations
might be able to produce more metals than those in low-mass galaxies
(higher yield); metals produced by low-mass galaxies might be more
easily lost into the IGM; or finally, low-mass galaxies might have
been originally higher mass objects, whose SF was truncated due to
loss of residual gas into the IGM. It is then interesting to investigate
the existence of a MZR of AC114 galaxies, and see what it can tell
us about the evolutionary history of the cluster.

\subsubsection{The MZR and its Interpretation}

\begin{figure*}
\begin{centering}
\includegraphics[width=0.8\textwidth]{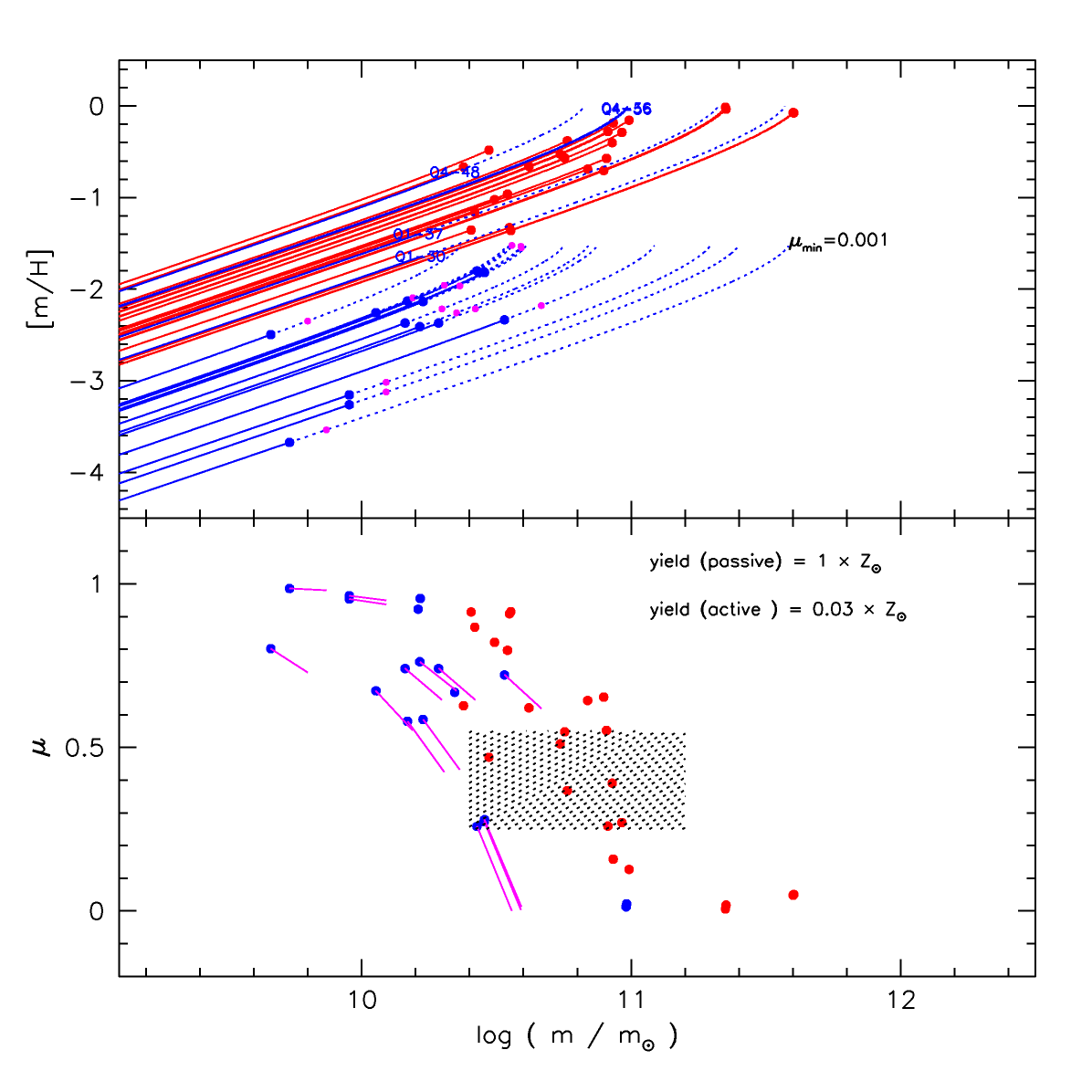} 
\par\end{centering}
\caption{The top panel show the location of passive galaxies (red symbols)
and active galaxies (blue symbols) in the {[}m/H{]} vs. stellar mass
plane. The tracks are close-box models as explained in the text: for
passive galaxies the tracks just reach their representative point,
while for active galaxies we also plot the expected evolution until
close to terminal gas fraction $\mu=0.001$. Tracks are not plotted
for active galaxies located in the passive galaxy region, which are
identified by their quadrant-slit combination. The magenta points
on active tracks represent the expected evolution until the present
time for a linear $\mu(t)$ function set by past evolution. The lower
panel shows the calculated $\mu$ as a function of the galaxy mass
and metallicity, assuming the yield values displayed in the plot.
The grey area shows the typical value of $\mu$ for local galaxies,
taken from Fig. \ref{fig:Distribution-of-local-mu}. Magenta segments
represent the expected evolution until the present time. .\label{fig:tracks}}
\end{figure*}

\begin{figure}
\noindent \begin{centering}
\includegraphics[width=0.7\columnwidth]{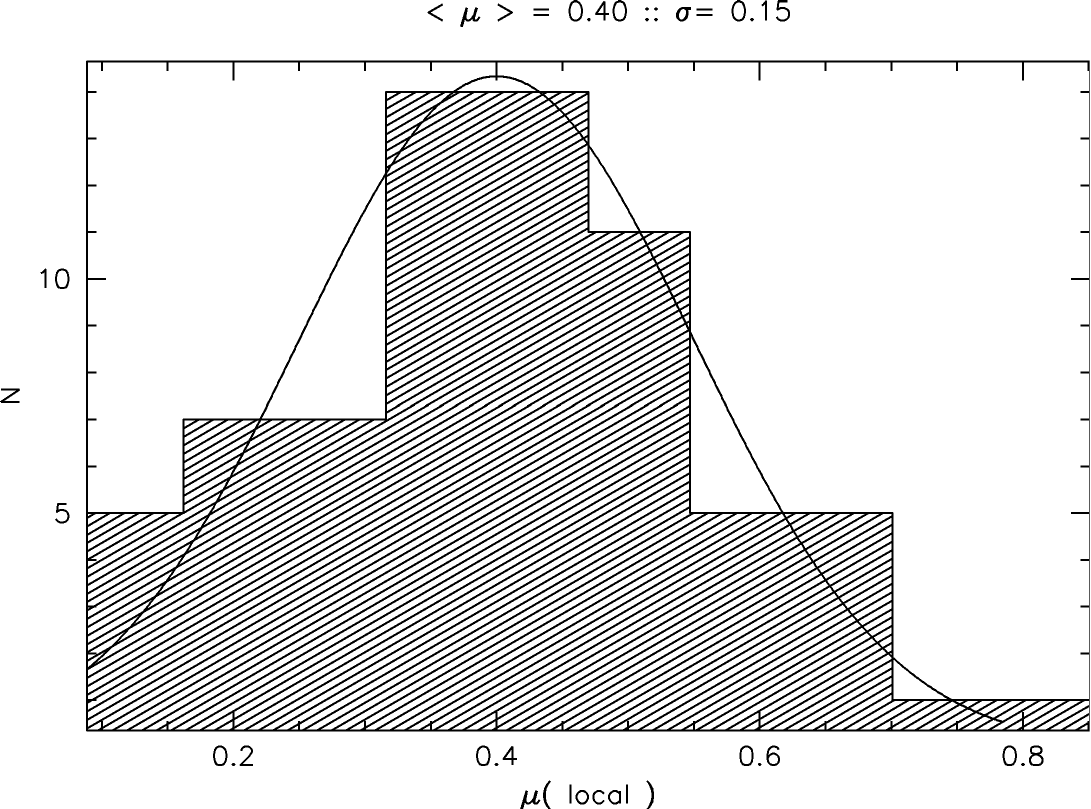} 
\par\end{centering}
\caption{Gas mas fraction distribution for local spiral galaxies, with data
from \citet{McGaugh1997GasMassFractionsandtheEvolutionofSpiralGalaxies}.
The curve is a Gaussian with average and dispersion as displayed in
the plot. \label{fig:Distribution-of-local-mu}}
\end{figure}

When $D_{n}(4000)$ is converted into a metallicity, Fig. \ref{fig:d4000-galaxies}
is converted into Fig. \ref{fig:tracks}, which shows the location
of AC114 galaxies in the {[}m/H{]} vs. mass plane. As expected, the
trend of $D_{n}(4000)$ with mass converts into a mass-metallicity
relation, for the two galaxy classes. To interpret the MZR, we make
use of the so-called closed-box model \citep{Searle1972}, which,
despite its simplicity, allows to draw interesting conclusions on
the evolutionary status of the galaxy cluster.

In the close box approximation, the evolution of a galaxy is parameterised
by the gas mass fraction $\mu=m_{{\rm gas}}/m_{{\rm tot}}=m_{{\rm gas}}/(m_{{\rm gas}}+m_{{\rm stars}})$:
a galaxy starts as a mass of pristine gas ($\mu=1$), which is converted
into stars thereby producing metals, until $\mu=0$. All along, the
total mass remains constant. In the same framework, according to \citet{Pagel1997_NucleosynthesisandChemicalEvolutionofGalaxies}
the average abundance of a stellar population starts at zero and it
evolves with $\mu$ according to the equation:

\begin{equation}
<z>=1+\frac{\mu\,\ln\mu}{1-\mu}\label{eq:<z>}
\end{equation}

\noindent where the metallicity is expressed as $z=Z/p$, and $p$
is the metal yield of a stellar generation\footnote{The yield of metal $i$ is given by $p_{i}=[\int_{m(t)}^{m_{U}}m\,q_{i}(m)\,\phi(m)\,dm]/S$,
where $q_{i}(m)$ is the yield per stellar mass, and $\phi(m)$ is
the initial mass function; the global yield is then $p=\Sigma p_{i}$.
$S$ is the mass in stars and remnants of each generation, such that
$p\times S$ is the mass of metals returned to the ISM.}: therefore $<Z>\rightarrow p$ when $\mu\rightarrow0$. At the end
of its SF history a galaxy will then have an average metallicity similar
to the yield, which can then be estimated in this way. Thus, if passive
galaxies have reached the end of their SF history, then Fig.~\ref{fig:tracks}
is telling us that their yield depends on galaxy mass. And because
the stellar yield only depends on the IMF, what must be changing is
the ability of a galaxy to return its metals to the ISM: indeed a
common explanation for the MZR is that a less massive object has a
shallower gravitational potential from which metals can escape more
easily \citep[e.g., ][]{Dekel1986c}.

\noindent If instead the yield is constant, then $\mu$ must depend
on the galaxy mass: this is shown in the lower panel of Fig.~\ref{fig:tracks},
where $\mu$ has been calculated assuming that the constant yield
is close to that of the most metal rich galaxies, i.e. equal to $Z_{\odot}$.
The plot shows that larger galaxies would have been able to convert
a higher fraction of their initial mass into stars. Because passive
galaxies do not have associated large gas masses, the unprocessed
gas must have been lost into the IGM, thus truncating their SF: again,
the gas dispersal would be easier for less massive objects. Of course
this would be a stochastic process which could be also influenced
by the galaxy location in a cluster and a MZR might be not expected,
but the close-box tracks of Fig.~\ref{fig:tracks} run almost parallel
to the MZR, so even a random truncation of the SF would naturally
lead to a relationship. Therefore this is our preferred explanation
for the distribution of passive galaxies in Fig.~\ref{fig:tracks}.

If active galaxies had the same yield as passive ones, their low metallicities
would be explained by a very slow chemical evolution: the close-box
model predicts that only 2\% of the initial gas mass needs to be processed
($\mu=0.98$), to reach a metallicity {[}m/H{]}=$-2$. The age of
the cluster is $\sim10$~Gyr, so if active galaxies continued their
evolution at the same pace, today ($\sim3.7$~Gyr later) they would
have reached {[}m/H{]}=$-1.86$, and consumed only an additional 1\%
of gas mass.

This means that we should observe a mass of gas associated to active
galaxies $\sim50$ times the mass of stars, and that their total masses
would be more than an order of magnitude larger than those of passive
galaxies. Both deductions are not in agreement with observations,
so we must conclude that active galaxies have a lower yield compared
to passive ones. Indeed this is also the conclusion of \citet{Koeppen2007APossibleOriginoftheMassMetallicityRelationofGalaxies},
based on the idea that stars in less massive galaxies are generated
in less massive stellar clusters: these are not able to form massive
stars, thereby skewing the IMF to lower masses and reducing yields.
More massive galaxies can produce more massive clusters because they
compress their gas more efficiently, as predicted by the Schmidt-Kennicutt
law \citep{Schmidt1959TheRateofStarFormation.,Kennicutt1998TheGlobalSchmidtLawinStarFormingGalaxies}.

To put some constraints to low-mass galaxy yields, we imposed that
the total masses of active galaxies should not be larger than those
of passive ones: thus we plotted the track that reaches the higher
mass at end of evolution, and changed the yield until its total mass
was similar to that of the most massive passive galaxy. The result
is $p=0.03\,Z_{\odot}$, which we adopted for all active galaxies:
this value is well within the yield reduction computed by \citet{Koeppen2007APossibleOriginoftheMassMetallicityRelationofGalaxies}.

As for passive galaxies, the location of active ones along their tracks
constrains their mass fractions, which are plotted in the lower panel
of Fig.~\ref{fig:tracks}. But because these objects are still evolving,
we can expect that nowadays they will have reached higher masses and
metallicities and lower $\mu$ values: therefore we also computed
the expected evolution from an age of the universe $t_{10}\simeq10$~Gyr
to the present time, assuming a linear dependence of $\mu(t)=1-(1-\mu_{10})/t_{10}\times t$.
This evolution is shown by the magenta symbols and magenta segments
in the upper and lower panels of Fig.~\ref{fig:tracks}, respectively.

The lower panel of the figure also shows the $\pm1\,\sigma$ range
of gas fractions in local spiral galaxies, taken from \citet{McGaugh1997GasMassFractionsandtheEvolutionofSpiralGalaxies}
(see Fig.~ \ref{fig:Distribution-of-local-mu}). The comparison tells
us that, at the current pace of evolution, most active galaxies in
AC114 will not be able to reach the same level of gas processing as
galaxies in the local universe.

\noindent Some active galaxies in the upper panel of Fig.~\ref{fig:tracks}
are located in the region of passive galaxies, so a comparable yield
was adopted to compute their evolutionary tracks. These are galaxies
Q1-30, Q1-37, Q4-36, Q4-48, and Q4-56: from Fig.~\ref{fig:Images-of-galaxies-active},
it appears that the first three objects are relatively large galaxies
with a disk/bulge morphology, therefore their high $D_{n}(4000)$
values might be due to the fact that our VIMOS slits include a large
fraction of the central, older and more metal-rich regions. To some
extent this also applies to Q4-48, while the case of Q4-36 requires
more investigation, because its classification as a high luminosity
galaxy does not seem granted by its appearance.

\subsubsection{Relationships for passive and active galaxies}

\noindent 
\begin{figure}
\includegraphics[width=1\columnwidth]{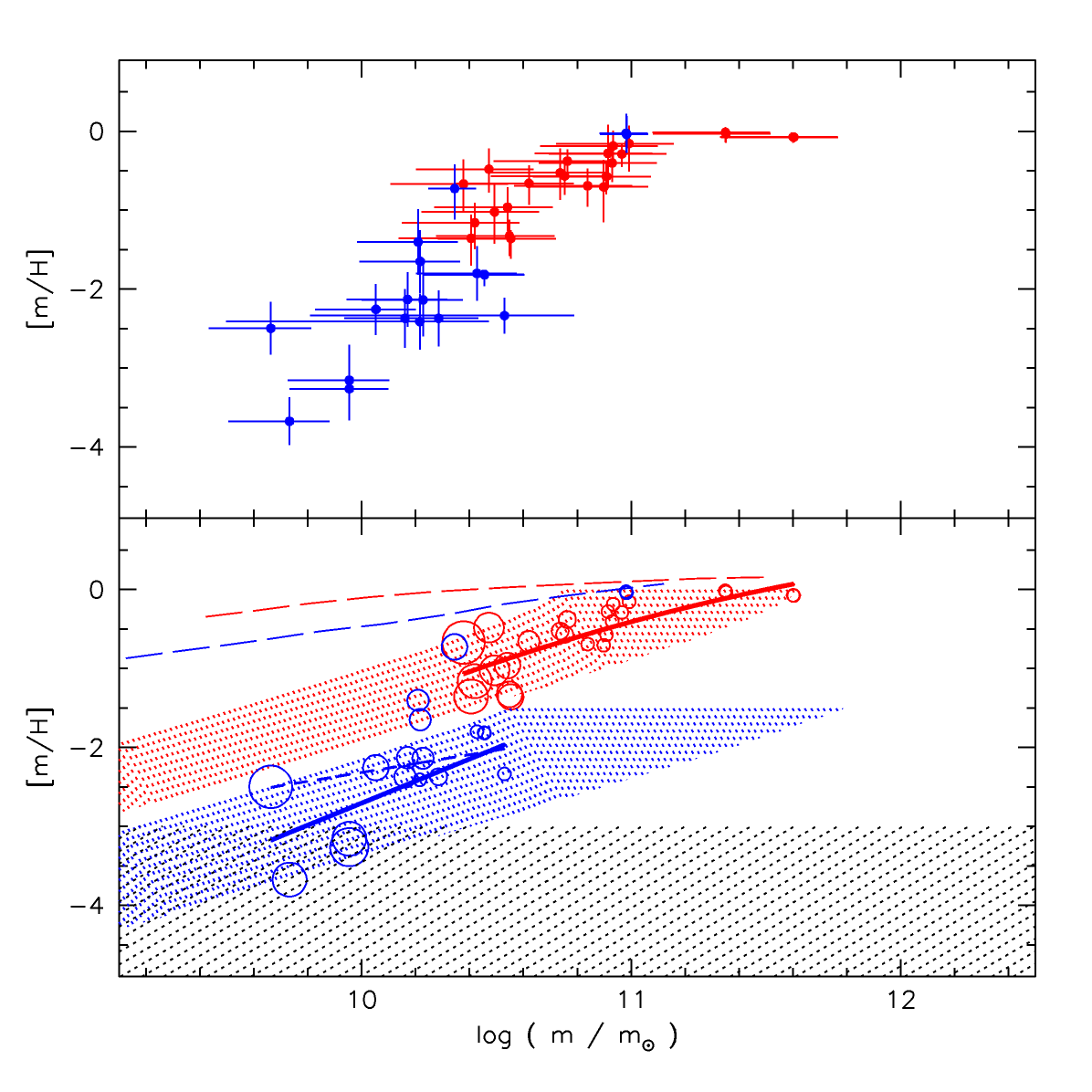}

\caption{The upper panel shows the location of active and passive galaxies
in the {[}m/H{]} vs. mass plane, with the two classes identified by
blue and red colour, respectively. A general trend of metallicity
increasing with mass can be seen. In the lower panel, symbols are
inversely proportional to the measurement errors: the red and blue
solid curves are weighted linear fit to points representing passive
and active galaxies. The blue dashed curve is the relation obtained
after excluding galaxies with metallicities lower than -3. The red
and blue shaded areas mark the location of the tracks taken from Fig.
\ref{fig:tracks}. The dashed curves on top of the panel represent
mass-metallicity relations from \citet{Peng2015StrangulationAsthePrimaryMechanismforShuttingdownStarFormationinGalaxies},
with red and blue colour distinguishing passive and active galaxies.
\label{fig:Mass-metallicity-relation.}}
\end{figure}

To compare the MZR of AC114 galaxies with other relations from the
literature, they are plotted again in Fig.~\ref{fig:Mass-metallicity-relation.},
without model tracks. Weighted linear fits to the data run nearly
parallel to the model tracks discussed in the previous section, both
for passive and active galaxies. Therefore in our scenario, mass-metallicity
relations for AC114 objects are naturally generated by galaxy evolution.

In the lower panel of the figure, dashed blue and red lines are the
relationships for active and passive galaxies found by \citet{Peng2015StrangulationAsthePrimaryMechanismforShuttingdownStarFormationinGalaxies},
who analysed the Sloan Digital Sky Survey (SDSS) spectra of local
galaxies ($z\approx0.05$). Clearly, galaxies in the AC114 cluster
have much lower metallicities than field ones, and for the case of
passive galaxies, we must conclude that the dense cluster environment
curtailed star formation of its more massive members in a short time
after their formation.

Active galaxies are evolving along different tracks than passive ones,
probably because of lower metal yields. And they are also evolving
slower, as it can be inferred from their younger ages and larger mass
fractions. We also expect that, even if allowed to reach the end of
their chemical evolution, they would still be $\sim1$ dex more metal
poor than local active galaxies. The likely explanation is that the
cluster environment facilitates the loss of metals from these lower
mass objects, thus reducing their effective yields even further.

An evolutionary scenario would then be that the first objects to form
got the largest masses, formed significant amounts of metals in SF
bursts that lasted only a few Gyr, cut short through their mutual
interactions. The stripped gas stayed in the cluster potential and
is now facilitating the loss of metals from less massive galaxies
that formed later via ram pressure stripping.

\section{Summary}

In this paper, we investigate the mass-metallicity relation for galaxies
in the Abell cluster~AC114 from 7~hours of VIMOS/MR data collected
at the ESO-VLT telescope in 2009. The dynamical analysis of Proust
\etal (2015) allowed us to select cluster members, whose spectra
are here analyzed with models from Bruzual \& Charlot (2003).

Active and passive galaxies are identified based on the presence/absence
of the {[}\ion{O}{II}{]}$\lambda3727$, and/or {[}\ion{O}{III}{]}$\lambda\lambda4959,5007$
and/or H$\beta$ emission lines, depending on the galaxy redshift,
and we conclude that active galaxies have lower average masses and
metallicities than passive ones.

We establish that the mass-metallicity relation of the cluster is
found to be steeper than that for galaxies in the local universe.
In a forthcoming paper, the MZR of active galaxies, based on the oxygen
abundance of their gaseous component, will be presented and discussed
in light of the present results. 

Fig. \ref{fig:Three-dimensional-representation} showed that passive
and active galaxies tend to occupy central and external regions respectively.
Therefore the results above show the well known fact that galaxy evolution
in the centre of galaxy clusters is faster than that in their periphery.
In the central regions of AC114, and in the course of the last 9.8~Gyr,
galaxies reached masses and metallicities that are comparable to present-day
galaxies, while in the outskirts galaxies are still forming stars,
and in their past they have built substantially less mass than central
objects.

The case of large active galaxies highlights the fact that conclusions
about their global properties are influenced by population gradients
within their bodies. Therefore a conclusive study could only be realised
by a two-dimensional mapping of their SEDs, such as can be afforded
by IFU instruments: in this respect, MUSE at the VLT would be an obvious
choice.

\section*{Acknowledgements}

DP thanks ESO in the context of the \textit{Visiting Scientists program}
for its hospitality at Santiago (Chile). We thank the anonymous referee
for a careful reading of the manuscript, which greatly improved the
presentation of our work. Alain Andrade is also thanked for helping
in the revision of the paper. 


\section*{Data Availability}

The complete redshift values of the 163~galaxies of this work are
listed in Table~5 and the description of each column can be found
in subsection 2.1.




\bibliographystyle{mnras}
\bibliography{AC114}

\begin{table*}
\caption{Positions, photometric data and velocities for galaxies of AC114.
The R column is the correlation peak height to the amplitude of the
asymmetric noise (see text for explanation).\label{tab:Positions,-photometric-data}}
{\small{}{}{}}%
\begin{tabular}{cccccccccl}
\hline 
\textbf{\small{}{}{}Gal.}{\small{}{}{}}  & \textbf{\small{}{}{}R.A.}{\small{}{}{}}  & \textbf{\small{}{}{}Dec.}{\small{}{}{}}  & \textbf{\small{}{}{}$B_{j}$}{\small{}{}{}}  & \textbf{\small{}{}{}$R_{f}$}{\small{}{}{}}  & \textbf{\small{}{}{}R$_{{\rm comp}}$}{\small{}{}{}}  & \textbf{\small{}{}{}redshift}{\small{}{}{}}  & \textbf{\small{}{}{}error}{\small{}{}{}}  & \textbf{\small{}{}{}R}{\small{}{}{}}  & \textbf{\small{}{}{}notes}{\small{}{}{}}\tabularnewline
 & \textbf{\small{}{}{}(J2000)}{\small{}{}{}}  & \textbf{\small{}{}{}(J2000)}{\small{}{}{}}  & \textbf{\small{}{}{}mag}{\small{}{}{}}  & \textbf{\small{}{}{}mag}{\small{}{}{}}  & \textbf{\small{}{}{}mag}{\small{}{}{}}  &  &  &  & \tabularnewline
\hline 
{\small{}{}{}Quadrant 1}  &  &  &  &  &  &  &  &  & \tabularnewline
{\small{}{}{}1}  & {\small{}{}{}22 59 20.70}  & {\small{}{}{}-34 42 24.6}  &  &  & {\small{}{}{}20.0}  & {\small{}{}{}0.39852}  & {\small{}{}{}0.00030}  & {\small{}{}{}3.50}  & {\small{}{}{}em:{H$\alpha$},S1}\tabularnewline
{\small{}{}{}2}  & {\small{}{}{}22 59 31.37}  & {\small{}{}{}-34 42 28.4}  &  &  & {\small{}{}{}20.5}  & {\small{}{}{}0.35306}  & {\small{}{}{}0.00018}  &  & {\small{}{}{}em:OII,{H$\beta$},2OIII,{H$\alpha$},S1}\tabularnewline
{\small{}{}{}3}  & {\small{}{}{}22 59 33.46}  & {\small{}{}{}-34 42 39.8}  &  &  & {\small{}{}{}18.0}  & {\small{}{}{}0.05717}  & {\small{}{}{}0.00011}  &  & {\small{}{}{}em:{H$\beta$},2OIII,{H$\alpha$},S1}\tabularnewline
{\small{}{}{}5}  & {\small{}{}{}22 59 32.31}  & {\small{}{}{}-34 42 58.2}  & {\small{}{}{}20.2}  &  & {\small{}{}{}20.1}  & {\small{}{}{}0.05842}  & {\small{}{}{}0.00009}  &  & {\small{}{}{}em:{H$\beta$},2OIII,{H$\alpha$},S1}\tabularnewline
{\small{}{}{}8}  & {\small{}{}{}22 59 36.51}  & {\small{}{}{}-34 43 25.4}  & {\small{}{}{}20.2}  &  & {\small{}{}{}20.3}  & {\small{}{}{}0.38058}  & {\small{}{}{}0.00022}  &  & {\small{}{}{}em:{H$\alpha$},S1}\tabularnewline
{\small{}{}{}9}  & {\small{}{}{}22 59 25.99}  & {\small{}{}{}-34 43 31.9}  &  &  &  & {\small{}{}{}0.21032?}  &  &  & {\small{}{}{}very uncertain}\tabularnewline
{\small{}{}{}10}  & {\small{}{}{}22 59 18.60}  & {\small{}{}{}-34 43 43.6}  & {\small{}{}{}20.5}  & {\small{}{}{}18.8}  & {\small{}{}{}18.9}  & {\small{}{}{}0.21940}  & {\small{}{}{}0.00024}  & {\small{}{}{}4.76}  & {\small{}{}{}em: {H$\alpha$},S1 65840 \kms}\tabularnewline
{\small{}{}{}11}  & {\small{}{}{}22 59 17.61}  & {\small{}{}{}-34 43 57.1}  &  &  & {\small{}{}{}19.3}  & {\small{}{}{}0.62605}  & {\small{}{}{}0.00031}  &  & {\small{}{}{}measured on OII,H,K (CaI)}\tabularnewline
{\small{}{}{}12}  & {\small{}{}{}22 59 20.05}  & {\small{}{}{}-34 44 09.4}  & {\small{}{}{}20.8}  & {\small{}{}{}19.5}  & {\small{}{}{}18.2}  & {\small{}{}{}0.22127}  & {\small{}{}{}0.00012}  & {\small{}{}{}3.56}  & {\small{}{}{}em:{H$\beta$},{H$\alpha$} 66509 \kms}\tabularnewline
{\small{}{}{}13}  & {\small{}{}{}22 59 20.00}  & {\small{}{}{}-34 44 20.3}  &  &  & {\small{}{}{}20.0}  & {\small{}{}{}0.40975}  & {\small{}{}{}0.00022}  & {\small{}{}{}4.79}  & \tabularnewline
{\small{}{}{}15}  & {\small{}{}{}22 59 28.09}  & {\small{}{}{}-34 44 29.0}  &  &  & {\small{}{}{}19.4}  & {\small{}{}{}0.07309}  & {\small{}{}{}0.00045}  &  & {\small{}{}{}em:{H$\beta$},2OIII,{H$\alpha$}}\tabularnewline
{\small{}{}{}16}  & {\small{}{}{}22 59 28.82}  & {\small{}{}{}-34 44 44.1}  &  &  & {\small{}{}{}18.6}  & {\small{}{}{}0.40819}  & {\small{}{}{}0.00039}  & {\small{}{}{}3.57}  & \tabularnewline
{\small{}{}{}17}  & {\small{}{}{}22 59 38.42}  & {\small{}{}{}-34 44 52.4}  &  &  & {\small{}{}{}20.5}  & {\small{}{}{}0.63163}  & {\small{}{}{}0.00045}  & {\small{}{}{}3.09}  & {\small{}{}{}very weak}\tabularnewline
{\small{}{}{}18}  & {\small{}{}{}22 59 23.15}  & {\small{}{}{}-34 45 03.0}  &  &  & {\small{}{}{}19.7}  & {\small{}{}{}0.31301}  & {\small{}{}{}0.00007}  &  & {\small{}{}{}em:{H$\alpha$},S1}\tabularnewline
{\small{}{}{}19}  & {\small{}{}{}22 59 18.51}  & {\small{}{}{}-34 45 13.1}  & {\small{}{}{}21.2}  & {\small{}{}{}20.3}  & {\small{}{}{}19.6}  & {\small{}{}{}0.40876}  & {\small{}{}{}0.00022}  &  & {\small{}{}{}em:OII,{H$\beta$},2OIII,{H$\alpha$},S1}\tabularnewline
{\small{}{}{}20}  & {\small{}{}{}22 59 40.37}  & {\small{}{}{}-34 45 21.8}  &  &  & {\small{}{}{}18.1}  & {\small{}{}{}0.23192}  & {\small{}{}{}0.00018}  & {\small{}{}{}5.23}  & \tabularnewline
{\small{}{}{}22}  & {\small{}{}{}22 59 24.80}  & {\small{}{}{}-34 45 36.6}  & {\small{}{}{}22.3}  & {\small{}{}{}19.6}  & {\small{}{}{}18.8}  & {\small{}{}{}0.41015}  & {\small{}{}{}0.00025}  & {\small{}{}{}5.19}  & {\small{}{}{}em:{H$\beta$},{H$\alpha$}}\tabularnewline
{\small{}{}{}24}  & {\small{}{}{}22 59 28.59}  & {\small{}{}{}-34 45 49.8}  & {\small{}{}{}21.6}  & {\small{}{}{}19.6}  & {\small{}{}{}18.9}  & {\small{}{}{}0.33369}  & {\small{}{}{}0.00018}  & {\small{}{}{}4.84}  & {\small{}{}{}em:OII}\tabularnewline
{\small{}{}{}25}  & {\small{}{}{}22 59 35.08}  & {\small{}{}{}-34 45 57.0}  & {\small{}{}{}22.4}  &  & {\small{}{}{}20.9}  & {\small{}{}{}0.48600}  & {\small{}{}{}0.00055}  &  & {\small{}{}{}measured on H,K (CaI)}\tabularnewline
{\small{}{}{}26}  & {\small{}{}{}22 59 32.09}  & {\small{}{}{}-34 46 13.6}  &  &  & {\small{}{}{}19.6}  & {\small{}{}{}0.53351}  & {\small{}{}{}0.00056}  &  & {\small{}{}{}measured on OII,H,K (CaI)}\tabularnewline
{\small{}{}{}27}  & {\small{}{}{}22 59 32.34}  & {\small{}{}{}-34 46 23.3}  &  &  & {\small{}{}{}19.9}  & {\small{}{}{}0.46615}  & {\small{}{}{}0.00022}  & {\small{}{}{}3.22}  & \tabularnewline
{\small{}{}{}28}  & {\small{}{}{}22 59 26.16}  & {\small{}{}{}-34 46 32.5}  & {\small{}{}{}20.2}  & {\small{}{}{}19.2}  & {\small{}{}{}19.0}  & {\small{}{}{}0.35118}  & {\small{}{}{}0.00028}  & {\small{}{}{}3.11}  & {\small{}{}{}weak confirmed on H,K (CaI)}\tabularnewline
{\small{}{}{}30}  & {\small{}{}{}22 59 36.18}  & {\small{}{}{}-34 46 43.1}  & {\small{}{}{}20.8}  & {\small{}{}{}18.8}  & {\small{}{}{}19.6}  & {\small{}{}{}0.31860}  & {\small{}{}{}0.00033}  & {\small{}{}{}3.15}  & \tabularnewline
{\small{}{}{}31}  & {\small{}{}{}22 59 33.65}  & {\small{}{}{}-34 46 49.2}  &  &  & {\small{}{}{}19.9}  & {\small{}{}{}0.33342}  &  &  & {\small{}{}{}measured on {H$\alpha$}}\tabularnewline
{\small{}{}{}34}  & {\small{}{}{}22 59 25.25}  & {\small{}{}{}-34 47 05.9}  &  &  & {\small{}{}{}20.0}  & {\small{}{}{}0.63182}  & {\small{}{}{}0.00068}  &  & {\small{}{}{}measured on H,K (CaI)}\tabularnewline
{\small{}{}{}36}  & {\small{}{}{}22 59 34.22}  & {\small{}{}{}-34 47 19.9}  &  &  & {\small{}{}{}19.6}  & {\small{}{}{}0.49978}  &  &  & {\small{}{}{}measured on OII}\tabularnewline
{\small{}{}{}37}  & {\small{}{}{}22 59 23.76}  & {\small{}{}{}-34 47 27.4}  & {\small{}{}{}21.0}  & {\small{}{}{}19.8}  & {\small{}{}{}19.6}  & {\small{}{}{}0.31604}  & {\small{}{}{}0.00004}  &  & {\small{}{}{}em:{H$\beta$},2OIII,{H$\alpha$},S1}\tabularnewline
{\small{}{}{}38}  & {\small{}{}{}22 59 18.07}  & {\small{}{}{}-34 47 39.8}  &  &  & {\small{}{}{}19.9}  & {\small{}{}{}0.49805}  & {\small{}{}{}0.00064}  & {\small{}{}{}3.04}  & {\small{}{}{}very weak}\tabularnewline
{\small{}{}{}40}  & {\small{}{}{}22 59 23.41}  & {\small{}{}{}-34 47 57.6}  & {\small{}{}{}20.6}  & {\small{}{}{}18.9}  & {\small{}{}{}18.1}  & {\small{}{}{}0.32283}  & {\small{}{}{}0.00027}  & {\small{}{}{}3.37}  & {\small{}{}{}em:{H$\beta$},2OIII,{H$\alpha$},S1: 96857 \kms}\tabularnewline
{\small{}{}{}41}  & {\small{}{}{}22 59 16.77}  & {\small{}{}{}-34 48 06.8}  &  &  & {\small{}{}{}20.0}  & {\small{}{}{}0.47645}  & {\small{}{}{}0.00054}  &  & {\small{}{}{}measured on H,K (CaI)}\tabularnewline
{\small{}{}{}44}  & {\small{}{}{}22 59 15.92}  & {\small{}{}{}-34 48 43.8}  & {\small{}{}{}22.2}  & {\small{}{}{}21.1}  & {\small{}{}{}19.1}  & {\small{}{}{}0.57779}  & {\small{}{}{}0.00066}  &  & {\small{}{}{}measured on OII,H,K (CaI)}\tabularnewline
{\small{}{}{}45a}  & {\small{}{}{}22 59 31.40}  & {\small{}{}{}-34 48 50.3}  &  &  &  & {\small{}{}{}0.35159}  &  &  & {\small{}{}{}measured on {H$\alpha$},S1}\tabularnewline
{\small{}{}{}45b}  & {\small{}{}{}22 59 31.40}  & {\small{}{}{}-34 48 50.4}  &  &  &  & {\small{}{}{}0.34461}  &  &  & {\small{}{}{}measured on {H$\alpha$},S1}\tabularnewline
{\small{}{}{}Quadrant 2}  &  &  &  &  &  &  &  &  & \tabularnewline
{\small{}{}{}1a}  & {\small{}{}{}22 59 16.30}  & {\small{}{}{}-34 33 40.6}  & {\small{}{}{}20.8}  & {\small{}{}{}19.0}  & {\small{}{}{}19.6}  & {\small{}{}{}0.25782}  & {\small{}{}{}0.00023}  & {\small{}{}{}3.43}  & {\small{}{}{}em:{H$\beta$},2OIII,{H$\alpha$}}\tabularnewline
{\small{}{}{}1b}  & {\small{}{}{}22 59 16.30}  & {\small{}{}{}-34 33 40.6}  &  &  &  & {\small{}{}{}0.33069}  & {\small{}{}{}0.00024}  & {\small{}{}{}3.18}  & {\small{}{}{}em:{H$\alpha$}, 2 galaxies}\tabularnewline
{\small{}{}{}2}  & {\small{}{}{}22 59 18.23}  & {\small{}{}{}-34 33 50.7}  &  &  & {\small{}{}{}18.2}  & {\small{}{}{}0.49578}  & {\small{}{}{}0.00029}  &  & {\small{}{}{}em:OII,{H$\beta$},2OIII,{H$\alpha$},S1}\tabularnewline
{\small{}{}{}3}  & {\small{}{}{}22 59 18.00}  & {\small{}{}{}-34 33 58.5}  &  &  & {\small{}{}{}20.5}  & {\small{}{}{}0.41000}  & {\small{}{}{}0.00033}  &  & {\small{}{}{}em:OII,{H$\beta$},2OIII,{H$\alpha$}}\tabularnewline
{\small{}{}{}4}  & {\small{}{}{}22 59 13.67}  & {\small{}{}{}-34 34 11.1}  &  &  & {\small{}{}{}19.3}  & {\small{}{}{}0.40846}  & {\small{}{}{}0.00033}  &  & {\small{}{}{}em:{H$\beta$},2OIII}\tabularnewline
{\small{}{}{}5}  & {\small{}{}{}22 59 15.03}  & {\small{}{}{}-34 34 23.5}  & {\small{}{}{}22.5}  & {\small{}{}{}20.1}  & {\small{}{}{}18.8}  & {\small{}{}{}0.41440}  & {\small{}{}{}0.00034}  & {\small{}{}{}3.01}  & {\small{}{}{}very uncertain}\tabularnewline
{\small{}{}{}6}  & {\small{}{}{}22 59 23.91}  & {\small{}{}{}-34 34 32.7}  & {\small{}{}{}21.3}  & {\small{}{}{}19.2}  & {\small{}{}{}18.9}  & {\small{}{}{}0.42284}  & {\small{}{}{}0.00021}  & {\small{}{}{}6.45}  & \tabularnewline
{\small{}{}{}7}  & {\small{}{}{}22 59 21.45}  & {\small{}{}{}-34 34 41.8}  &  &  & {\small{}{}{}18.6}  & {\small{}{}{}0.56176}  & {\small{}{}{}0.00052}  &  & {\small{}{}{}measured on H,K (CaI)}\tabularnewline
{\small{}{}{}8}  & {\small{}{}{}22 59 12.89}  & {\small{}{}{}-34 34 48.4}  &  &  & {\small{}{}{}21.2}  & {\small{}{}{}0.25744}  & {\small{}{}{}0.00018}  &  & {\small{}{}{}measured on 2OIII}\tabularnewline
{\small{}{}{}9}  & {\small{}{}{}22 59 17.73}  & {\small{}{}{}-34 34 54.1}  &  &  & {\small{}{}{}20.9}  & {\small{}{}{}0.37079}  & {\small{}{}{}0.00063}  &  & {\small{}{}{}measured on 2OIII}\tabularnewline
{\small{}{}{}10}  & {\small{}{}{}22 59 18.77}  & {\small{}{}{}-34 35 02.2}  & {\small{}{}{}21.5}  & {\small{}{}{}19.9}  & {\small{}{}{}19.5}  & {\small{}{}{}0.40848}  & {\small{}{}{}0.00023}  & {\small{}{}{}3.36}  & \tabularnewline
{\small{}{}{}11}  & {\small{}{}{}22 59 16.88}  & {\small{}{}{}-34 35 08.5}  &  &  & {\small{}{}{}17.6}  & {\small{}{}{}0.34546}  &  &  & {\small{}{}{}measured on {H$\alpha$}}\tabularnewline
{\small{}{}{}12}  & {\small{}{}{}22 59 32.84}  & {\small{}{}{}-34 35 24.7}  & {\small{}{}{}21.9}  &  & {\small{}{}{}19.2}  & {\small{}{}{}0.59523}  &  &  & {\small{}{}{}measured on OII}\tabularnewline
{\small{}{}{}13}  & {\small{}{}{}22 59 26.31}  & {\small{}{}{}-34 35 32.2}  &  &  & {\small{}{}{}17.8}  & {\small{}{}{}0.27013}  & {\small{}{}{}0.00038}  & {\small{}{}{}3.52}  & {\small{}{}{}very weak}\tabularnewline
{\small{}{}{}14}  & {\small{}{}{}22 59 20.21}  & {\small{}{}{}-34 35 43.7}  &  &  & {\small{}{}{}18.7}  & {\small{}{}{}0.37370}  & {\small{}{}{}0.00009}  &  & {\small{}{}{}em:OII,{H$\alpha$},S1}\tabularnewline
{\small{}{}{}15}  & {\small{}{}{}22 59 15.61}  & {\small{}{}{}-34 35 51.1}  &  &  & {\small{}{}{}20.7}  & {\small{}{}{}0.41045}  & {\small{}{}{}0.00026}  & {\small{}{}{}4.01}  & \tabularnewline
{\small{}{}{}16}  & {\small{}{}{}22 59 19.18}  & {\small{}{}{}-34 35 59.7}  & {\small{}{}{}21.4}  & {\small{}{}{}20.4}  & {\small{}{}{}19.7}  & {\small{}{}{}0.56822}  & {\small{}{}{}0.00074}  &  & {\small{}{}{}measured on OII,H,K (CaI)}\tabularnewline
{\small{}{}{}18}  & {\small{}{}{}22 59 12.16}  & {\small{}{}{}-34 36 28.0}  & {\small{}{}{}21.0}  &  & {\small{}{}{}19.4}  & {\small{}{}{}0.38110}  & {\small{}{}{}0.00042}  & {\small{}{}{}4.03}  & \tabularnewline
{\small{}{}{}19}  & {\small{}{}{}22 59 19.31}  & {\small{}{}{}-34 36 38.6}  & {\small{}{}{}22.00}  & {\small{}{}{}20.1}  & {\small{}{}{}18.8}  & {\small{}{}{}0.37706}  & {\small{}{}{}0.00021}  & {\small{}{}{}6.71}  & \tabularnewline
{\small{}{}{}21}  & {\small{}{}{}22 59 21.69}  & {\small{}{}{}-34 36 53.5}  &  &  & {\small{}{}{}19.2}  & {\small{}{}{}0.41304}  & {\small{}{}{}0.00020}  & {\small{}{}{}6.52}  & \tabularnewline
{\small{}{}{}22}  & {\small{}{}{}22 59 36.79}  & {\small{}{}{}-34 36 53.5}  &  &  & {\small{}{}{}21.3}  & {\small{}{}{}0.39104}  &  &  & {\small{}{}{}em:{H$\alpha$}}\tabularnewline
{\small{}{}{}23}  & {\small{}{}{}22 59 21.57}  & {\small{}{}{}-34 37 07.8}  & {\small{}{}{}19.3}  & {\small{}{}{}18.4}  & {\small{}{}{}18.3}  & {\small{}{}{}0.41310}  & {\small{}{}{}0.00022}  & {\small{}{}{}6.81}  & \tabularnewline
{\small{}{}{}24}  & {\small{}{}{}22 59 36.81}  & {\small{}{}{}-34 37 21.2}  &  &  & {\small{}{}{}19.2}  & {\small{}{}{}0.37972}  & {\small{}{}{}0.00052}  &  & {\small{}{}{}em:OII,{H$\beta$},2OIII}\tabularnewline
\hline 
\end{tabular}
\end{table*}

\begin{table*}
\begin{tabular}{cccccccccl}
\hline 
\textbf{Gal.}  & \textbf{R.A.}  & \textbf{Dec.}  & \textbf{$B_{j}$}  & \textbf{$R_{f}$}  & \textbf{R$_{{\rm comp}}$}  & \textbf{redshift}  & \textbf{error}  & \textbf{R}  & \textbf{notes}\tabularnewline
 & \textbf{(J2000)}  & \textbf{(J2000)}  & \textbf{mag}  & \textbf{mag}  & \textbf{mag}  &  &  &  & \tabularnewline
\hline 
26  & 22 59 12.34  & -34 37 37.5  & 22.3  & 20.2  & 19.4  & 0.48676  & 0.00018  &  & em:OII,{H$\beta$},2OIII\tabularnewline
27  & 22 59 21.00  & -34 37 46.0  & 20.6  & 20.0  & 19.9  & 0.49642  & 0.00024  & 4.53  & \tabularnewline
28  & 22 59 15.56  & -34 37 57.2  & 20.8  & 19.4  & 19.7  & 0.61361  & 0.00047  &  & measured on OII,{H$\beta$}\tabularnewline
29  & 22 59 17.62  & -34 38 03.8  &  &  & 20.9  & 0.52230  &  &  & measured on OII\tabularnewline
31  & 22 59 24.67  & -34 38 18.4  & 21.4  & 20.6  & 18.5  & 0.49717  & 0.00016  &  & em:OII,{H$\beta$},2OIII,{H$\alpha$}\tabularnewline
32  & 22 59 25.00  & -34 38 35.6  & 20.8  & 19.5  & 19.5  & 0.30913  & 0.00028  &  & em:{H$\beta$},2OIII,{H$\alpha$}\tabularnewline
33  & 22 59 16.65  & -34 38 43.7  & 20.1  & 18.1  & 20.4  & 0.46922  & 0.00051  & 3.17  & em:{H$\beta$} 140472 \kms\tabularnewline
35  & 22 59 40.57  & -34 39 05.1  &  &  & 19.7  & 0.31290  & 0.00016  & 7.72  & \tabularnewline
36  & 22 59 25.79  & -34 39 11.8  & 21.4  & 20.0  & 19.8  & 0.46367  & 0.00020  & 6.71  & \tabularnewline
37  & 22 59 13.94  & -34 39 33.0  & 21.8  & 20.7  & 18.7  & 0.13963  & 0.00063  &  & measured on {H$\beta$},2OIII\tabularnewline
38  & 22 59 23.08  & -34 39 47.1  &  &  & 20.1  & 0.29869  & 0.00017  &  & em:{H$\beta$},2OIII,{H$\alpha$},S1\tabularnewline
39  & 22 59 39.01  & -34 39 53.1  &  &  & 20.6  & 0.37203  & 0.00012  &  & em:OII,{H$\beta$},2OIII,{H$\alpha$}\tabularnewline
40  & 22 59 36.39  & -34 39 58.4  &  &  & 20.3  & 0.41913  & 0.00023  & 4.96  & \tabularnewline
41  & 22 59 43.80  & -34 40 06.7  &  &  & 20.5  & 0.54027  &  &  & measured on OII\tabularnewline
Quadrant 3  &  &  &  &  &  &  &  &  & \tabularnewline
1  & 22 58 41.57  & -34 33 18.1  & 20.3  & 18.8  & 19.8  & 0.32951  & 0.00015  &  & em:{H$\beta$},2OIII,{H$\alpha$},S1\tabularnewline
2  & 22 58 41.21  & -34 33 27.8  & 20.5  & 18.8  & 18.7  & 0.33307  & 0.00011  &  & em:{H$\beta$},2OIII,{H$\alpha$},S1\tabularnewline
3  & 22 58 40.11  & -34 33 34.6  & 21.9  & 19.2  & 19.4  & 0.33127  & 0.00024  & 5.48  & \tabularnewline
4  & 22 58 40.08  & -34 33 42.2  & 21.3  & 19.9  & 20.0  & 0.21259  & 0.00019  &  & em:{H$\beta$},2OIII,{H$\alpha$},S1\tabularnewline
5  & 22 58 41.10  & -34 33 49.0  &  &  & 20.5  & 0.33189  & 0.00020  &  & em:{H$\beta$},2OIII,{H$\alpha$},S1\tabularnewline
7  & 22 58 48.40  & -34 34 01.6  &  &  &  & 0.31092  & 0.00054  &  & measured on H,K (CaI)\tabularnewline
8  & 22 58 42.73  & -34 34 23.0  & 22.0  &  & 19.7  & 0.39969  & 0.00028  & 3.48  & em:{H$\alpha$}:119724 \kms\tabularnewline
10  & 22 58 22.64  & -34 34 33.3  & 19.8  & 18.7  & 18.8  & 0.31365  & 0.00012  & 4.63  & \tabularnewline
11  & 22 58 35.95  & -34 34 45.6  & 22.4  & 20.6  & 20.1  & 0.21245  & 0.00026  &  & em:{H$\beta$},2OIII,{H$\alpha$},S1\tabularnewline
12  & 22 58 36.88  & -34 34 58.2  & 21.5  & 19.6  & 19.0  & 0.30939  & 0.00009  &  & em:{H$\beta$},2OIII,{H$\alpha$},S1\tabularnewline
13  & 22 58 51.29  & -34 35 07.6  &  &  & 19.3  & 0.65153  & 0.00048  & 3.03  & very weak\tabularnewline
15  & 22 58 41.00  & -34 35 21.4  &  &  & 20.3  & 0.47562  & 0.00038  & 3.01  & very weak\tabularnewline
16  & 22 58 45.84  & -34 35 30.4  & 20.9  & 20.4  & 20.0  & 0.17739  & 0.00012  &  & em:{H$\beta$},2OIII,{H$\alpha$},S1\tabularnewline
18  & 22 58 56.50  & -34 35 48.2  & 20.8  & 19.3  & 19.5  & 0.22176  & 0.00009  &  & em:{H$\beta$},2OIII,{H$\alpha$},S1\tabularnewline
19  & 22 58 23.26  & -34 35 57.0  & 20.0  & 18.6  & 18.9  & 0.15860  & 0.00015  & 4.84  & \tabularnewline
20  & 22 58 32.46  & -34 36 07.9  &  &  & 19.2  & 0.35487  & 0.00038  & 3.07  & confirmed on OII\tabularnewline
21  & 22 58 32.39  & -34 36 15.5  & 21.1  & 19.2  & 20.8  & 0.35231  & 0.00021  &  & em:{H$\beta$},{H$\alpha$},S1,S2\tabularnewline
23  & 22 58 22.99  & -34 36 26.7  & 22.1  & 20.5  & 20.5  & 0.35371  & 0.00035  &  & measured on H,K (CaI)\tabularnewline
24  & 22 58 28.94  & -34 36 34.7  &  &  & 18.2  & 0.30386  & 0.00045  & 3.01  & very weak\tabularnewline
25  & 22 58 40.69  & -34 36 41.1  &  &  & 20.1  & 0.59540  & 0.00068  &  & measured on H,K (CaI)\tabularnewline
26  & 22 58 28.24  & -34 36 48.7  &  &  & 20.2  & 0.31477  & 0.00024  & 4.07  & \tabularnewline
28  & 22 58 24.28  & -34 37 02.1  &  &  & 18.4  & 0.25761  & 0.00018  & 5.51  & \tabularnewline
29  & 22 58 32.32  & -34 37 07.5  & 21.6  & 20.1  & 20.6  & 0.47283  & 0.00038  & 3.15  & \tabularnewline
30  & 22 58 27.93  & -34 37 16.9  & 21.2  & 20.1  & 21.1  & 0.40221  & 0.00023  &  & em:OII, {H$\beta$},2OIII,{H$\alpha$}\tabularnewline
31  & 22 58 44.38  & -34 37 24.2  & 22.2  & 20.2  & 20.1  & 0.39764  & 0.00040  & 3.04  & em:{H$\beta$},{H$\alpha$}\tabularnewline
33  & 22 58 39.09  & -34 37 52.3  &  &  & 18.8  & 0.34480  &  &  & measured on {H$\alpha$}\tabularnewline
34  & 22 58 39.66  & -34 38 03.8  &  &  & 20.1  & 0.35179  & 0.00011  &  & em:OII,{H$\beta$},2OIII,{H$\alpha$},S1\tabularnewline
35  & 22 58 33.67  & -34 38 18.9  & 21.4  & 20.1  & 18.9  & 0.43091  & 0.00014  &  & em:{H$\beta$},2OIII,{H$\alpha$},S1\tabularnewline
36  & 22 58 31.51  & -34 38 24.2  & 22.3  & 20.7  & 20.5  & 0.20326  & 0.00031  &  & em:{H$\alpha$},S1\tabularnewline
38  & 22 58 41.21  & -34 38 40.5  &  &  & 19.3  & 0.43205  & 0.00029  & 3.58  & em: OII\tabularnewline
39  & 22 58 46.63  & -34 38 45.7  & 21.9  & 20.6  & 20.2  & 0.39935  & 0.00011  &  & em:OII,{H$\beta$},2OIII,{H$\alpha$},S1\tabularnewline
40  & 22 58 34.07  & -34 38 53.1  & 21.9  & 19.9  & 19.3  & 0.31828  & 0.00031  & 4.12  & \tabularnewline
43  & 22 58 39.42  & -34 39 13.5  &  &  & 20.2  & 0.51016  & 0.00028  & 3.32  & \tabularnewline
44  & 22 58 39.45  & -34 39 24.1  & 22.2  & 19.4  & 19.5  & 0.22464  & 0.00017  & 3.34  & em: {H$\alpha$},S1 67344 \kms\tabularnewline
45  & 22 58 43.50  & -34 39 32.3  & 21.6  & 20.6  & 20.1  & 0.62790  & 0.00061  &  & measured on OII,H,K (CaI)\tabularnewline
46  & 22 58 42.70  & -34 39 40.6  &  &  & 20.4  & 0.52153  & 0.00028  & 3.21  & \tabularnewline
49  & 22 58 37.09  & -34 40 06.8  & 19.4  & 18.2  & 18.9  & 0.20657  & 0.00019  & 4.82  & em:{H$\alpha$},S1 61753 \kms\tabularnewline
Quadrant 4  &  &  &  &  &  &  &  &  & \tabularnewline
2  & 22 58 34.99  & -34 42 24.0  &  &  & 19.5  & 0.31982  & 0.00035  & 5.82  & em:{H$\alpha$} weak\tabularnewline
3  & 22 58 45.17  & -34 42 28.9  & 22.6  & 19.1  & 20.1  & 0.41385  & 0.00042  &  & measured on OII,H,K (CaI)\tabularnewline
4  & 22 58 35.53  & -34 42 35.8  &  &  & 19.7  & 0.31108  & 0.00014  & 3.42  & \tabularnewline
\hline 
\end{tabular}
\end{table*}

\begin{table*}
\begin{tabular}{cccccccccl}
\hline 
\textbf{Gal.}  & \textbf{R.A.}  & \textbf{Dec.}  & \textbf{$B_{j}$}  & \textbf{$R_{f}$}  & \textbf{R$_{{\rm comp}}$}  & \textbf{redshift}  & \textbf{error}  & \textbf{R}  & \textbf{notes}\tabularnewline
 & \textbf{(J2000)}  & \textbf{(J2000)}  & \textbf{mag}  & \textbf{mag}  & \textbf{mag}  &  &  &  & \tabularnewline
\hline 
5  & 22 58 42.57  & -34 42 42.5  &  &  & 20.6  & 0.42558  & 0.00012  &  & em:{H$\beta$},2OIII,{H$\alpha$},S1\tabularnewline
7  & 22 58 32.37  & -34 42 55.1  & 20.4  & 19.7  & 20.0  & 0.16952  & 0.00009  &  & em:{H$\beta$},2OIII,{H$\alpha$},S1\tabularnewline
8  & 22 58 32.75  & -34 43 02.0  &  &  & 19.5  & 0.17061  & 0.00022  &  & em:{H$\beta$},2OIII,{H$\alpha$},S1\tabularnewline
9a  & 22 58 31.70  & -34 43 18.8  & 21.0  & 18.9  & 18.4  & 0.29095  & 0.00009  & 10.95  & \tabularnewline
9b  & 22 58 31.70  & -34 43 18.8  &  &  &  & 0.25836  & 0.00035  &  & measured on H,K (CaI)\tabularnewline
10  & 22 58 25.31  & -34 43 25.5  & 20.5  & 18.8  & 19.5  & 0.31228  &  &  & em:{H$\alpha$} very weak\tabularnewline
11  & 22 58 32.43  & -34 43 30.3  & 19.0  & 17.4  & 20.0  & 0.30473  & 0.00008  & 3.02  & em:{H$\alpha$}\tabularnewline
12  & 22 58 26.84  & -34 43 34.1  & 22.1  & 21.0  & 19.7  & 0.31166  & 0.00026  & 3.05  & \tabularnewline
13  & 22 58 26.64  & -34 43 44.0  & 22.4  & 20.5  & 20.9  & 0.47564  & 0.00016  & 3.40  & \tabularnewline
14  & 22 58 41.60  & -34 43 46.5  & 20.5  & 18.9  & 20.7  & 0.30460  & 0.00007  &  & em:OII,{H$\beta$},2OIII,{H$\alpha$},S1\tabularnewline
15  & 22 58 25.91  & -34 43 52.0  & 21.2  & 19.8  & 20.9  & 0.30799  & 0.00030  &  & em:{H$\beta$},2OIII\tabularnewline
16  & 22 58 29.50  & -34 43 55.5  & 22.2  & 19.8  & 20.4  & 0.31854  & 0.00035  & 3.52  & \tabularnewline
18  & 22 58 25.81  & -34 44 06.4  &  &  & 20.6  & 0.32247  & 0.00029  & 3.19  & \tabularnewline
19  & 22 58 26.24  & -34 44 12.1  &  &  & 20.2  & 0.31070  & 0.00011  &  & em:{H$\beta$},2OIII,{H$\alpha$},S1\tabularnewline
20  & 22 58 39.44  & -34 44 18.6  &  &  & 19.6  & 0.30863  & 0.00011  &  & em:{H$\beta$},2OIII,{H$\alpha$},S1\tabularnewline
21  & 22 58 50.72  & -34 44 16.7  &  &  & 19.5  & 0.30605  & 0.00024  & 4.32  & \tabularnewline
22  & 22 58 39.87  & -34 44 28.0  &  &  & 19.4  & 0.31445  & 0.00022  & 5.69  & \tabularnewline
23  & 22 58 35.11  & -34 44 34.8  & 21.9  &  & 20.6  & 0.32268  & 0.00030  & 3.09  & \tabularnewline
24  & 22 58 33.19  & -34 44 39.7  &  &  & 20.5  & 0.31448  & 0.00028  & 3.12  & \tabularnewline
25  & 22 58 30.22  & -34 44 42.4  &  &  & 19.4  & 0.30132  & 0.00035  & 3.20  & \tabularnewline
26  & 22 58 42.00  & -34 44 51.2  & 20.6  & 18.9  & 18.9  & 0.32462  & 0.00012  &  & em:OII,{H$\beta$},2OIII,{H$\alpha$},S1\tabularnewline
27  & 22 58 29.86  & -34 44 55.5  &  &  & 21.4  & 0.31265  & 0.00023  & 3.08  & \tabularnewline
28  & 22 58 26.83  & -34 45 02.6  &  &  & 20.0  & 0.31467  & 0.00032  & 3.53  & \tabularnewline
31  & 22 58 36.47  & -34 45 17.4  & 21.0  & 19.3  & 19.1  & 0.31779  & 0.00024  & 5.33  & \tabularnewline
32  & 22 58 38.00  & -34 45 24.1  & 20.2  & 17.7  & 18.2  & 0.31708  & 0.00024  & 5.65  & \tabularnewline
33  & 22 58 41.30  & -34 45 30.6  & 21.1  & 19.6  & 19.8  & 0.10193  & 0.00029  &  & em:OII,{H$\beta$},2OIII,{H$\alpha$},S1\tabularnewline
35  & 22 58 47.92  & -34 45 45.0  & 21.7  & 19.5  & 20.2  & 0.17040  & 0.00011  &  & em:{H$\beta$},2OIII,{H$\alpha$},S1\tabularnewline
36  & 22 58 35.91  & -34 45 50.1  & 19.8  & 17.7  & 18.4  & 0.31307  & 0.00019  & 5.65  & \tabularnewline
37  & 22 58 29.65  & -34 45 57.5  &  &  & 19.1  & 0.17051  & 0.00018  & 3.01  & uncertain\tabularnewline
38  & 22 58 28.42  & -34 46 02.3  & 22.0  & 19.8  & 19.2  & 0.30100  & 0.00026  & 5.63  & \tabularnewline
40  & 22 58 24.92  & -34 46 18.7  & 20.8  & 19.0  & 19.0  & 0.31779  & 0.00004  &  & em:{H$\alpha$}\tabularnewline
41  & 22 58 27.85  & -34 46 26.3  &  &  & 19.7  & 0.31816  & 0.00033  & 4.55  & \tabularnewline
42  & 22 58 29.75  & -34 46 32.1  & 22.2  & 20.5  & 19.9  & 0.50228  & 0.00035  & 3.04  & \tabularnewline
43  & 22 58 28.88  & -34 46 40.3  &  &  & 20.2  & 0.31717  & 0.00033  & 3.68  & \tabularnewline
44  & 22 58 40.93  & -34 46 45.7  & 20.6  & 18.6  & 18.3  & 0.31523  & 0.00030  & 4.73  & \tabularnewline
45  & 22 58 36.47  & -34 46 51.2  &  &  & 20.6  & 0.31389  & 0.00019  & 4.66  & \tabularnewline
46  & 22 58 51.00  & -34 47 00.6  & 20.8  & 19.5  & 20.1  & 0.32953  & 0.00028  & 3.10  & \tabularnewline
47  & 22 58 34.80  & -34 47 07.1  & 22.2  &  & 19.3  & 0.31674  & 0.00019  & 6.24  & \tabularnewline
48  & 22 58 29.14  & -34 47 13.1  & 21.6  & 20.0  & 20.0  & 0.31462  & 0.00022  & 5.29  & \tabularnewline
49  & 22 58 39.56  & -34 47 17.4  & 21.8  & 19.1  & 20.0  & 0.31818  & 0.00024  & 4.89  & \tabularnewline
50  & 22 58 48.11  & -34 47 21.3  &  &  & 20.2  & 0.33032  & 0.00039  & 3.02  & very weak\tabularnewline
51  & 22 58 46.31  & -34 47 31.5  & 20.8  & 18.8  & 19.1  & 0.30954  & 0.00019  & 6.32  & \tabularnewline
52  & 22 58 26.11  & -34 47 37.5  & 22.6  & 20.9  & 18.2  & 0.31789  & 0.00042  & 3.04  & confirmed on H,K (CaI)\tabularnewline
54  & 22 58 33.80  & -34 47 46.6  & 22.3  & 19.5  & 18.9  & 0.34779  & 0.00030  & 3.06  & weak\tabularnewline
55a  & 22 58 33.96  & -34 47 53.3  &  &  & 20.2  & 0.31776  & 0.00031  & 3.26  & \tabularnewline
55b  & 22 58 33.96  & -34 47 53.3  &  &  &  & 0.17199  &  &  & em:{H$\beta$},2OIII,{H$\alpha$},S1\tabularnewline
56  & 22 58 44.05  & -34 47 59.4  &  &  & 18.4  & 0.31380  & 0.00015  &  & measured on H,K (CaI)\tabularnewline
57  & 22 58 49.74  & -34 48 01.9  & 21.6  & 19.7  & 19.1  & 0.31451  & 0.00028  & 3.35  & \tabularnewline
58  & 22 58 41.65  & -34 48 06.7  & 21.9  & 19.2  & 19.2  & 0.31082  & 0.00028  & 6.10  & \tabularnewline
60  & 22 58 44.45  & -34 48 21.3  &  &  & 19.0  & 0.32473  & 0.00031  & 4.42  & \tabularnewline
61  & 22 58 38.22  & -34 48 27.8  & 20.7  & 18.6  & 19.7  & 0.72072  & 0.00014  & 3.05  & \tabularnewline
62  & 22 58 42.81  & -34 48 33.4  & 21.0  & 18.9  & 29.9  & 0.31090  & 0.00047  & 5.43  & \tabularnewline
63  & 22 58 52.67  & -34 48 39.1  &  &  & 20.2  & 0.31874  & 0.00024  & 4.98  & \tabularnewline
64  & 22 58 31.46  & -34 48 53.3  &  &  & 18.3  & 0.31447  & 0.00030  & 5.28  & \tabularnewline
65  & 22 58 34.69  & -34 49 01.3  &  &  & 20.6  & 0.47788  & 0.00023  & 5.33  & \tabularnewline
\hline 
\end{tabular}\bigskip{}
 
\end{table*}





\clearpage{}

\appendix




\section{Spectra of galaxies identified as active based on {[}\ion{O}{II}{]}$\lambda3727$\label{sec:Spectra-of-galaxies}}

\begin{figure}
\includegraphics[width=1\columnwidth]{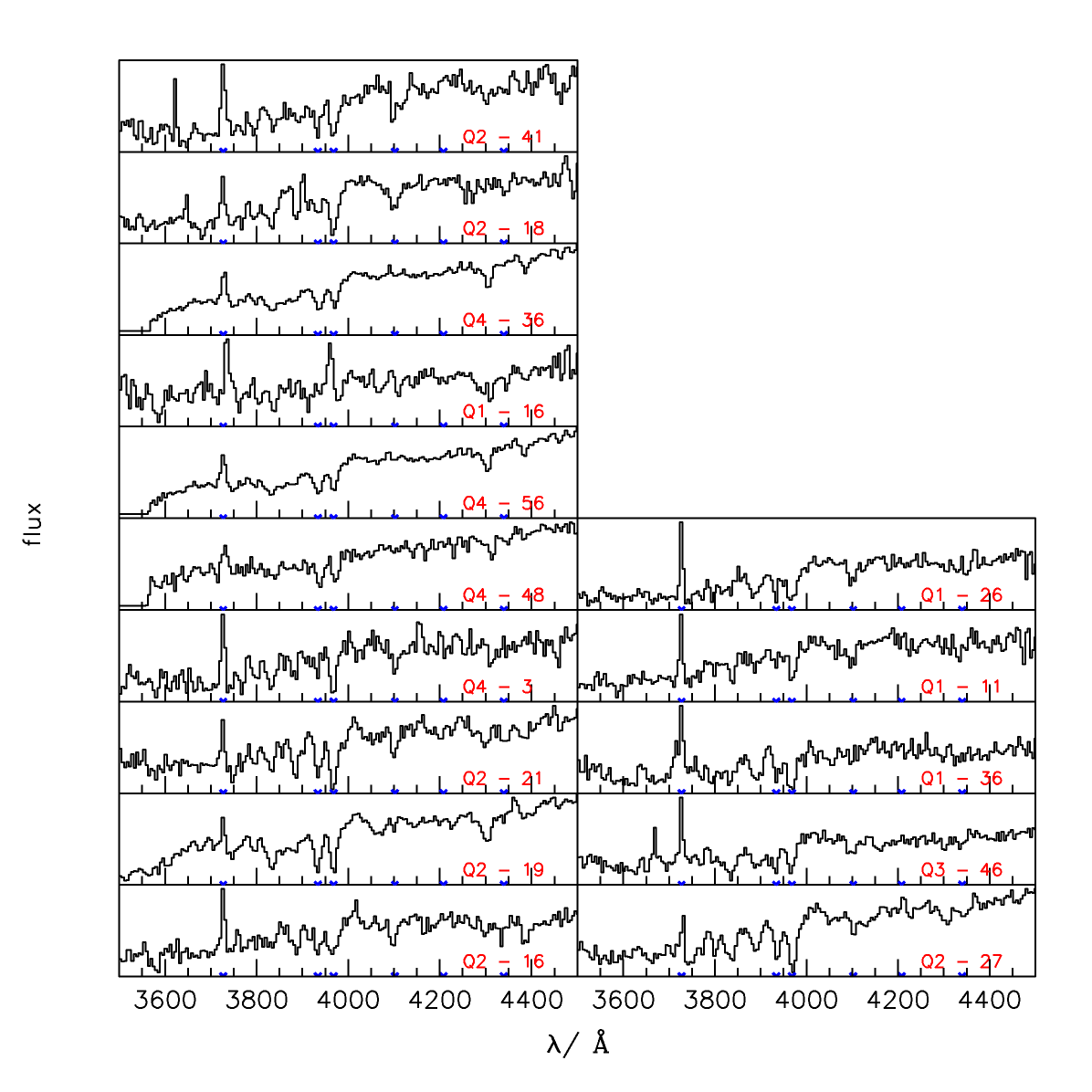}

\caption{Spectra of galaxies identified as active based on {[}\ion{O}{II}{]}$\lambda3727$.
\label{fig:Spectra-of-galaxies-3727}}
\end{figure}

\clearpage{}

\section{Postage stamps\label{sec:Postage-stamps}}

\begin{figure*}
\begin{centering}
\includegraphics[width=1\textwidth]{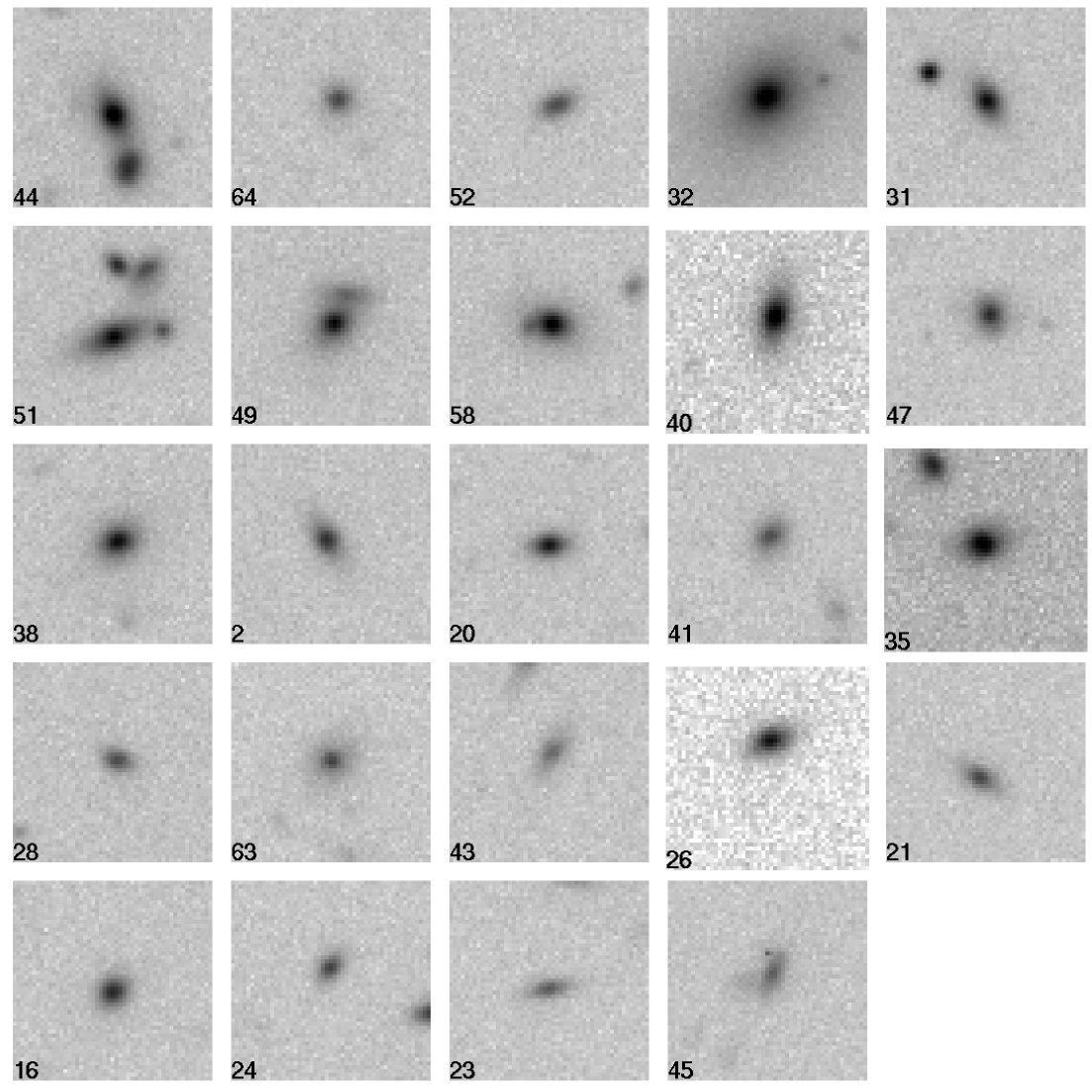} 
\par\end{centering}
\caption{Images of galaxies classified as passive, from the VIMOS preimaging,
sorted in luminosity from top to bottom. The number identifies the
slit, while the quadrant can be evinced from Table~\ref{tab:Main-characteristics-of-passive}.
Each frame is 10.5$\arcsec\times$ 10.5$\arcsec$, and the image quality
is $\sim1^{\prime\prime}$~FWHM. Most galaxies have regular shapes,
as expected from E/SO types, but notable exceptions are galaxies 44,
51, 49, 53, and 58, which appear to be interacting with smaller neighbours.
\label{fig:Images-of-galaxies-passive}}
\end{figure*}

\begin{figure*}
\begin{centering}
\includegraphics[width=1\textwidth]{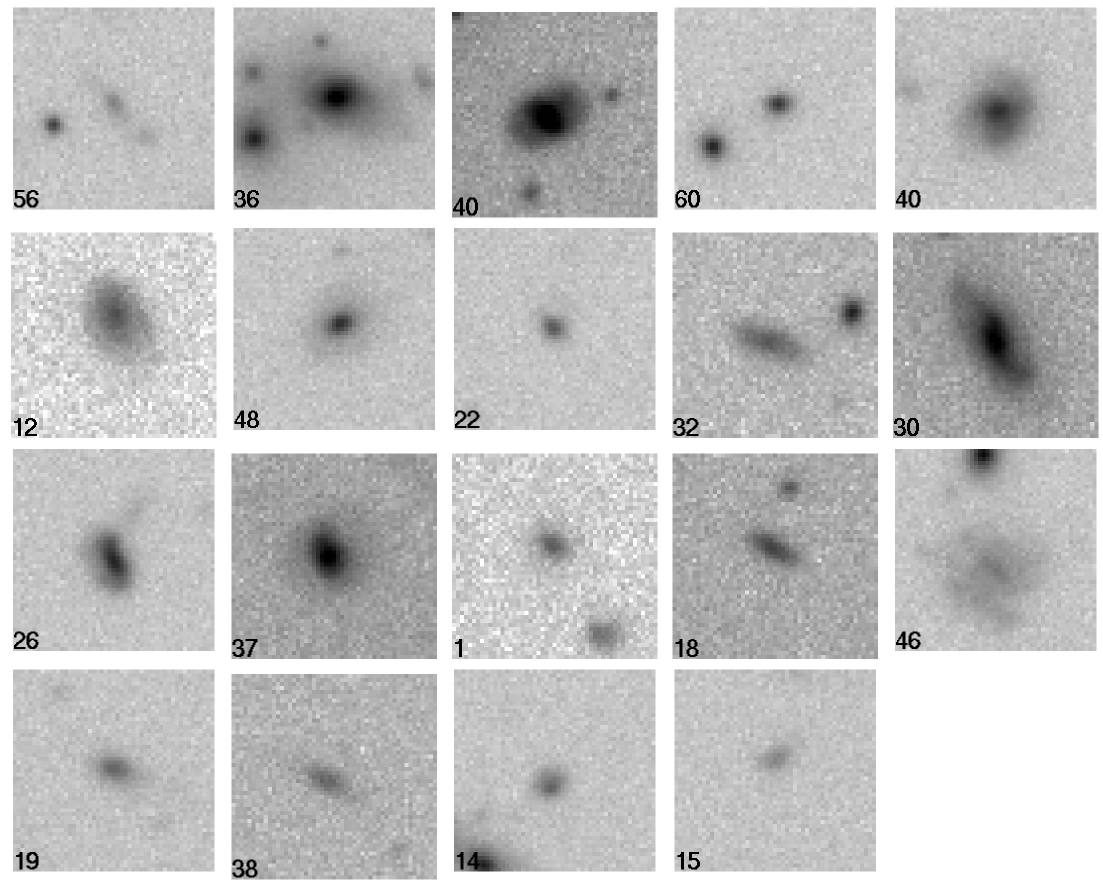} 
\par\end{centering}
\caption{Same as Fig.~\ref{fig:Images-of-galaxies-passive} for galaxies classified
as active. \label{fig:Images-of-galaxies-active}}
\end{figure*}

\clearpage{}

\section{Galaxies in common with CS87\label{sec:Galaxies-in-common-cs87}}

\begin{figure}
\includegraphics[width=1\columnwidth]{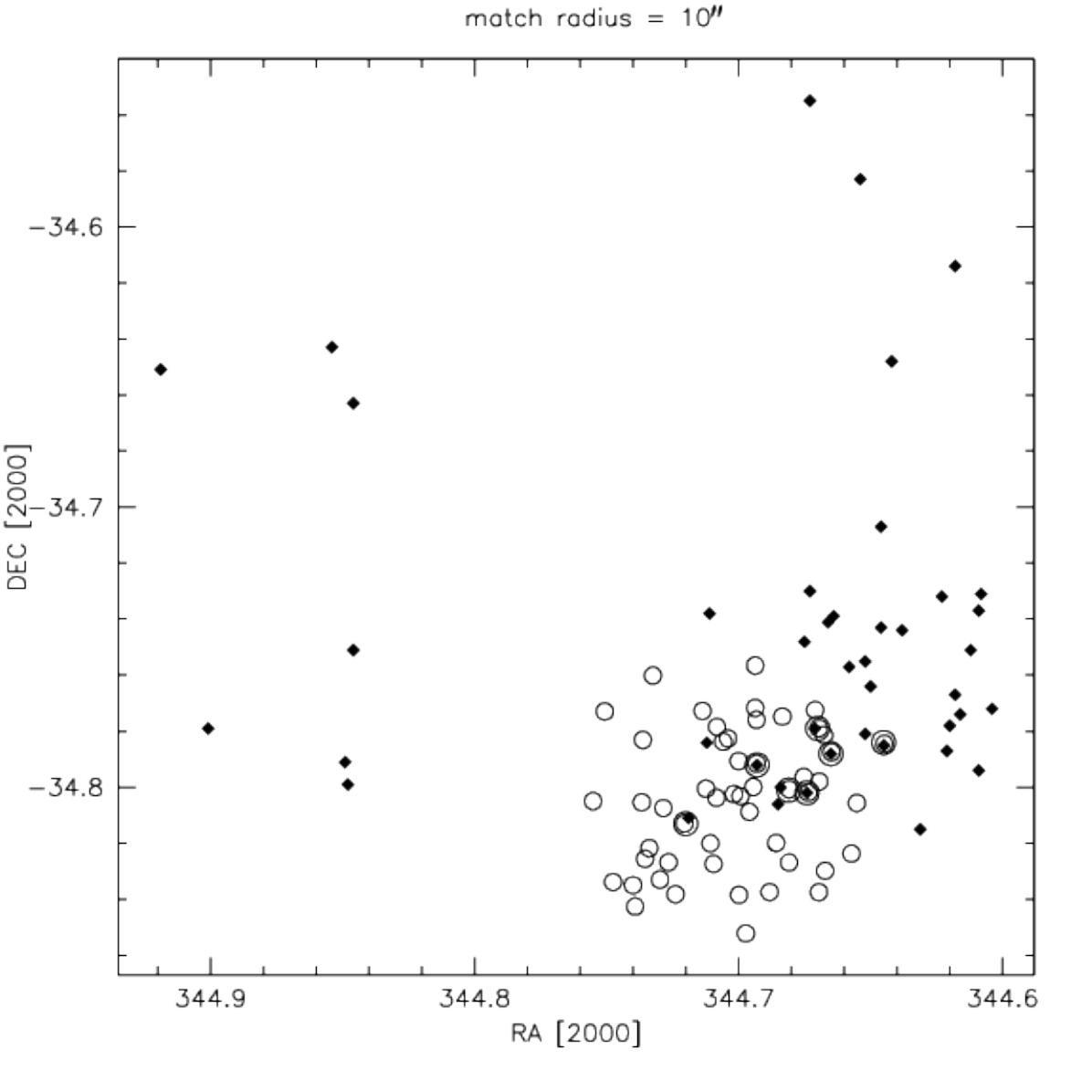}

\caption{Representation on sky of galaxies from this work (black symbols),
and from Couch and Sharples (1987; open circles). After transforming
coordinates from B1950.0 to J2000.0, objects were matched with a radius
of $10^{\prime\prime}$. Matching galaxies are marked with a double
circle. \label{fig:Representation-on-sky-common}}

\end{figure}

\begin{table*}
\caption{Parameters of galaxies in common between this work and \citet{Couch1987a}.
Columns 1 and 2 are the identifications in the respective works; columns
3, 4 and 5, are coordinates in this work and the separation in arcsec;
columns 6 and 7 are the redshift measured in this work and the difference
with CS87; columns 8, 9, and 10 are the colours and magnitude from
CS87 and the difference in magnitude with the present work; column
11 is the cross-correlation quality from CS87; columns 12 and 13 are
the {[}O\textsc{ii}{]}$\lambda3727$ and H$\delta$ lines measurements
from CS87; columns 14 and 15 are the metallicity and error from this
work; and finally the last column is our galaxy classification, as
(p)assive or (a)ctive. An empty line separates possible matches from
unlikely ones. \label{tab:Parameters-of-galaxies-common}}

\begin{tabular}{llllrlr@{\extracolsep{0pt}.}lllr@{\extracolsep{0pt}.}lllllll}
 &  &  &  &  &  & \multicolumn{2}{c}{} &  &  & \multicolumn{2}{c}{} &  &  &  &  &  & \tabularnewline
\hline 
\hline 
S23 & CS87 & RA{[}2000{]} & DEC{[}2000{]} & d$^{\prime\prime}$ & z & \multicolumn{2}{c}{$\Delta z$} & B$_{{\rm J}}-$R$_{{\rm F}}$ & R$_{{\rm F}}$ & \multicolumn{2}{c}{$\Delta R$} & r & {[}O\textsc{ii}{]} & H$\delta$ & {[}m/H{]} & $\epsilon${[}m/H{]} & flag\tabularnewline
1 & 2 & 3 & 4 & 5 & 6 & \multicolumn{2}{c}{7} & 8 & 9 & \multicolumn{2}{c}{10} & 11 & 12 & 13 & 14 & 15 & 16\tabularnewline
\hline 
 &  &  &  &  &  & \multicolumn{2}{c}{} &  &  & \multicolumn{2}{c}{} &  &  &  &  &  & \tabularnewline
Q4-51 & C5 & 344.693 & -34.792 & 0.0 & 0.310 & -0&0011 & 2.20 & 19.19 & -0&09 & 6.0 &  &  & -0.29 & 0.36 & p\tabularnewline
Q4-58 & C145 & 344.674 & -34.802 & 0.0 & 0.311 & -0&0015 & 2.37 & 19.18 & 0&02 & 7.1 &  & 1.3 & -0.40 & 0.24 & p\tabularnewline
Q4-49 & C247 & 344.665 & -34.788 & 0.0 & 0.318 & -0&0015 & 2.46 & 19.12 & 0&88 & 7.6 &  & 2.8 & -0.18 & 0.26 & p\tabularnewline
Q4-47 & C832 & 344.645 & -34.784 & 3.6 & 0.317 & -0&0011 & 2.25 & 19.98 & -0&68 & 5.3 &  & 3.6 & -0.57 & 0.27 & p\tabularnewline
 &  &  &  &  &  & \multicolumn{2}{c}{} &  &  & \multicolumn{2}{c}{} &  &  &  &  &  & \tabularnewline
Q4-44 & C243 & 344.670 & -34.779 & 2.9 & 0.315 & -0&0100 & 1.58 & 19.57 & -1&27 & 2.2 & 19.4 & 5.2 & -0.08 & 0.08 & p\tabularnewline
Q4-63 & C191 & 344.720 & -34.813 & 7.8 & 0.319 & 0&0144 & 1.63 & 19.69 & 0&51 & 4.7 &  & 5.6 & -1.36 & 0.35 & p\tabularnewline
Q4-56 & C134 & 344.681 & -34.801 & 9.5 & 0.314 & 0&2172 & 0.86 & 19.66 & -1&26 &  &  &  & -0.04 & 0.15 & a\tabularnewline
 &  &  &  &  &  & \multicolumn{2}{c}{} &  &  & \multicolumn{2}{c}{} &  &  &  &  &  & \tabularnewline
\hline 
\end{tabular}
\end{table*}

\begin{figure}
\includegraphics[width=1\columnwidth]{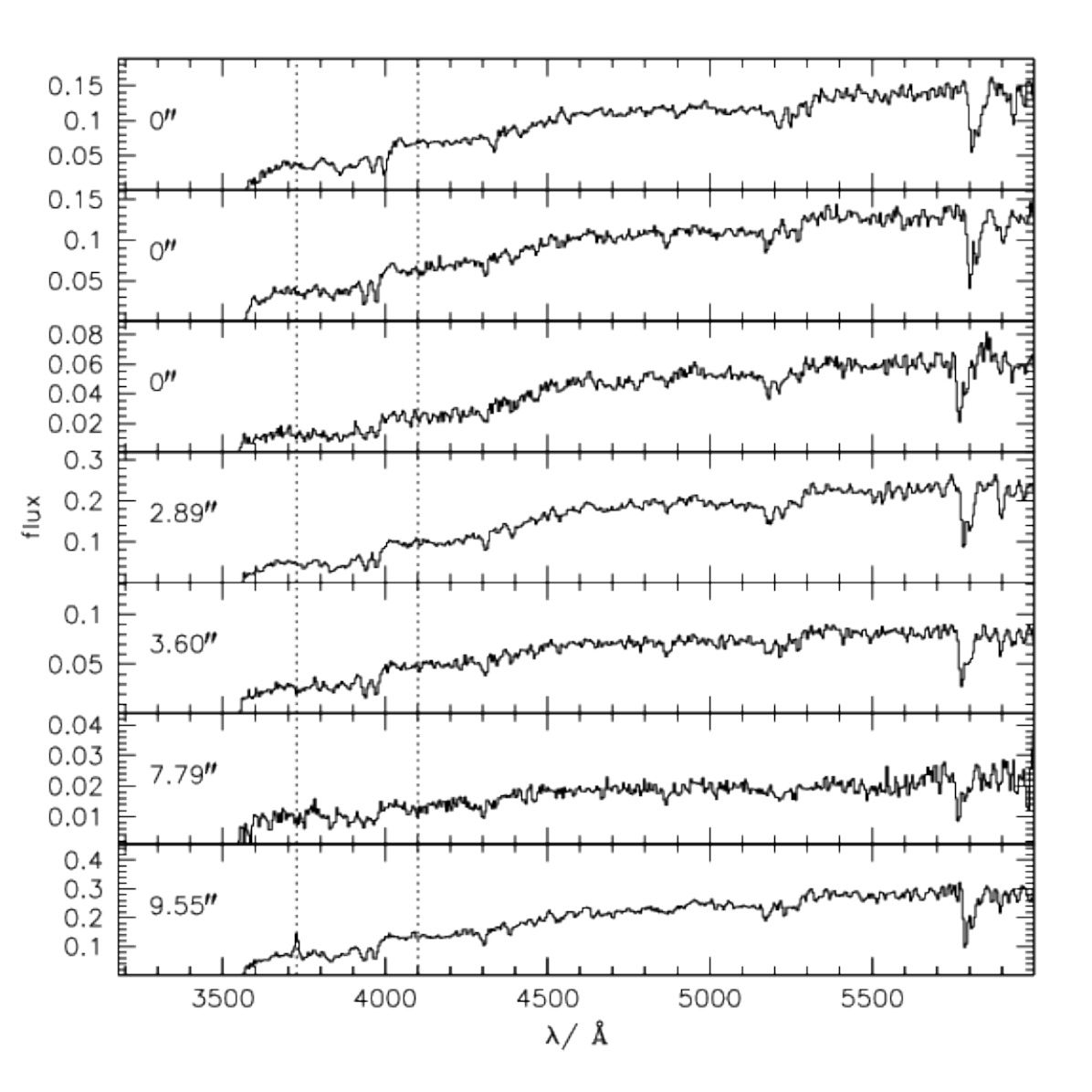}

\caption{Spectra of galaxies in common with CS87. The separation on sky is
displayed in each panel, with separation increasing from top to bottom.
Vertical dotted lines mark the position of {[}O\textsc{ii}{]}$\lambda3727$
and H$\delta$. \label{fig:Spectra-of-galaxies-common}}

\end{figure}

After transforming from the B1950.0 to the J2000.0\footnote{Using the NED coordinate calculator https://ned.ipac.caltech.edu/}
system, coordinates were used together with a matching radius of 10
arcsec, to find common objects between this work and \citet{Couch1987a}.
The result is shown in Fig.~\ref{fig:Representation-on-sky-common}
and is summarised in Table~\ref{tab:Parameters-of-galaxies-common},
where galaxies are listed from top to bottom in order of increasing
separation on sky. There are only a few galaxies in common, and in
particular only three have null separation on sky; galaxy Q4-47 might
also be a possible match, because the difference in redshift is comparable
to the first three. The other three galaxies are likely not the same
object in the two works, because of the large difference in redshift.
Besides, {[}O\textsc{ii}{]}$\lambda3727$ is detected in galaxy C243
but it is absent in our Q4-44; and conversely, it is not present in
C134 but we detect it in Q4-56, as Fig.~\ref{fig:Spectra-of-galaxies-common}
shows. Indeed, Q4-56 is classified as `active' as we have shown
above. 

Based on their colours, all four common galaxies belong to the `red
sequence' in CS87 and have normal H$\delta$ absorption, with the
exception of C832 which has a marginally stronger absorption. Figure~\ref{fig:Spectra-of-galaxies-common}
shows the spectra of the seven galaxies, and it confirms the absence
of emission lines in the four best matches.

\clearpage{}

\section{Summary of classification of all galaxies\label{sec:Summary-of-classification}}

The results of Section~\ref{subsec:Classification-of-galaxies} are
summarised in Table~\ref{tab:Classification-of-galaxies}. In the
table, and from left to right, columns contain an ordinal number,
the VIMOS quadrant and slit, the redshift, the classification flag
(a)ctive or (p)assive, and notes: this scheme is repeated in three
blocks. Notes are the following: {[}Oii{]} = galaxy identified by
the presence of {[}\ion{O}{II}{]}$\lambda3727$; OI = bad sky
subtraction of {[}\ion{O}{I}{]}$\lambda5579$ oxygen line, which
creates a spurious {[}\ion{O}{II}{]}$\lambda3727$ emission line
for objects at redshift $\simeq0.4967$; M-dwarf = object is actually
a local star; ::~=~spectrum is uncertain and is not used in this
paper.

\begin{table*}
\caption{Classification of galaxies as described in Section \ref{subsec:Classification-of-galaxies}.
\label{tab:Classification-of-galaxies}}

\begin{tabular}{llllllllllllllllll}
 &  &  &  &  &  &  &  &  &  &  &  &  &  &  &  &  & \tabularnewline
\hline 
\hline 
n & q & sl & z & a/p & notes & n & q & sl & z & a/p & notes & n & q & sl & z & a/p & notes\tabularnewline
\hline 
 &  &  &  &  &  &  &  &  &  &  &  &  &  &  &  &  & \tabularnewline
1 & 1 & 1 & 0.3985 & a & {[}Oii{]} & 56 & 2 & 29 & 0.523 & a &  & 110 & 4 & 7 & 0.1695 & a & \tabularnewline
2 & 1 & 2 & 0.3531 & a &  & 57 & 2 & 31 & 0.4972 & a &  & 111 & 4 & 8 & 0.1706 & a & \tabularnewline
3 & 1 & 3 & 0.05717 & a &  & 58 & 2 & 32 & 0.3091 & a &  & 112 & 4 & 9 & 0.2584 & p & \tabularnewline
4 & 1 & 5 & 0.05842 & a &  & 59 & 2 & 33 & 0.5797 & a &  & 113 & 4 & 10 & 0.3123 & p & \tabularnewline
5 & 1 & 8 & 0.3806 & a &  & 60 & 2 & 35 & 0.3129 & p &  & 114 & 4 & 11 & 0.3047 & p & \tabularnewline
6 & 1 & 10 & 0.2194 & p &  & 61 & 2 & 36 & 0.4637 & a &  & 115 & 4 & 12 & 0.3117 & p & \tabularnewline
7 & 1 & 11 & 0.626 & a & {[}Oii{]} & 62 & 2 & 37 & 0.1396 & a &  & 116 & 4 & 13 & 0.4756 & p & \tabularnewline
8 & 1 & 12 & 0.2213 & a &  & 63 & 2 & 38 & 0.2987 & a &  & 117 & 4 & 14 & 0.3046 & a & \tabularnewline
9 & 1 & 13 & 0.4098 & p &  & 64 & 2 & 39 & 0.372 & a &  & 118 & 4 & 15 & 0.308 & a & \tabularnewline
10 & 1 & 15 & 0.07309 & a &  & 65 & 2 & 40 & 0.4191 & p &  & 119 & 4 & 16 & 0.3185 & a & ::\tabularnewline
11 & 1 & 16 & 0.4082 & a & {[}Oii{]} & 66 & 2 & 41 & 0.5403 & a & {[}Oii{]} & 120 & 4 & 18 & 0.527 & p & \tabularnewline
12 & 1 & 17 & 0.6316 & p &  & 67 & 3 & 1 & 0.3295 & a &  & 121 & 4 & 19 & 0.3107 & a & \tabularnewline
13 & 1 & 18 & 0.313 & a &  & 68 & 3 & 2 & 0.3331 & a &  & 122 & 4 & 20 & 0.3086 & p & \tabularnewline
14 & 1 & 19 & 0.4088 & a &  & 69 & 3 & 3 & 0.3313 & p &  & 123 & 4 & 21 & 0.3061 & p & \tabularnewline
15 & 1 & 20 & 0.2319 & p &  & 70 & 3 & 4 & 0.2126 & a &  & 124 & 4 & 22 & 0.3145 & a & \tabularnewline
16 & 1 & 22 & 0.4102 & a &  & 71 & 3 & 5 & 0.3319 & a &  & 125 & 4 & 23 & 0.3227 & p & \tabularnewline
17 & 1 & 24 & 0.3337 & p &  & 72 & 3 & 6 & 0.4968 & p & OI & 126 & 4 & 24 & 0.3145 & p & \tabularnewline
18 & 1 & 25 & 0.486 & p &  & 73 & 3 & 8 & 0.3997 & p & :: & 127 & 4 & 25 & 0.3013 & p & \tabularnewline
19 & 1 & 26 & 0.5335 & a & {[}Oii{]} & 74 & 3 & 10 & 0.3137 & p & :: & 128 & 4 & 26 & 0.3246 & a & \tabularnewline
20 & 1 & 27 & 0.4662 & p &  & 75 & 3 & 11 & 0.2125 & a &  & 129 & 4 & 27 & 0.3127 & p & \tabularnewline
21 & 1 & 28 & 0.3512 & p &  & 76 & 3 & 12 & 0.3094 & a &  & 130 & 4 & 28 & 0.3147 & p & \tabularnewline
22 & 1 & 30 & 0.3186 & a & {[}Oii{]} & 77 & 3 & 13 & 0.6515 & a &  & 131 & 4 & 31 & 0.3178 & p & \tabularnewline
23 & 1 & 31 & 0.3334 & p &  & 78 & 3 & 15 & 0.4756 & p &  & 132 & 4 & 32 & 0.3171 & p & \tabularnewline
24 & 1 & 34 & 0.6319 & p &  & 79 & 3 & 16 & 0.1774 & a &  & 133 & 4 & 33 & 0.1019 & a & \tabularnewline
25 & 1 & 36 & 0.4998 & a & {[}Oii{]} & 80 & 3 & 17 & 0.4967 & p & OI & 134 & 4 & 35 & 0.1704 & a & \tabularnewline
26 & 1 & 37 & 0.316 & a &  & 81 & 3 & 18 & 0.2218 & a &  & 135 & 4 & 36 & 0.3131 & a & {[}Oii{]}\tabularnewline
27 & 1 & 38 & 0.498 & p &  & 82 & 3 & 19 & 0.1586 & p &  & 136 & 4 & 37 & 0.1705 & p & \tabularnewline
28 & 1 & 40 & 0.3228 & a &  & 83 & 3 & 20 & 0.3549 & a &  & 137 & 4 & 38 & 0.301 & p & \tabularnewline
29 & 1 & 41 & 0.4764 & p &  & 84 & 3 & 21 & 0.3523 & p & :: & 138 & 4 & 39 & 0.3259 & p & M-dwarf\tabularnewline
30 & 1 & 44 & 0.5778 & a &  & 85 & 3 & 23 & 0.3537 & p &  & 139 & 4 & 40 & 0.3178 & a & \tabularnewline
31 & 2 & 1 & 0.2578 & a &  & 86 & 3 & 24 & 0.3039 & p & :: & 140 & 4 & 41 & 0.3182 & p & \tabularnewline
32 & 2 & 2 & 0.4958 & a &  & 87 & 3 & 25 & 0.5954 & p &  & 141 & 4 & 42 & 0.5023 & p & \tabularnewline
33 & 2 & 3 & 0.41 & a &  & 88 & 3 & 26 & 0.3148 & p &  & 142 & 4 & 43 & 0.3172 & p & \tabularnewline
34 & 2 & 4 & 0.4085 & a &  & 89 & 3 & 27 & 0.4969 & p & OI & 143 & 4 & 44 & 0.3152 & p & \tabularnewline
35 & 2 & 5 & 0.4144 & p &  & 90 & 3 & 28 & 0.2576 & p &  & 144 & 4 & 45 & 0.3139 & p & \tabularnewline
36 & 2 & 6 & 0.4228 & p &  & 91 & 3 & 29 & 0.4728 & a & :: & 145 & 4 & 46 & 0.3295 & a & {[}Oii{]}\tabularnewline
37 & 2 & 7 & 0.5618 & p &  & 92 & 3 & 30 & 0.4022 & a &  & 146 & 4 & 47 & 0.3167 & p & \tabularnewline
38 & 2 & 8 & 0.2574 & a &  & 93 & 3 & 31 & 0.332 & a &  & 147 & 4 & 48 & 0.3146 & a & {[}Oii{]}\tabularnewline
39 & 2 & 9 & 0.3708 & a &  & 94 & 3 & 33 & 0.3448 & p &  & 148 & 4 & 49 & 0.3182 & p & \tabularnewline
40 & 2 & 10 & 0.4085 & p & :: & 95 & 3 & 34 & 0.3518 & a &  & 149 & 4 & 50 & 0.3303 & p & \tabularnewline
41 & 2 & 11 & 0.3455 & p & :: & 96 & 3 & 35 & 0.4309 & a &  & 150 & 4 & 51 & 0.3095 & p & \tabularnewline
42 & 2 & 12 & 0.5952 & p &  & 97 & 3 & 36 & 0.2033 & a &  & 151 & 4 & 52 & 0.3179 & p & \tabularnewline
43 & 2 & 13 & 0.2701 & p &  & 98 & 3 & 38 & 0.4321 & p & :: & 152 & 4 & 54 & 0.3478 & p & \tabularnewline
44 & 2 & 14 & 0.3737 & a & {[}Oii{]} & 99 & 3 & 39 & 0.3994 & a &  & 153 & 4 & 55 & 0.172 & a & \tabularnewline
45 & 2 & 15 & 0.4105 & p &  & 100 & 3 & 40 & 0.3183 & p &  & 154 & 4 & 56 & 0.3138 & a & {[}Oii{]}\tabularnewline
46 & 2 & 16 & 0.5682 & a & {[}Oii{]} & 101 & 3 & 43 & 0.5101 & p & :: & 155 & 4 & 57 & 0.3145 & a & ::\tabularnewline
47 & 2 & 18 & 0.6152 & a & {[}Oii{]} & 102 & 3 & 44 & 0.2246 & p & :: & 156 & 4 & 58 & 0.3108 & p & \tabularnewline
48 & 2 & 19 & 0.3771 & a & {[}Oii{]} & 103 & 3 & 45 & 0.6279 & p &  & 157 & 4 & 59 & 0.3183 & p & M-dwarf\tabularnewline
49 & 2 & 21 & 0.413 & a & {[}Oii{]} & 104 & 3 & 46 & 0.5215 & a & {[}Oii{]} & 158 & 4 & 60 & 0.3183 & a & \tabularnewline
50 & 2 & 22 & 0.391 & a &  & 105 & 3 & 49 & 0.2066 & p &  & 159 & 4 & 61 & 0.7207 & p & \tabularnewline
51 & 2 & 23 & 0.4131 & p &  & 106 & 4 & 2 & 0.3198 & p &  & 160 & 4 & 62 & 0.2838 & p & \tabularnewline
52 & 2 & 24 & 0.3797 & a &  & 107 & 4 & 3 & 0.4139 & a & {[}Oii{]} & 161 & 4 & 63 & 0.3187 & p & \tabularnewline
53 & 2 & 26 & 0.4868 & a &  & 108 & 4 & 4 & 0.3111 & p &  & 162 & 4 & 64 & 0.3145 & p & \tabularnewline
54 & 2 & 27 & 0.4964 & a & {[}Oii{]} & 109 & 4 & 5 & 0.4135 & a &  & 163 & 4 & 65 & 0.4779 & p & \tabularnewline
55 & 2 & 28 & 0.6136 & p & :: &  &  &  &  &  &  &  &  &  &  &  & \tabularnewline
 &  &  &  &  &  &  &  &  &  &  &  &  &  &  &  &  & \tabularnewline
\hline 
\end{tabular}

\end{table*}

\label{lastpage} 
\end{document}